\documentclass[12pt]{article}
\usepackage{scicite,graphicx}
\usepackage{times}
\usepackage{sci_macros}
\usepackage[normalem]{ulem}
\usepackage{soul}
\usepackage{multirow}
\usepackage{array}
\usepackage{booktabs}
\usepackage{lineno}
\usepackage{amsmath}
\usepackage{bm}

\usepackage[colorlinks]{hyperref}
\usepackage{subfig}
\usepackage[font=footnotesize]{caption}

\usepackage{scicite,graphicx}
\usepackage[hyphenbreaks]{breakurl}

\usepackage[symbol]{footmisc}

\newenvironment{sciabstract}{%
\begin{quote} }
{\end{quote}}

\title{Very high energy gamma-ray emission beyond 10 TeV  from GRB 221009A}

\date{}

\begin{document}


\baselineskip 18pt


\maketitle

\centerline{\author{\large The LHAASO Collaboration\footnotemark[1]\footnotemark[2]}}
\vspace{10pt}

\footnotetext[1]{Corresponding authors: S.Z.~Chen (chensz@ihep.ac.cn), X.J.~Bi (bixj@ihep.ac.cn), \\ \indent\indent S.C.~Hu (hushicong@ihep.ac.cn),  X.Y.~Wang (xywang@nju.edu.cn)}
\footnotetext[2]{The LHAASO Collaboration
authors and affiliations are listed in the supplementary materials}

\begin{sciabstract}
\bf
The highest energy gamma-rays from gamma-ray bursts (GRBs) have important implications for their radiation mechanism. Here we report for the first time the detection of gamma-rays up to 13 TeV from the brightest GRB 221009A by the Large High Altitude Air-shower Observatory (LHAASO). The LHAASO-KM2A detector registered more than 140 gamma-rays with energies above 3 TeV during 230$-$900s after the trigger.  The intrinsic energy spectrum of gamma-rays can be described by a power-law after correcting for extragalactic background light (EBL) absorption. Such a hard spectrum challenges the synchrotron self-Compton (SSC) scenario of relativistic electrons for the afterglow emission above several TeV. Observations of gamma-rays up to 13 TeV from a source with a measured redshift of z=0.151   hints more transparency in intergalactic space than previously expected. Alternatively, one may invoke new physics such as Lorentz Invariance Violation (LIV) or an axion origin of very high energy (VHE) signals.
\end{sciabstract}

{\bf One sentence summary}: LHAASO detected  the highest energy gamma-rays, beyond  10 TeV, from  the brightest gamma-ray burst GRB 221009A.

\section{Introduction}
Gamma-ray bursts (GRBs) are sudden explosions of gamma-rays that occur in random directions from cosmological distances. They are the most luminous fireworks in the Universe and are thought to originate from collapsing massive stars and compact star mergers. Prompt flashes and long-lasting afterglows have been thoroughly studied, with thousands of GRBs observed in a wide range of energies from radio waves to megaelectronvolt (MeV) gamma-rays. A fraction of GRBs are also observed with gigaelectronvolt (GeV) gamma-rays. Recently, a handful of GRBs have been observed in teraelectronvolt (TeV)  gamma-rays during the afterglow period \cite{2019Natur.575..455M,2019Natur.575..459M,2019Natur.575..464A,2021Sci...372.1081H}. The synchrotron self-Compton (SSC) process of relativistic electrons in the afterglow has been proposed to explain the  origin of the TeV emission\cite{2019ApJ...880L..27D,2019ApJ...884..117W,2019Natur.575..459M}. However,  challenges \cite{2021Sci...372.1081H} have been reported for such a scenario. Observation of the highest energy emission from GRBs is important to probe the gamma-ray emission mechanism and particle acceleration processes in GRBs.

The exceptionally bright GRB 221009A (RA=288.264$^{\circ}$, Dec=19.768$^{\circ}$) was detected by Fermi-GBM on October 9, 2022, at 13:16:59.99 UT (denoted as T$_{0}$ hereafter)\cite{2023arXiv230314172L}, about one hour earlier than the Swift trigger\cite{2022GCN.32635....1K}. The GBM light curve consists of two emission episodes: a single isolated peak from T$_{0}$ to T$_{0}$+20s, followed by a longer, extremely bright, multi-pulsed emission episode from about T$_{0}$+220s to T$_{0}$+550s\cite{2023arXiv230314172L}. It was also detected by the Large Area Telescope (Fermi-LAT) with the highest-energy gamma-ray of 99.3 GeV observed 240 seconds after the GBM trigger \cite{2022GCN.32658....1P}. This GRB was also clearly observed by many other detectors, such as Insight-HXMT, GECAM-C \cite{2023arXiv230301203A}, and Konus-Wind \cite{2023arXiv230213383F}. Follow-up optical observations revealed that the redshift of this GRB is $z$=0.151 (corresponding to a distance $\sim$753 Mpc)\cite{2022GCN.32648....1D}. The estimated isotropic energy release is about $10^{55}$ ergs\cite{2023arXiv230314172L,2023arXiv230301203A,2023arXiv230213383F}, which denotes it as an extremely energetic GRB.

\section{Results}

LHAASO observed GRB 221009A at the highest energy band. LHAASO\cite{2022ChPhC..46c0001M}
consists of three interconnected detectors located at 4,410 m above sea level in Sichuan Province, China. The sub-arrays, a 78,000 m$^2$ Water Cherenkov Detector Array (WCDA) and a 1.3 km$^2$ Kilometer Squared Array (KM2A), are dedicated to gamma-ray observations with a wide field-of-view (FOV). Both arrays can monitor the sky for zenith angles less than 50$^{\circ}$.
GRB 221009A entered the FOV of LHAASO at around T$_{0}$-20000s, culminating at a zenith angle of 9.5$^{\circ}$ at T$_{0}$-7000s, and left the FOV at T$_{0}$+6000s.
For the first time for a TeV-detected GRB, the observations cover both the prompt period and the afterglow period.
The zenith angles were 28$^{\circ}$ at T$_{0}$ and 31.5$^{\circ}$ at T$_{0}$+1000s, which are favorable positions for LHAASO observation. Within 2000s after T$_{0}$, more than 5000 gamma-rays at $>$0.5 TeV were detected by WCDA from  GRB 221009A with significance above 100 s.d.\cite{2022GCN.32677....1H}.
A previous report has focused on the temporal characteristics of the TeV emission based on WCDA data at lower energies, which revealed the earliest TeV afterglow from GRB 221009A\cite{2023ScienceGRB}.
The light curve includes a sharp rise at T$_{0}$+230s reaching the peak at about T$_{0}$+245s and a following smooth decay with the jet break.
Here, we  focus on the spectral results extending to energies above 10 TeV, as measured by KM2A.

\begin{figure}
\centering
\includegraphics[width=10cm,height=5cm]{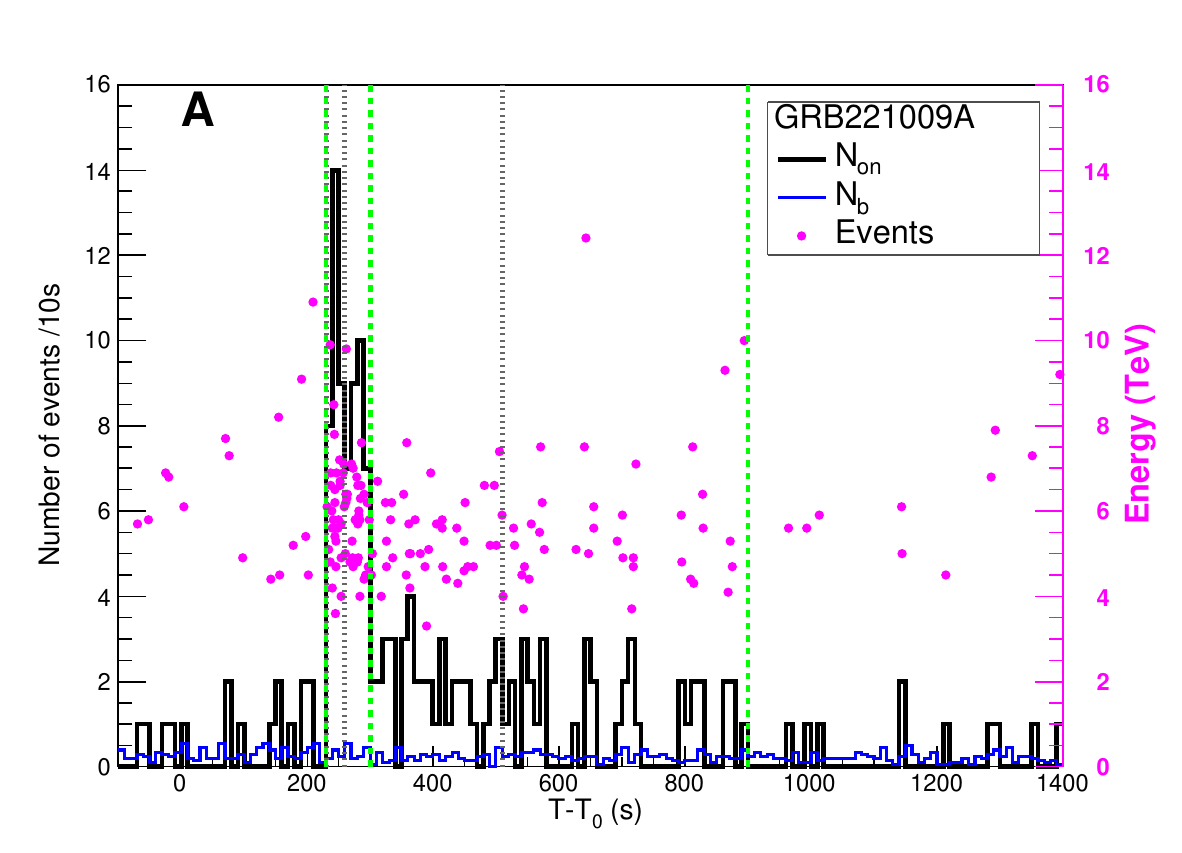}
\includegraphics[width=5cm,height=5cm]{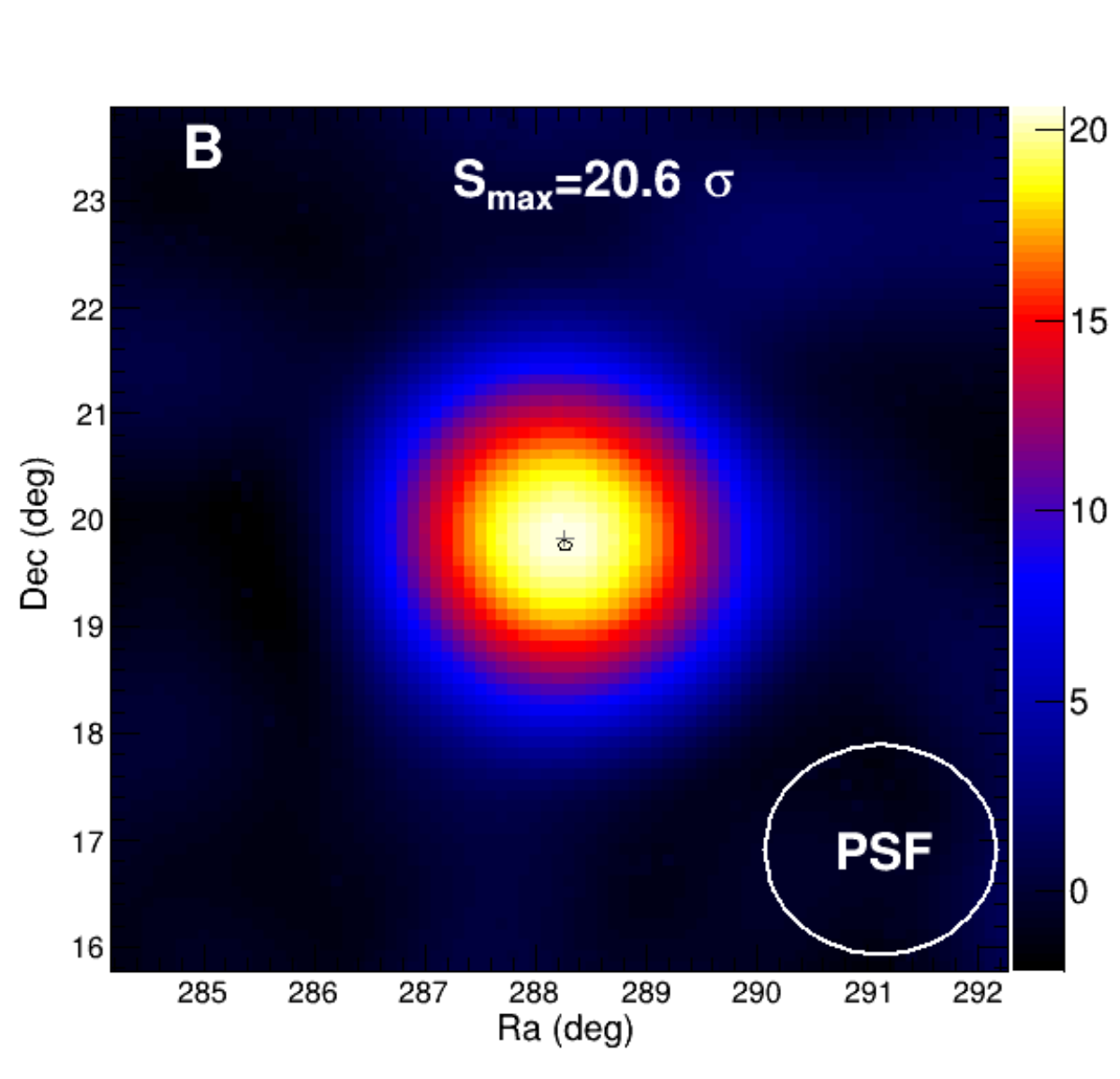}
\caption{
{\bf The   light curve and significance map of GRB 221009A obtained by  KM2A.} (A) The gamma-ray-count light curve obtained by  KM2A with each time-bin of 10s. The black curve indicates the events from the angular cone centered on the GRB, and the blue curve indicates the number of events due to cosmic ray background estimated from 20 similar  angular cones at off-source directions with the same zenith angle. The gray dashed lines indicate the peak times of the multi-pulsed emission observed by GECAM-C \cite{2023arXiv230301203A} in the MeV band. The green dashed lines indicate the times of T$_{0}$+230s, T$_{0}$+300s, and T$_{0}$+900s. The pink points indicate the energy marked by the right label and the arrival time of each event.  The energies of each event were reconstructed assuming the spectra shown in panel B of Figure \ref{fig:sed}. (B)
The significance map around GRB 221009A as observed by KM2A. The plus sign and corresponding length denote the position and error determined by KM2A. The black circle denotes the position of the GRB reported by Fermi-LAT. The white circle shows the size of the PSF that contains 68\% of the events.}
\label{fig:lc}
\end{figure}

KM2A is optimized for gamma-rays with energies from 10 TeV to 10 PeV\cite{2021ChPhC..45b5002A}. It also has sensitivity to detect gamma-rays below 10 TeV with an effective area of 10,000 m$^2$ at 4 TeV and 100,000 m$^2$ at 7 TeV, which overlaps with WCDA. Below 10 TeV, KM2A can still remove 98\% of the cosmic ray background using a ``muon-less" content criterion.  The angular resolution is about 1$^{\circ}$ and the  energy resolution is around 40\%.
The count-rate light curve observed by KM2A is shown in the left panel of Figure \ref{fig:lc}.  The emission started at T$_{0}$+230s with flux peak around T$_{0}$+245s. The emission gradually decreased since T$_{0}$+300s and faded out after T$_{0}$+900s. During the period from T$_{0}$+230s to T$_{0}$+900s, 142 events with energies above 3 TeV from the direction of GRB 221009A were detected, and the cosmic ray background was estimated to be 16.7. Therefore, the gamma-ray emission was detected with a significance of 20.6 $\sigma$. The position of the GRB was estimated to be RA=288.26$^{\circ}$ $\pm$0.07$^{\circ}$, Dec=19.83$^{\circ}\pm$0.07$^{\circ}$ using the KM2A data, which is consistent with that of Fermi-LAT. The significance map can be found in the right panel of Figure \ref{fig:lc}.

The KM2A data are divided into two  intervals for spectral analysis, according to the light curve shown in Figure \ref{fig:lc}, i.e., interval 1 from T$_{0}$+230s to T$_{0}$+300s and interval 2 from T$_{0}$+300s to T$_{0}$+900s. We combined the WCDA and KM2A data using a joint forward-folded fit to determine the spectral energy distribution (SED) of GRB 221009A, which is shown in Figure \ref{fig:sed}, assuming a  log-parabola (denoted as LP)  or  a power-law with exponential cutoff (denoted as PLEC) function. WCDA covers the energy range from 0.2 TeV to 7 TeV, and KM2A covers the energy range from 3 TeV to 20 TeV.
   Details about the analysis of KM2A data are presented in the Materials and methods section.
More details about the analysis of WCDA data can be found elsewhere\cite{2023ScienceGRB}. The two measurements are consistent with each other in the overlapping energies from 3 TeV to 7 TeV.
Fitting an LP function, the yielded  $\chi^2/ndf$, where $ndf$ is the number of degrees of freedom, are 14.1/9 and 37.3/10, respectively. The probability for the second interval is very low which indicate the LP function is not favored.
If fitting an PLEC function, the yielded $\chi^2/ndf$ become much better, with 10.1/9 and 18.3/10 for the two time intervals, respectively. Therefore, the fitting using the PLEC function is better than that using the LP function.
The detailed parameters yielded by these two functions for the two intervals are listed   in Table \ref{tab:fit}. The spectrum at interval 2 is slightly harder than that of interval 1.
Besides the statistic uncertainty, an addition 7\% systematic uncertainty should also exist for the flux according to \cite{2021ChPhC..45b5002A}.
It is worth noting that the precise measurement by LHAASO is the first detection of a cutoff at the high-energy end of a gamma-ray spectrum from a GRB.

 With the KM2A energy resolution of about 40\% at 10 TeV, the reconstruction of energy of an event is non-trivial,  in particular for very steep spectra, e.g. in the region of an energy cutoff. To correctly estimate the true energy and corresponding errors of each high energy event, the probability function of the true energy is constructed for each event, which  strongly depends on the assumption of the spectral function. More details can be found in the  material and method section.

Using the   LP function spectra shown in panel A of Figure \ref{fig:sed}, eight events with reconstructed energy above 10 TeV were observed during the period from T$_{0}$+230s to T$_{0}$+900s. The event with the maximum energy is at 17.8$_{-5.1}^{+7.4}$ TeV.  This energy is similar with that  reported in the GCN \cite{2022GCN.32677....1H}, which is preliminary estimated using the  KM2A alone data assuming a power-law function spectrum. However, if using the PLEC function spectra  shown in panel B of Figure \ref{fig:sed}, the energy of the maximum energy event is 12.2$_{-2.4}^{+3.5}$ TeV.
 If the spectral shape is governed by the strongly energy-dependent absorption of gamma-rays in collisions with photons of the Extragalactic Background Light (EBL),  which will be discussed later, the highest energy photon is reconstructed at 12.5$_{-2.4}^{+3.2}$  TeV.
The energies of each event shown in Figure \ref{fig:lc} were obtained under the PLEC function spectra.  The energies of the nine gamma-ray-like events with the highest energies have been estimated under three spectra, i.e., LP, PLEC, and an EBL model.  It is worth to note that the LP spectrum is not favored by the data.

 There is furthermore a small probability that an event is a misidentified cosmic-ray event; the chance probability has been estimated for each event individually and varies for the nine events between 4.5\% and 17\%.
More details can be found from the Table \ref{tab:event} in the Supplementary Materials  section.
Gamma-ray events with energy above 100 TeV were searched for during a long period from T$_{0}$ to T$_{0}$+6000s; however, no event was detected.


\subsection{EBL absorption of emission from GRB 221009A}

It is known that very high-energy (VHE) gamma-rays emitted from distant astronomical sources are affected by the EBL absorption  through photon-photon interaction, inducing a cutoff in the energy spectrum in the form of $e^{-\tau(E)}$ with $\tau(E)$ being the energy-dependent optical depth. For GRB 221009A at a redshift of z=0.151, the absorption at 1 TeV is modest with a survival fraction ranging from 18\% to 21\%, while it becomes very heavy at 10 TeV ranging from 0.5\% to 0.05\%, depending on the EBL models (see  Figure \ref{fig:ebls}).
The cutoff energies in the two intervals measured by LHAASO are consistent with each other within a $2.1\sigma$ error range, which is expected due to EBL absorption. Therefore, the spectra of GRB 221009A measured by LHAASO  provide an excellent test of EBL models.

To quantitatively test different EBL models, we utilize a log-parabolic function to describe the intrinsic GRB spectrum and then fit the EBL-attenuated gamma-ray spectrum in the form dN/dE = J$_0$E$^{a+b.log(E)}e^{-\tau(E)}$ to the data. The fitting results are given in the Supplementary Materials section. The goodness of fit combining the two intervals is comparable for different EBL models, i.e., Saldana-Lopez et al. 2021\cite{2021MNRAS.507.5144S}, Gilmore et al. 2012\cite{2012MNRAS.422.3189G},  Dominguez et al. 2011\cite{2011MNRAS.410.2556D}, and Finke et al. 2010\cite{2010ApJ...712..238F}, as shown in Table \ref{tab:ebls}.
The intrinsic GRB spectrum obtained using the recent EBL model of Saldana-Lopez et al. 2021\cite{2021MNRAS.507.5144S} is shown in panel A of Figure \ref{fig:intrinsic}.
 The spectral curvature is insignificant and therefore a simple power-law (PL) function is used. The detailed parameters yielded by the fitting for the two intervals are listed   in Table \ref{tab:fit}.

The intrinsic SED of the two intervals follows a single power-law model with an index of -(2.35$\pm$0.03) for interval 1 and -(2.26$\pm$0.02) for interval 2. For interval 1, the $\chi^2$/ndf of the fit is 11.0/10, and the maximum deviation is at 5.7 TeV, which deviates from the fit line by 2.3 $\sigma$. For interval 2, the $\chi^2$/ndf of the fit is 6.6/11, while a clear deviation appears at the highest energy point.  The measured flux is 11 times higher than the fit line, however, the deviation is not statistically significant due to large flux error (panel A of Figure \ref{fig:intrinsic}).
This situation is similar to that obtained for the blazar Mrk501 during its huge flare in 1997\cite{1999A&A...349...11A}. Even though some hypotheses were proposed to produce a non-classical sharp pile-up spectrum\cite{2002A&A...384..834A}, a straightforward approach is to reduce the deviation at the highest energy point by decreasing the optical depth at high energy. An empirical  method would be to decrease the EBL intensity,  the measurement of which still contains large uncertainties due to the heavy contamination caused by foregrounds of different origin, predominantly by the zodiacal (interplanetary dust) light.

The absorption of gamma-rays at a specific energy $E$ is mainly due to the EBL photons within a narrow interval with $\Delta\lambda \sim (1\pm1/2)\lambda$ centered on $\lambda^\approx 1.5(E/1TeV) \mu m$\cite{2001ICRC...27I.250A}. Therefore, the measured gamma-ray spectra from GRB 221009A can be used to constrain the EBL intensities at different wavelengths. For this purpose, the EBL model of  Saldana-Lopez et al. 2021\cite{2021MNRAS.507.5144S} is independently re-scaled at three wavelength ranges, i.e., $\lambda <8\mu m$, 8$\mu m<\lambda<28\mu m$, and $\lambda>28\mu m$, which make the dominant contribution to the optical depth for gamma-rays below 5 TeV, between 5 to 10 TeV, and above 10 TeV, respectively.
Assuming a log-parabolic intrinsic spectral function by fitting to the LHAASO data, we obtain the re-scaled factors for the three EBL ranges  as
1.30$_{-0.20}^{+0.33}$, 1.20$_{-0.20}^{+0.23}$, and 0.40$_{-0.16}^{+0.44}$,
respectively.  The constrained EBL intensity is shown in  Figure \ref{fig:EBL-fit}.

To check the effect of  possible absolute energy uncertainty, we have  shifted the energy by $\pm$10\%. Then the scale factor of the last EBL bin changes from 0.4 to 0.35 and 0.60, respectively.  It should be emphasized that the EBL intensity at large wavelengths is now close to the lower bound set by galaxy count observations \cite{Berta:2010rc}.

The intrinsic spectra using the re-scaled best-fit EBL model are shown in panel B of Figure \ref{fig:intrinsic}.  The detailed parameters yielded by the fitting for the two intervals are listed   in Table \ref{tab:fit}.
For interval 1, the intrinsic SED follows a single power-law model with an index of -(2.12$\pm$0.03), with a $\chi^2$/ndf of 5.9/10 yielded by the fitting. For interval 2, the intrinsic SED favors a log-parabolic form with an index of -(2.03$\pm$0.02)-(0.15$\pm$0.06)log(E/TeV). The corresponding $\chi^2$/ndf of the fit is 5.5/10. We note that the flux at the highest energy point is much lower now and consistent with the fit spectrum since the EBL absorption is reduced after re-scaling EBL for wavelengths $\lambda>28\mu m$.
Using the LHAASO constrained EBL model and assuming the intrinsic SED shown in panel B of Figure \ref{fig:intrinsic}, the energy of the maximum energy event is estimated to be 12.5$_{-2.4}^{+3.2}$ TeV.

\subsection{EBL absorption corrected light curve  from GRB 221009A}

\begin{figure}
\centering
\includegraphics[width=0.7\textwidth,height=7cm]{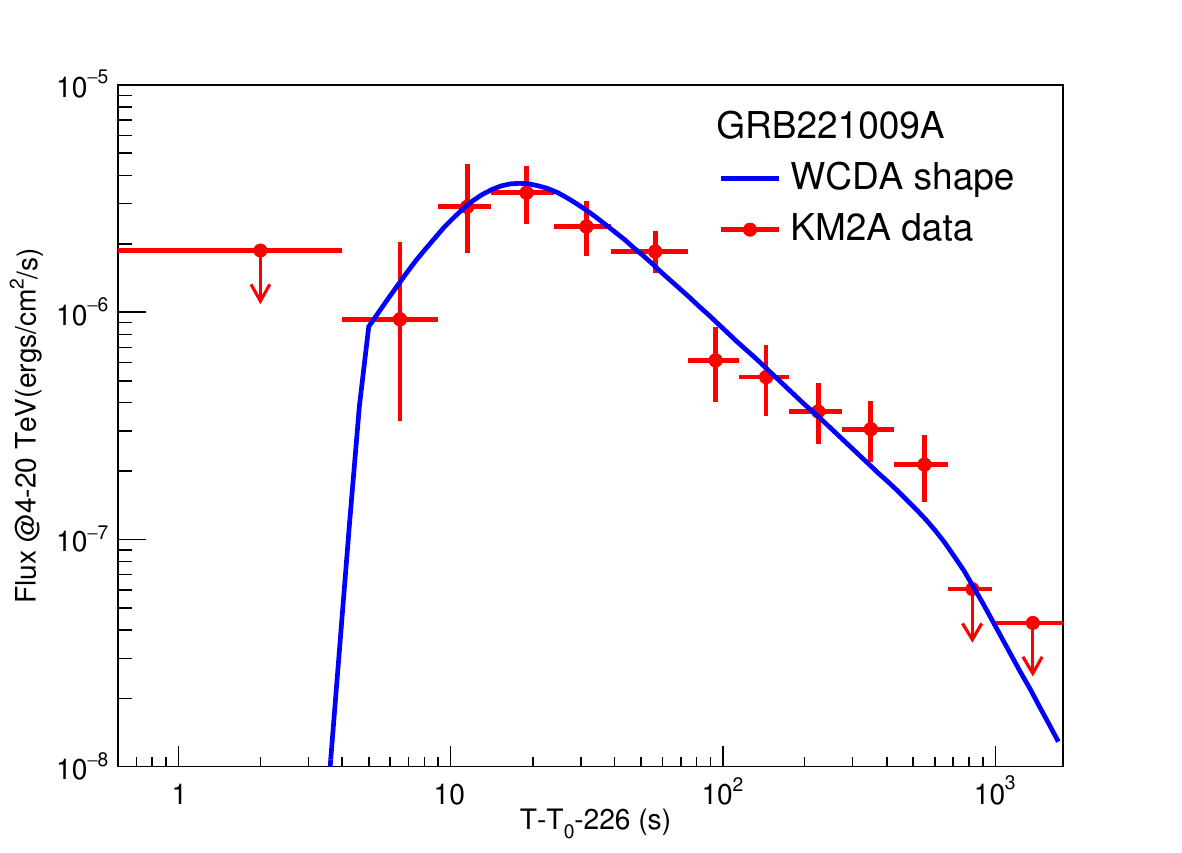}
\caption{ {\bf Flux light curve measured by KM2A in the 4$-$20 TeV band for GRB 221009A. }  Red points indicate the observations, with error bars indicating the $\sim 1 \sigma$ statistical uncertainty and  upper limits achieved with 95\% confidence level. The solid blue curve shows the fitted model adopted in ~\cite{2023ScienceGRB} to fit WCDA data, which consists of four joint power-laws that
describe the four-segment features: rapid rise, slow rise, slow decay, and steep decay. The parameters yielded in ~\cite{2023ScienceGRB} is directly adopted here while the scaling factor is achieved by fitting the KM2A data points. The $\chi^2$/ndf of the fit is 5.6/9.
}
\label{fig:fluence}
\end{figure}

The light curve of GRB 221009A in 0.2-5 TeV, as observed by WCDA, is characterized by a rapid rise to a peak, followed by a decay that persists for at least 3000 s after the peak~\cite{2023ScienceGRB}. The smooth temporal profile suggests an external shock origin for the  emission,  caused by
the interaction of the GRB ejecta with the ambient medium.
 To compare the temporal variation between WCDA and KM2A emissions, the count rate light curve shown in Figure \ref{fig:lc} is converted to the energy flux light curve assuming the spectra shown in the right panel of Figure \ref{fig:intrinsic}.  A direct comparison between the  energy flux light curves observed by WCDA  and that of KM2A is shown in Figure \ref{fig:fluence}.  The solid line indicates the best-fit four-segment   light curve observed by WCDA. Obviously,
the light curve observed by KM2A is similar to that of WCDA, indicating that the bulk of KM2A emission   should have the same origin as that of the WCDA.

\subsection{Limits on new physics using the observation of GRB 221009A}

The detection of TeV $\gamma$-rays from a cosmological distance has important implications for possible new physics beyond the Standard Model (SM) of particle physics. New physics scenarios, such as axions and LIV, may affect the process of EBL absorption. To simplify the discussion, we adopt the EBL model of Saldana-Lopez  et al. 2021 \cite{2021MNRAS.507.5144S}  here.

The axion and axion-like particles (ALPs) are hypothetical particles introduced by theoretical models developed beyond the SM. They are among the most attractive cold dark matter candidates. The most distinctive feature of axion is its coupling vertex with two photons, leading to axion-gamma-ray conversion in external magnetic fields for astrophysical detection. The axion-gamma-ray conversion leads to a suppression of EBL absorption of high-energy gamma-rays and higher transparency of space.
We still utilize a log-parabolic form to describe the intrinsic GRB spectrum in this scenario. The $\chi^{2}$ of the best fitting for the spectra during the two intervals  reduces by only  1.7 compared to  that without axion-gamma-ray conversion. Therefore, we give constraints on the parameter space of axion coupling with 95\% confidence level (C.L.).
Details of the calculation are given in the Materials and methods section. In Figure \ref{axion}, the constraint on the axion-gamma-ray coupling constant is shown. It shows that the constraint on the coupling constant by LHAASO observation of GRB 221009A is improved compared with the constraint set by CAST\cite{CAST:2017uph} (CERN Axion Solar Telescope)
and is comparable to that derived from the observation of PKS 2155-304 energy spectrum by HESS \cite{HESS:2013udx} and
a long-term monitoring of Mrk 421 by ARGO-YBJ \cite{Li:2020pcn}.

Lorentz invariance is a fundamental principle in modern physics. However, some theories that attempt to unify quantum mechanics and general relativity may suggest LIV at the energy scale approaching the Planck scale M$_\mathrm{Pl}$ = $1.22\times 10^{28}$eV. This may allow many interesting phenomena which are otherwise forbidden processes \cite{LHAASO:2021opi}.
One consequence  is that the threshold of an interaction may be altered. For  GRB gamma-rays, if the threshold of interacting with EBL is slightly increased, the process $\gamma\gamma \rightarrow e^-e^+$ is suppressed, and the universe becomes more transparent. In such a scenario,
the detection of gamma rays beyond 10 TeV is easier to understand.
However, the reduce of the $\chi^{2}$ of the fitting is only 2.6, which is turned to set a constraint on the LIV effect. We derive that for the first-order LIV, the effective energy scale of LIV should be greater than 1.5 M$_\mathrm{Pl}$. This constraint is comparable to that derived by exploring the energy dependence of the propagation speed of gamma-rays from GRB090510 by Fermi-LAT\cite{FermiGBMLAT:2009nfe}.

\section{Discussion}
In this work, we report the first detection of gamma-rays beyond 10 TeV from a gamma-ray burst, GRB 221009A, during the early afterglow period by LHAASO. The measured energy spectrum has a clear cutoff at the high-energy end, which is  consistent with the standard picture of gamma-rays propagating and interacting with EBL.
The derived intrinsic energy spectrum shows a power-law shape extending beyond 10 TeV, indicating a highly efficient particle acceleration mechanism working within the relativistic shock of the GRB. The LHAASO data prefer a suppression of the EBL flux at mid-IR wavelengths.
These observations also place strong constraints on  new physics parameters, such as an axion origin of the signal or LIV.
In the following, we will discussion the implication for GRB physics and the possible reasons account for  the highest energy gamma-ray.

\subsection{Implication for GRB physics}

Figure \ref{fig:intrinsic} shows that the intrinsic spectrum of GRB 221009A does not exhibit softening up to at least 10 TeV.  For previous TeV afterglows, the SSC process of relativistic electrons has been proposed to produce the TeV emission\cite{2019ApJ...880L..27D,2019ApJ...884..117W,2019Natur.575..459M}.  However, since the Klein-Nishina (KN) effect becomes increasingly significant at higher energies, one would expect a softening of the SSC spectrum towards higher energies\cite{2009ApJ...703..675N}. This can  be seen from the comparison between the model that explains the WCDA data\cite{2023ScienceGRB} and the data of KM2A of GRB 221009A (see  Figure \ref{fig:Comparison-KM2A}).  This may be similar to the challenge in explaining the TeV data of GRB 190829A with the SSC scenario \cite{2021Sci...372.1081H}. Such a discrepancy would suggest more complicated processes occurring during the early afterglow phase.

One possible solution to this discrepancy for GRB 221009A is to assume that an additional hard spectral component emerges at the highest energy end. In the "standard" paradigm of an external shock,   two types of shocks are formed when the ultrarelativistic  ejecta are decelerated by the swept-up ambient medium: a forward shock  propagating into the external
medium that produces the afterglow emission,  and a reverse shock propagating back into the ejecta shell.
In the external shocks, both electrons and protons are accelerated.
Proton synchrotron emission from accelerated ultra-high-energy cosmic rays (UHECRs) in the external reverse shock has been suggested to produce a hard spectral component in GRB 221009A\cite{2023ApJ...947L..14Z}. The magnetic field in the reverse shock could be larger than that of the forward shock, so protons could be accelerated to ultra-high energies by the reverse shock and thus produce synchrotron emission above 10 TeV\cite{2023ApJ...947L..14Z}.  For a proton spectrum $dN_{p}/dE\sim E^{-2}$, the spectral index of the proton synchrotron emission would be $-1.5$, much harder than the SSC emission. In addition, the intergalactic electromagnetic cascade due to the propagation of UHECRs \cite{2023A&A...670L..12D} can also produce a hard spectral component at the highest energy, because $>10$ TeV gamma-rays could be produced by UHECRs that have propagated to a nearby distance to us. To explain the observed flux of $>10$ TeV gamma-rays in the hadronic scenario, both models  require  efficient production of UHECRs from GRBs\cite{2023ApJ...947L..14Z,2023A&A...670L..12D}.

 An alternative explanation could be to invoke an additional hard leptonic component. This could be realized in the multi-zone models where the magnetic field is inhomogeneous
throughout the emitting volume\cite{2023ApJ...947...87K}. Synchrotron photons from the strong magnetic field zone provide the dominant target for cooling of the
electrons in the weak magnetic field zone, which could result in a formation of a hard electron distribution  due to
the Klein-Nishina scattering effect.   A hard electron spectral component could also be formed  by the hydrodynamical turbulence that is excited in the GRB  forward shock and  stochastically accelerates protons and electrons\cite{2016PhRvD..94b3005A}. The stochastic acceleration can yield a hard electron spectrum with $p < 2$,  which has also been  discussed as a possible electron acceleration mechanism  in AGN jets.

\subsection{Possible reasons account for  the highest energy gamma-ray event}

The detection of a gamma-ray event beyond 10 TeV from a source at a cosmological distance of z=0.151 seems unlikely, considering the heavy absorption by EBL. As shown above, it may hint a suppression of the EBL intensity at large wavelengths. Several possible reasons may account for the observation.

The first possibility is that the highest gamma-ray event is actually a  cosmic ray background event with unusually low muon content. The gamma-ray and cosmic ray events recorded by LHAASO are mainly discriminated by the muon content in the air shower. Therefore, a less muonic cosmic ray event could mimic a gamma-ray event. The chance of misidentification has been estimated to be 4.5\%, taking into account the arrival time, direction, and the muon content.
The second possibility is that this event is actually a lower energy gamma-ray but is reconstructed to above 10 TeV because of the poor energy resolution. If the intrinsic SED follows the fitting line shown in panel A of Figure \ref{fig:intrinsic} and taking the EBL model of Saldana-Lopez et al. 2021 \cite{2021MNRAS.507.5144S}, the expected number of events beyond the highest energy during the interval from T$_{0}$+300s to T$_{0}$+900s is 0.009,  which will be enlarged to 0.015 if using the interval from T0+230s to T0+900s.
The third possibility is that the absorption of EBL is weaker than the current models predict. As shown in panel B of Figure \ref{fig:intrinsic}, the situation improves after decreasing the EBL intensity at wavelengths above 28 $\mu$m.

Besides these, other interesting possibilities that avoid EBL absorption may be due to some ``exotic" physics, such as Lorentz invariance violation (LIV) or an axion origin of the signal. However, detailed discussions about these scenarios will be plagued with uncertainties of EBL models. Simple constraints on these new physics scenarios have been discussed.

\section{Materials and Methods}
KM2A is composed of 5216 electromagnetic particle detectors (EDs) and 1188 muon detectors (MDs), which are distributed in an area of 1.3 km$^2$. The EDs are designed to detect the gamma-ray/cosmic ray incident showers with determining their directions and energies. The MDs are designed to detect the muon component of showers, which is used to discriminate between gamma-ray and hadron-induced showers. The whole KM2A detector was completed and operational on July 19th, 2021, and the duty cycle is about 99\%. A trigger is generated when 20 EDs are fired within a 400 ns window, and the trigger rate is about 2.5 kHz. The performance, including angular resolution, energy resolution, and gamma-ray/cosmic-ray discrimination power, of KM2A for gamma-rays has been thoroughly tested using the observation of the Crab Nebula\cite{2021ChPhC..45b5002A}. The following will show some  key information and methods for the observation of GRB 221009A with KM2A.

\subsection{Effective area of KM2A for gamma-rays}
\begin{figure}
\centering
\includegraphics[width=7.5cm,height=6cm]{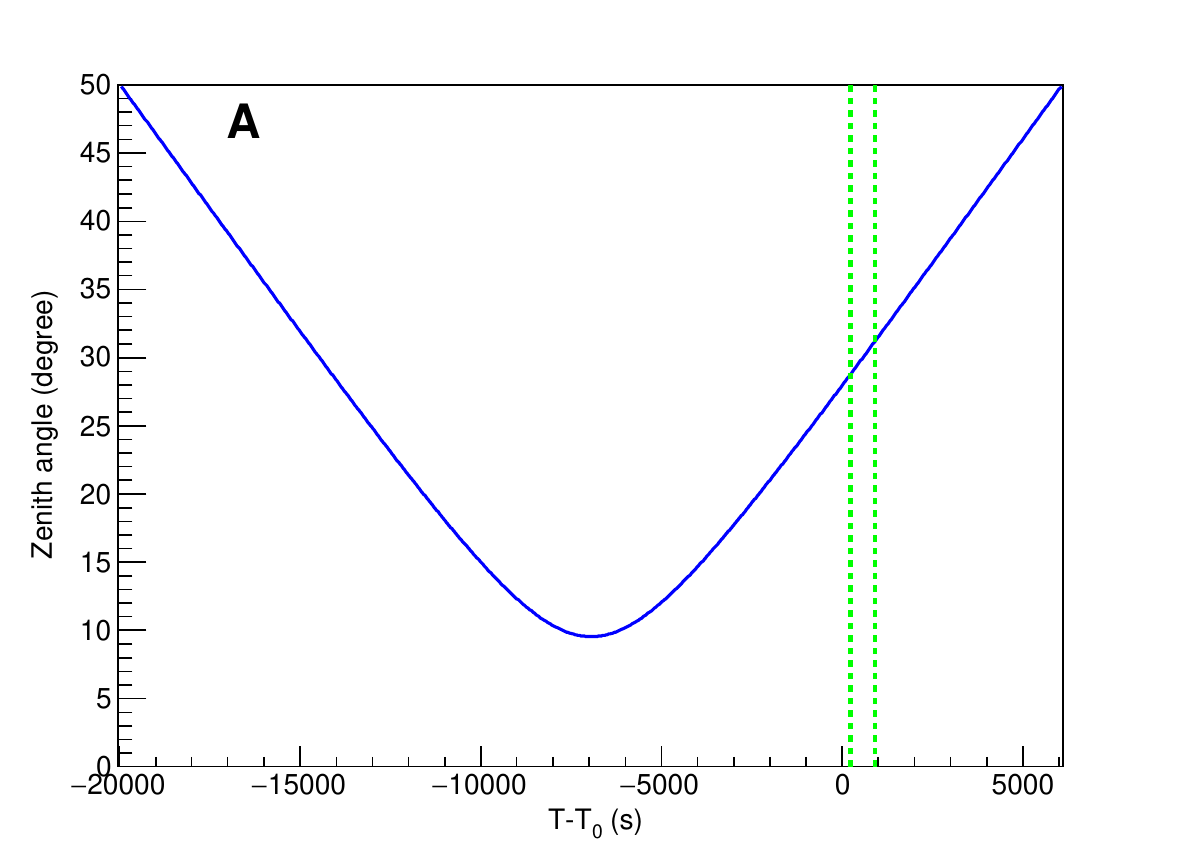}
\includegraphics[width=7.5cm,height=6cm]{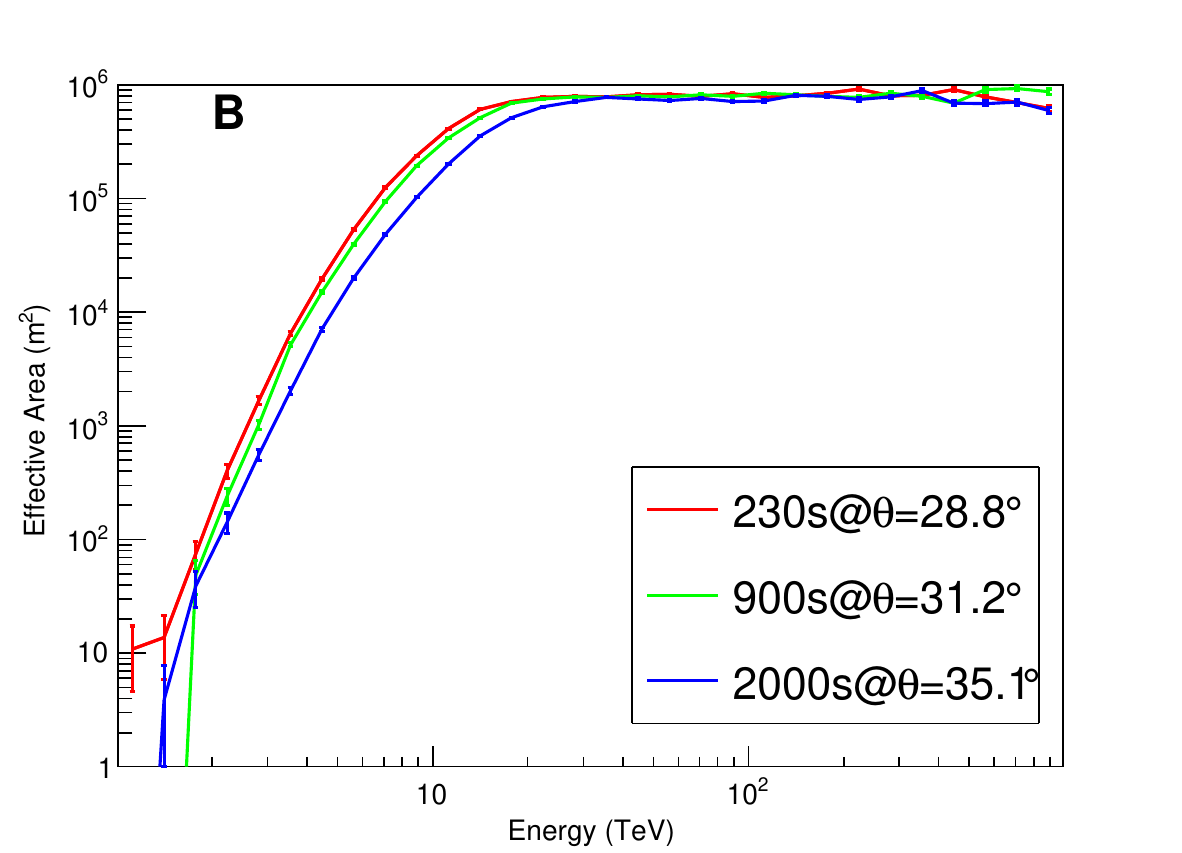}
\includegraphics[width=7.5cm,height=6cm]{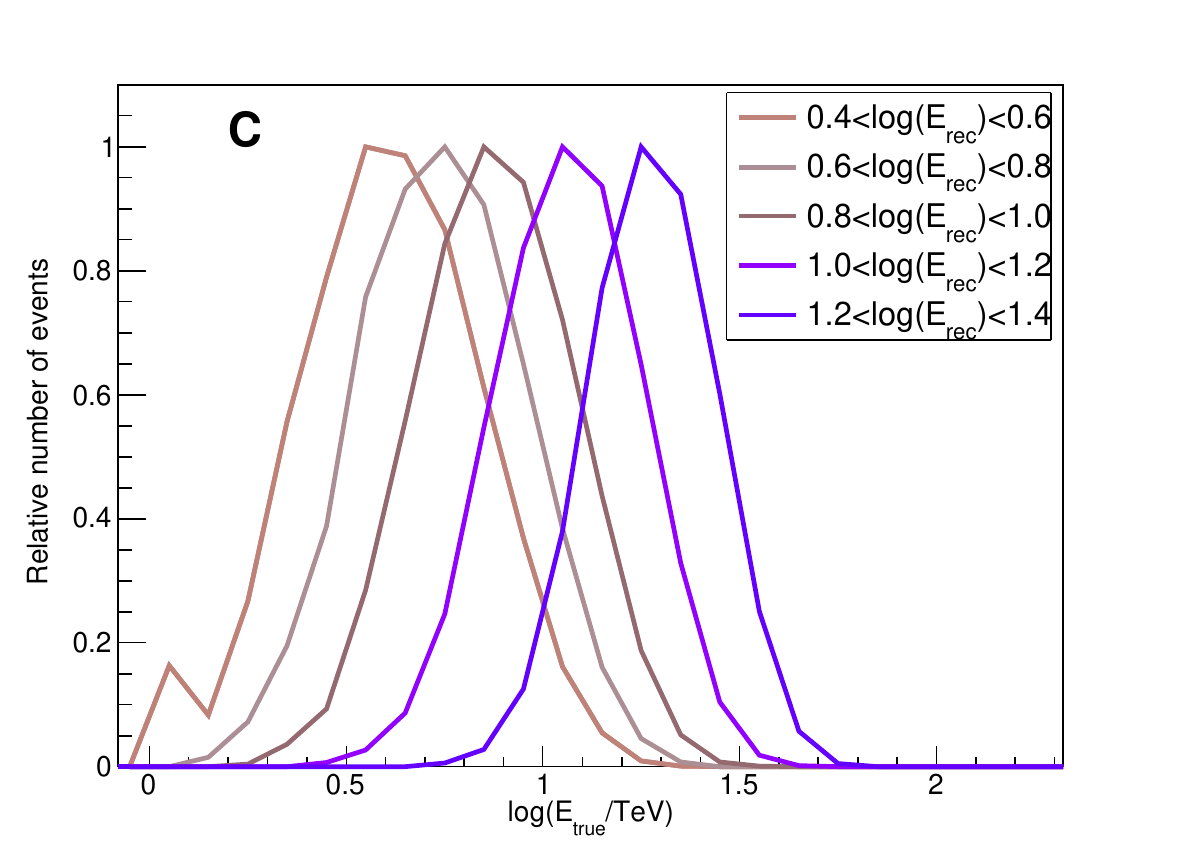}
\includegraphics[width=7.5cm,height=6cm]{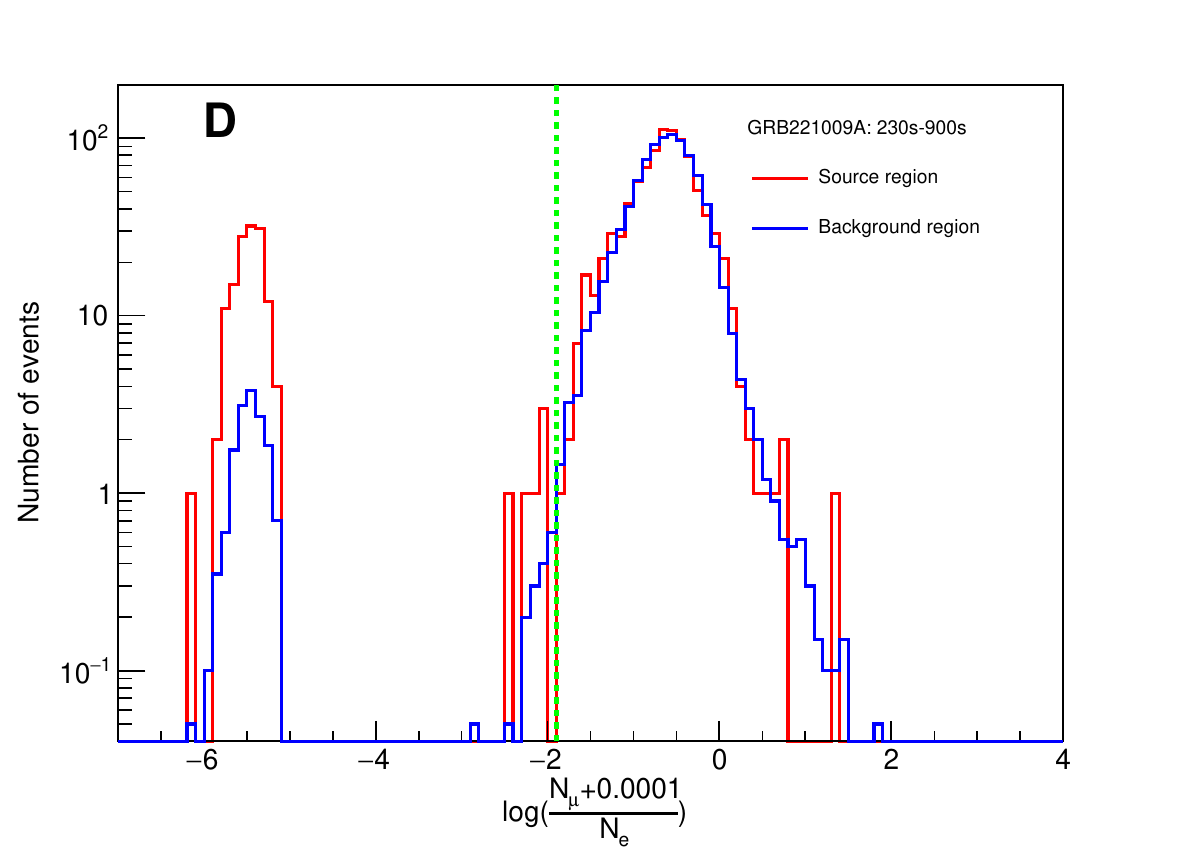}
\caption{ {\bf Some key information for the observation of GRB 221009A with KM2A. }
(A) The zenith angle of  GRB 221009A within the FOV of LHAASO as a function of the observation time. The dotted green lines mark the times of T$_0$+230s and T$_0$+900s, respectively. (B) The effective area of KM2A for gamma-rays as a function of the gamma-ray energy at three zenith angles, 28.8$^{\circ}$, 31.2$^{\circ}$, and 35.1$^{\circ}$, corresponding to the observation times T$_0$+230s, T$_0$+900s, T$_0$+2000s, respectively. (C) The distributions of simulated gamma-ray primary true energy (E$_{true}$) in each reconstructed energy (E$_{rec}$) bin in the energy range from 2.5 TeV to 25 TeV.  A power-law spectrum with index of -4.7 is used to derive the  E$_{rec}$. The number of events has been re-scaled to make the peak value 1.  The bumpy structure shown in the low energy tail of the lowest energy bin is mainly due to statistical fluctuation. (D) The distribution of the ratio R=$\mathrm{\log((N_{\mu}+0.0001)/N_e)}$, where N$_{u}$ is the number of muons measured by MDs and N$_{e}$ is the number of electromagnetic particles measured by EDs.
 The peak between -5 and -6 are events with N$_{u}$=0, i.e. $\mathrm{log(0.0001/N_e)}$, smeared out due to the variation of N$_{e}$.
The red curve represents the events from the GRB source region, while the blue curve represents the events from the 20 off-source regions, with a weight of 1/20 used for each event. The dotted green line marks the position of R=-1.9.
}
\label{fig:zen}
\end{figure}

The LHAASO observatory arrays can monitor the overhead sky with a zenith angle less than 50$^{\circ}$. The zenith angle of GRB 221009A as a function of time is shown in panel A of Figure \ref{fig:zen}. The duty cycle of KM2A was 100\% during this period. The zenith angles were 28.8$^{\circ}$ at T$_0$+230s, 31.2$^{\circ}$ at T$_0$+900s, and 35.1$^{\circ}$ at T$_0$+2000s. The effective area of KM2A for gamma-rays with different energies at these zenith angles is shown in  the panel B of Figure \ref{fig:zen}. At a zenith angle of 31.2$^{\circ}$, the effective area is 10,000 m$^2$ at 4 TeV and 100,000 m$^2$ at 7 TeV, reaching a roughly constant value of 900,000 m$^2$ around 20 TeV. For comparison, the detector area of WCDA is 78,000 m$^2$.

\subsection{Energy reconstruction for SED measurement}
In the normal KM2A data analysis pipeline, the particle density at 50 m (denoted as $\rho_{50}$) from the shower core location is used to evaluate the gamma-ray energy.  For a certain zenith angle $\theta$ and a certain $\rho_{50}$ value, the probability distribution of the true energy  can be achieved using the Bayes theorem:
\begin{equation}
P(E|(\rho_{50},\theta))=\frac{ f(E) A_{eff}(E,\theta) P(\rho_{50}|(E,\theta))}{\int f(E) A_{eff}(E,\theta) P(\rho_{50}|(E,\theta))dE}
\end{equation}
where $f(E)$ is spectral function, $A_{eff}(E,\theta)$ is the effective area at true energy $E$ and zenith angle $\theta$, $P(\rho_{50}|(E,\theta))$ is the probability to measure $\rho_{50}$ for a gamma-ray  event with energy $E$ and zenith angle $\theta$.
Obviously, the distribution of $P(E|(\rho_{50},\theta))$ depends on the assumption of  $f(E)$.  Usually, the median value of the distribution of $P(E|(\rho_{50},\theta))$ is used as the reconstructed energy $E_{rec}$ for a gamma-ray event with a value of $\rho_{50}$ at zenith angle $\theta$ assuming a power-law spectral function with index of -3.0 \cite{2021ChPhC..45b5002A}.
Taking into account the derived spectrum for GRB 221009A using the KM2A data, the relation function between $\rho_{50}$ and  E$_{rec}$ was renewed, assuming a power-law energy spectrum with an index of $\alpha$=-4.7.  This updating will significantly reduces  the value of E$_{rec}$. In this work, events with reconstructed energy (E$_{rec}$) above 2.5 TeV are divided into five bins per decade. The distributions of primary true energy (E$_{true}$) of the simulated gamma-ray events in each reconstructed energy bin are shown in panel C of Figure \ref{fig:zen}, which roughly presents the energy resolution of KM2A at energies below 25 TeV.

\subsection{Gamma-ray and cosmic ray background discrimination}
Most of the events recorded by KM2A are cosmic ray-induced showers, which constitute the major background for gamma-ray observations.
Considering that gamma-ray induced showers are muon-poor and cosmic ray-induced showers are muon-rich, the ratio R=$\mathrm{\log((N_{\mu}+0.0001)/N_e)}$ between the measured muons and electrons is used to discriminate primary gamma-rays from cosmic nuclei. The panel D of Figure \ref{fig:zen} shows the distribution of this ratio for  all events from the GRB direction and the background regions. The background regions are   20 off-source regions   with the same zenith angle as the GRB. The ratio distribution is about the same for the source region and background region at R$>$-1.9, which is dominated by the cosmic ray background. The source region is clearly higher than  the background region at R$<$-1.9, which is mainly due to gamma-ray signals. The events with R$<$-1.9 are used to select gamma-ray-like events to analyze the emission from the GRB in this work. This selection criterion can remove 98$\%$ of the cosmic ray background. The survival fraction of gamma-rays is 74$\%$ according to MC simulation.

\subsection{Spectrum and flux determination}
\begin{figure}
\centering
\includegraphics[width=0.48\textwidth,height=6.5cm]{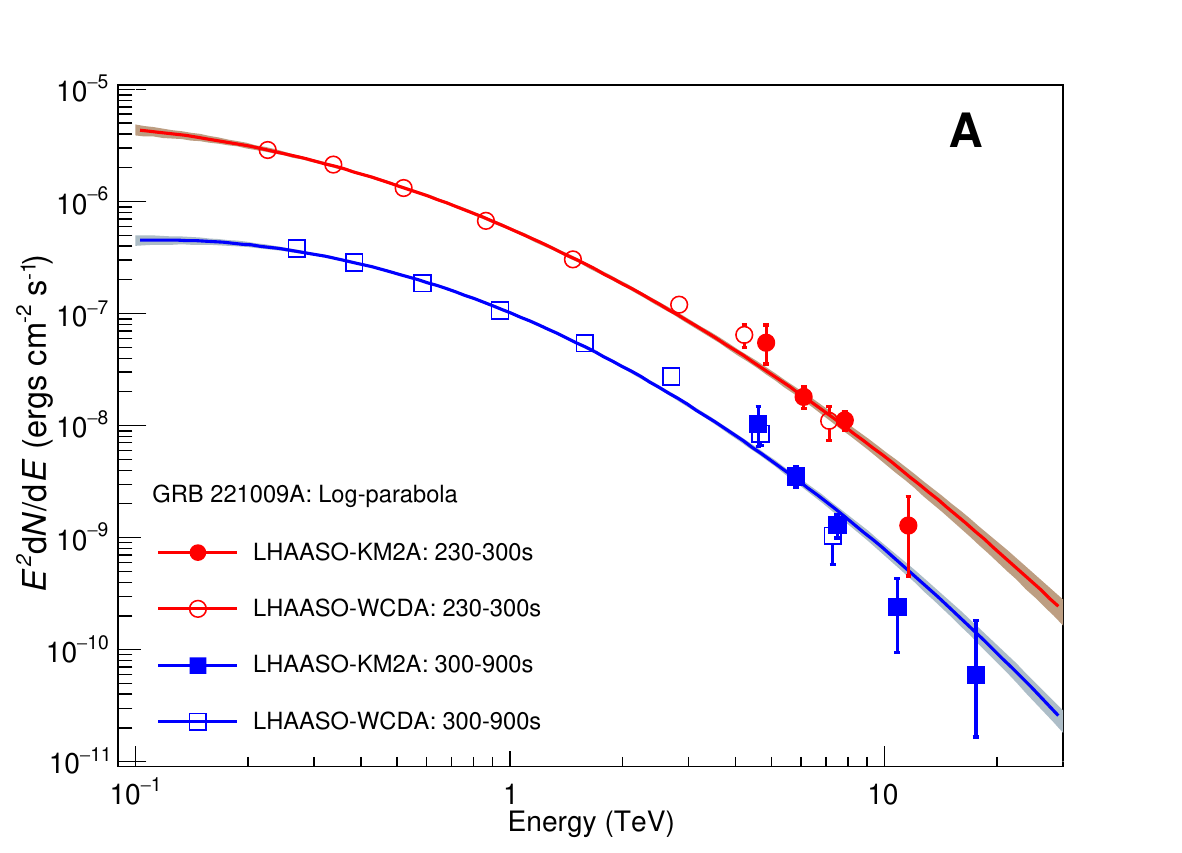}
\includegraphics[width=0.48\textwidth,height=6.5cm]{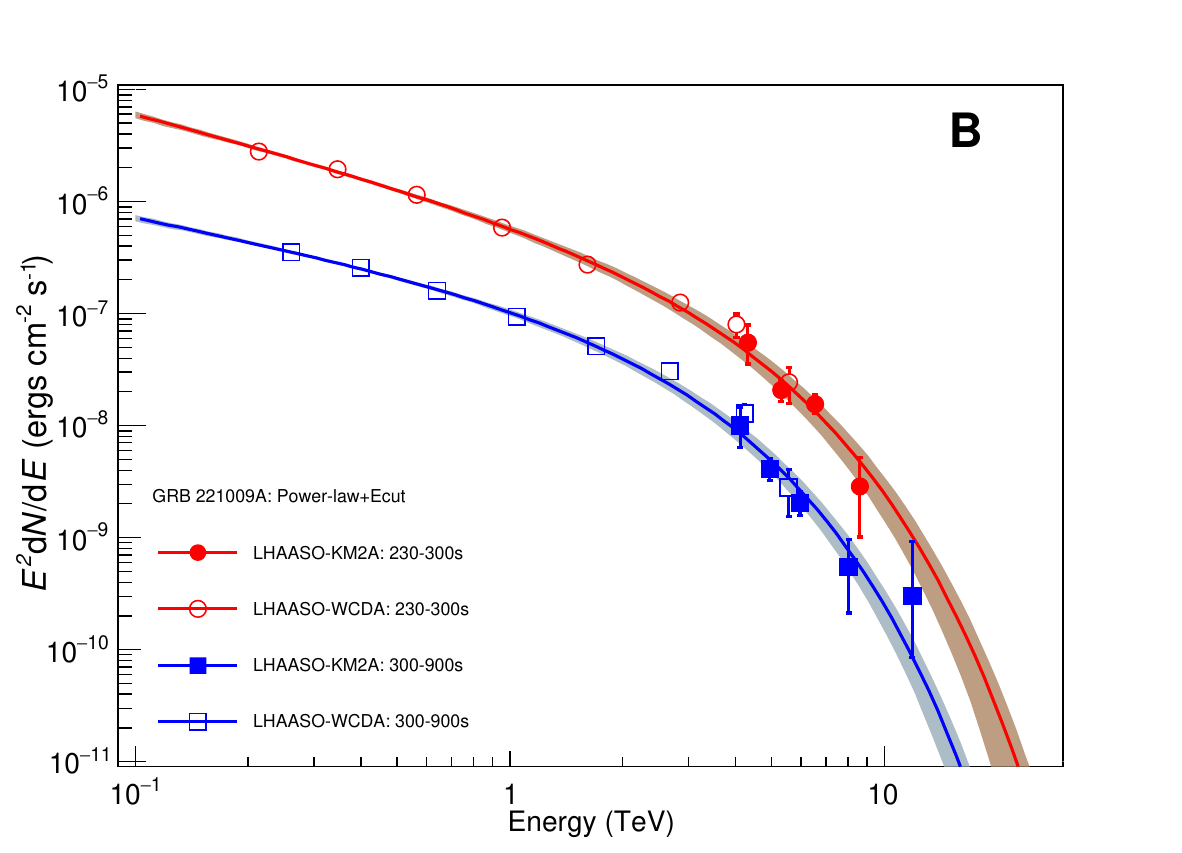}
\caption{{\bf Observed VHE spectra of GRB 221009A by LHAASO  for the two intervals.} Interval 1 is from T$_{0}$+230s to  T$_{0}$+300s (red points) and interval 2 is from  T$_{0}$+300s to T$_{0}$+900s (blue points). The solid lines indicate the best-fitting results, and the shaded regions indicate the 1-sigma error region. (A) The log-parabola function is used to fit the observational data. (B) The power-law with exponential cutoff function is adopted to fit the observational data.
 }
\label{fig:sed}
\end{figure}

The gamma-ray flux from GRB 221009A is estimated using the number of excess events and the corresponding statistical uncertainty in each energy bin. Combined with the WCDA measurement, the gamma-ray emission from the GRB is assumed to follow a log-parabola (LP) spectrum or a power-law  with an exponential cutoff (PLEC) spectrum. The response of the KM2A detector was simulated by tracing the trajectory of the GRB within the FOV of KM2A.  The expected energy distribution detected by detector $N(E_{rec})$ can be achieved:
\begin{equation}
N(E_{rec})=\int\int f(E) A_{eff}(E,\theta(t)) P(E_{rec}|(E,\theta(t)))dEdt
\end{equation}
where $A_{eff}(E,\theta)$ is effective area at true energy $E$ and zenith angle $\theta$, $P(E_{rec}|(E,\theta))$ is the probability to measure $E_{rec}$ for an event with energy $E$
and zenith angle $\theta$, $\theta(t)$ is the zenith angle of the GRB at time t.
With this energy distribution, the number of signals expected by the MC simulation $N_{MC_i}(f(E))$ using the spectrum $f(E)$ in each energy bin can be achieved.  Then,
the best-fit values of the spectrum $f(E)$ can be  obtained using a forward-folded method to minimize a $\chi^2$ function for all energy bins:
\begin{equation}
\chi^{2}=\sum_{i=1}^{n}
\begin{pmatrix}
\frac{N_{s_i}-N_{MC_i}(f(E))}{\sigma _{Ns_i}}
\end{pmatrix}^{2} ,
\end{equation}
where $N_{s_i}$ is the number of excess events, and $\sigma_{Ns_i}$ is the uncertainty of $N_{s_i}$ in the $i$th energy bin. $n$ denotes the number of energy bins. In this forward-folded method, the biases and energy resolution in the energy assignments shown in the  Figure \ref{fig:zen} are taken into account. The resulting differential flux has been shown in Figure \ref{fig:sed}. The detailed information about these results
are listed in  Table \ref{tab:sed}.   The median energy E$_{LP}$ and E$_{PLEC}$ is the median value of the probability distribution of the true energy $P(E|(E_1<E_{rec}<E_2))$  for a reconstructed energy bin  $[E_1,E_2]$ assuming corresponding LP and PLEC spectrum, respectively.
\begin{equation}
P(E|(E_1<E_{rec}<E_2))=\frac{\int \int_{E_1}^{E_2}  f(E) A_{eff}(E,\theta(t)) P(E_{rec}|(E,\theta(t))) dtdE_{rec}}{\int \int \int_{E_1}^{E_2} f(E) A_{eff}(E,\theta(t)) P(E_{rec}|(E,\theta(t)))dEdtdE_{rec}}
\end{equation}
The flux at corresponding median energy $E_{m}$ is achieved using
$\frac{N_{S_i}}{N_{MC_i}(f(E))}f(E_{m})$.

\subsection{Energy reconstruction  for the highest energy events}
Due to the poor energy resolution of KM2A at the low-energy band, the distribution of $P(E_{rec}|(E,\theta))$ is wide across different $E_{rec}$ for a given true energy $E$. Hence, the probability distribution of the true energy $E$  for a reconstructed energy $E_{rec}$ is also much wide, which can be achieved using the Bayes theorem:
\begin{equation}
P(E|(E_{rec},\theta))=\frac{ f(E) A_{eff}(E,\theta) P(E_{rec}|(E,\theta))}{\int f(E) A_{eff}(E,\theta) P(E_{rec}|(E,\theta))dE} \end{equation}
Obviously, the distribution of $P(E|(E_{rec},\theta))$ depends on the assumption of the spectral function $f(E)$.  To correctly estimate the true energies and corresponding energy errors of the highest energy events observed by KM2A, the $P(E|(E_{rec},\theta))$ for each event using different $f(E)$ is achieved. The median energy and corresponding errors can be achieved by integrating $P(E|(E_{rec},\theta))$ from 0 TeV to $E_{\xi}$ when the corresponding value equal to $\xi$:
\begin{equation}
\xi =\int_{0}^{E_{\xi}} P(E|(E_{rec},\theta)) dE
\end{equation}
The median energy is estimated using $\xi=0.5$. The corresponding errors are estimated using $\xi=0.16$ and $\xi=0.84$, respectively.

\newpage
\clearpage

\begin{figure}
\centering
\includegraphics[width=0.48\textwidth,height=6.5cm]{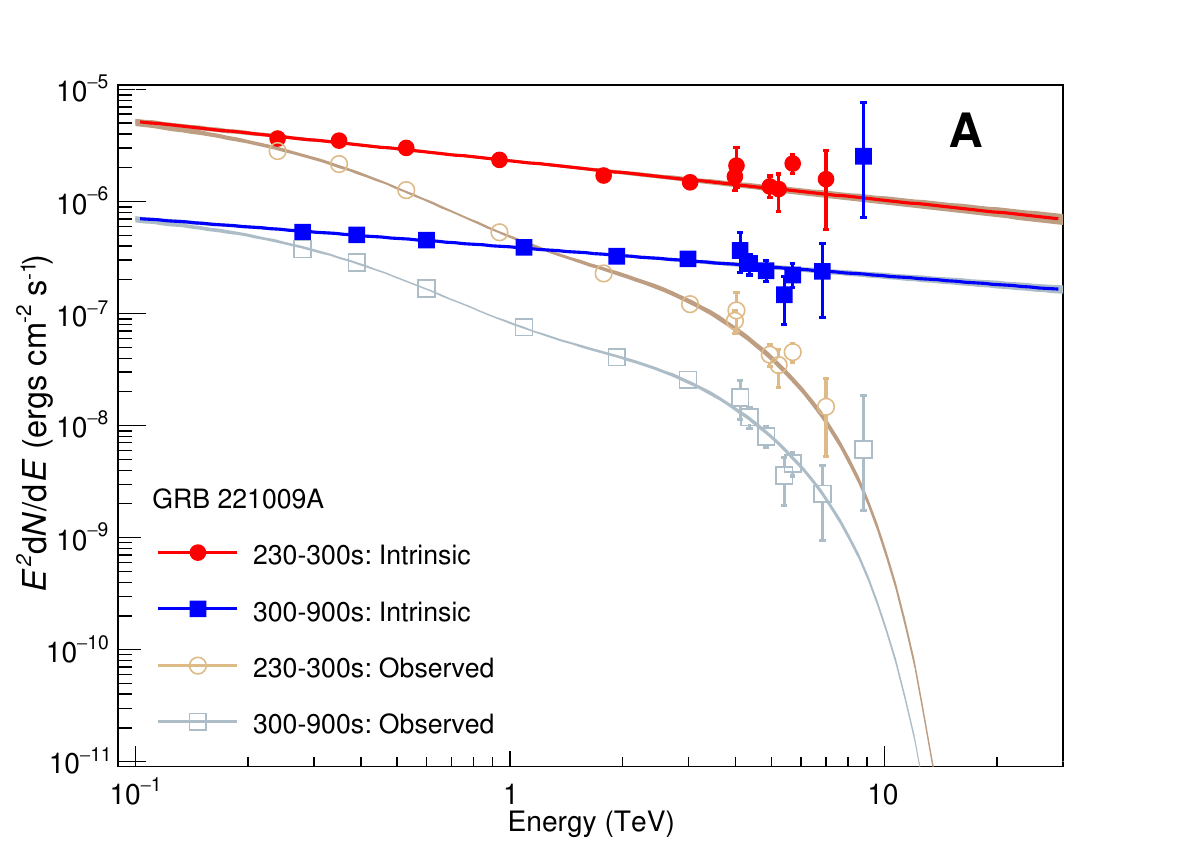}
\includegraphics[width=0.48\textwidth,height=6.5cm]{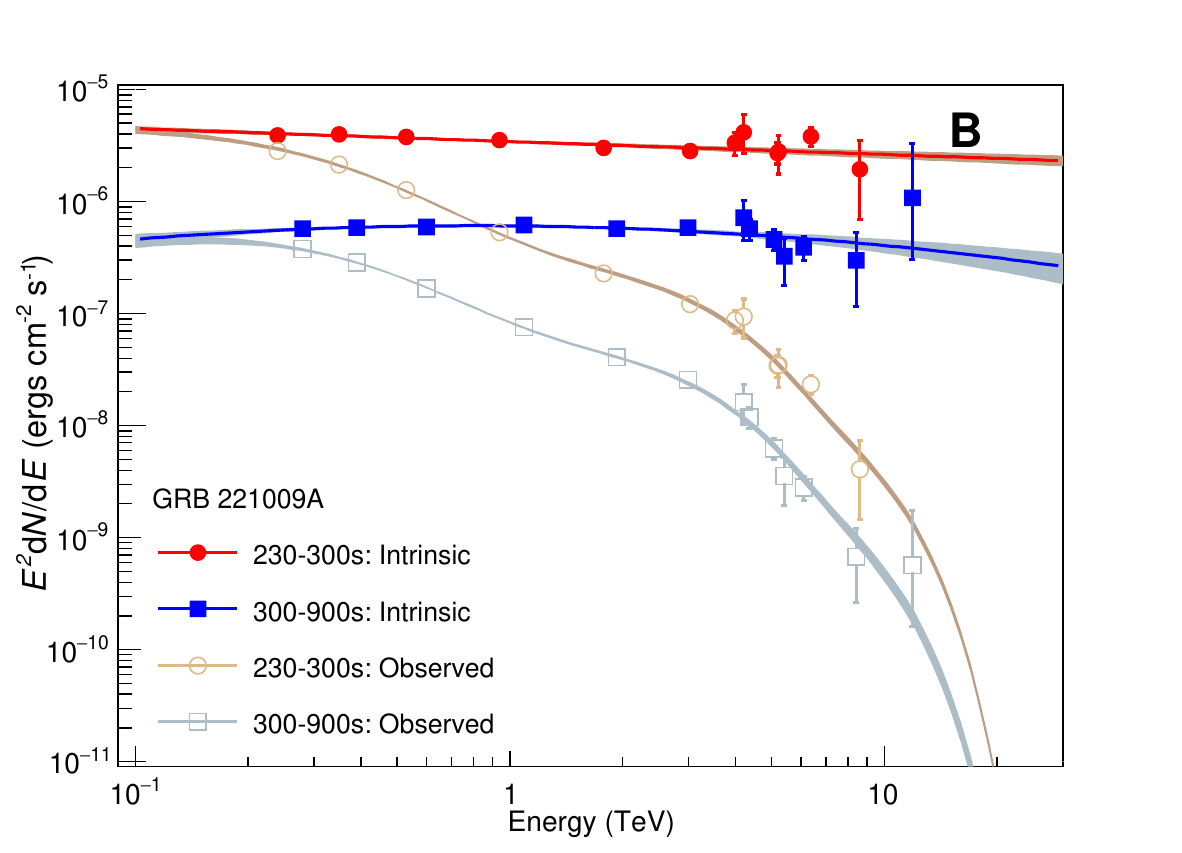}
\caption{
{\bf Intrinsic VHE spectra of GRB 221009A corrected for EBL absorption. }
(A) Filled points show the intrinsic  spectrum of GRB 221009A corrected for EBL absorption using the model of Saldana-Lopez et al. 2021\cite{2021MNRAS.507.5144S}. The red points are for  interval 1 from T$_{0}$+230s to T$_{0}$+300s, and the blue points are for  interval 2 from T$_{0}$+300s to T$_{0}$+900s.  The solid lines indicate the best-fitting results using the power-law function, and the shaded regions indicate the 1-sigma error region. The unfilled points and shaded regions  are corresponding observed spectra.
(B) Filled points show the intrinsic spectrum of GRB 221009A corrected for EBL absorption using the LHAASO-constrained  EBL model. The red solid line indicates the best-fitting result for interval 1, which is a power-law function, and the blue solid line indicates the best-fitting result for interval 2, which is a log-parabolic function. The points and shaded regions are similar to those in panel A.
}
\label{fig:intrinsic}
\end{figure}

\begin{figure}
\centering
\includegraphics[width=0.9\textwidth,height=8cm]{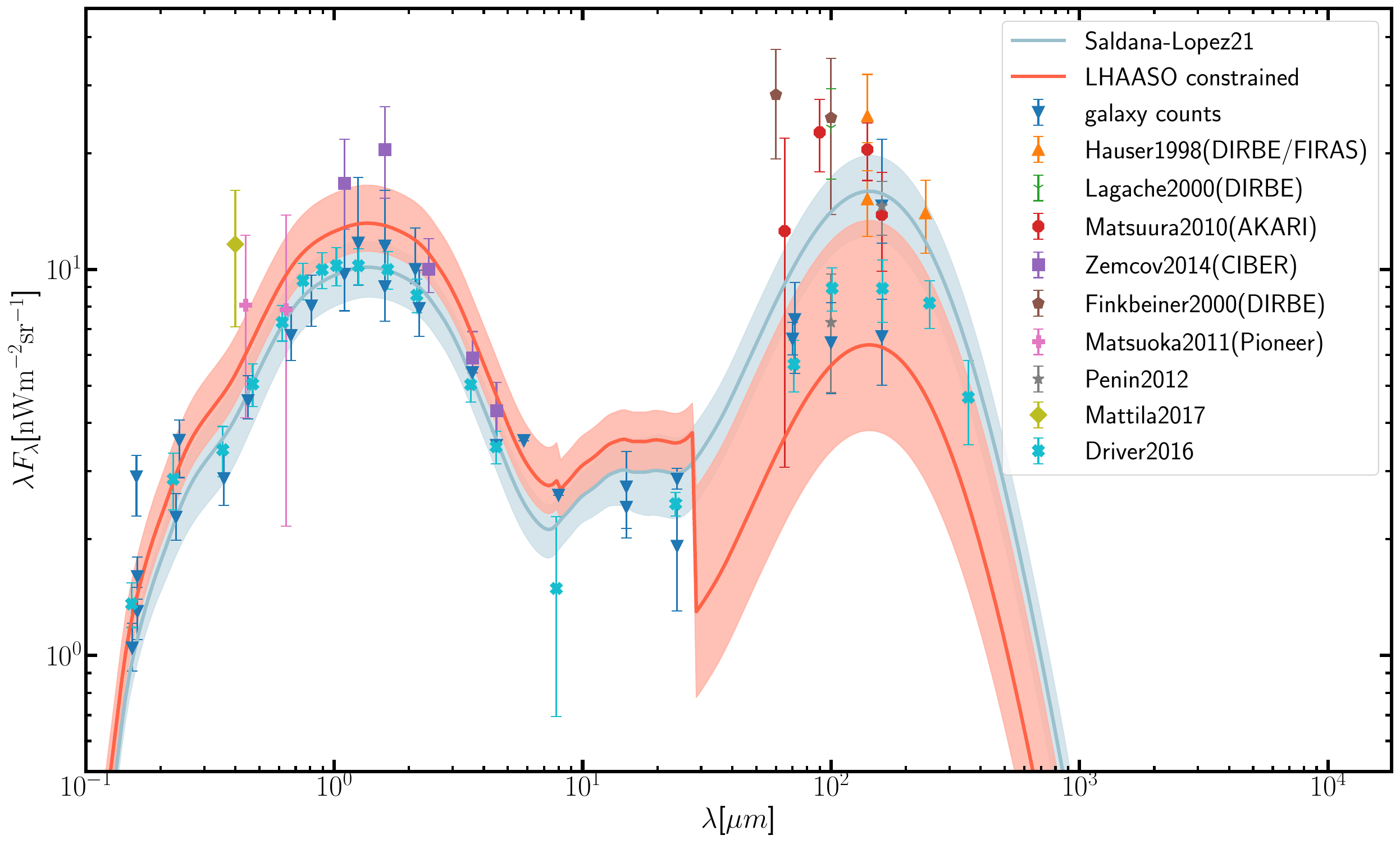}
\caption{ {\bf The density of EBL as function of wavelength at z = 0.} The best fit EBL from the GRB 221009A measurement is shown with red line and corresponding error region in three wavelength bins. The EBL of  the Saldana-Lopez et al. 2021  model \cite{2021MNRAS.507.5144S} is shown with blue line and corresponding error region.
 The data points represent direct EBL measurements, taken from  \cite{Hauser:1998ri,Lagache:1999ji,Gardner:1999jy,Elbaz:2002vd,Fazio:2004kx,Xu:2004zg,Bethermin:2010jb,Matsuura:2010rb,Voyer:2011mx,Zemcov:2014eca,Driver:2016krv,Finkbeiner:2000vr,Madau:1999yh,Metcalfe:2003zi,Papovich:2004vh,Frayer:2006qq,Berta:2010rc,Keenan:2010na,Matsuoka:2011hb,Penin:2011nq,10.1093/mnras/stx1296}.
Downward triangles   correspond to galaxy counts\cite{Berta:2010rc}, which are taken as the lower limits of EBL.
 }
\label{fig:EBL-fit}
\end{figure}

\begin{figure}
\centering
\includegraphics[width=0.58\textwidth]{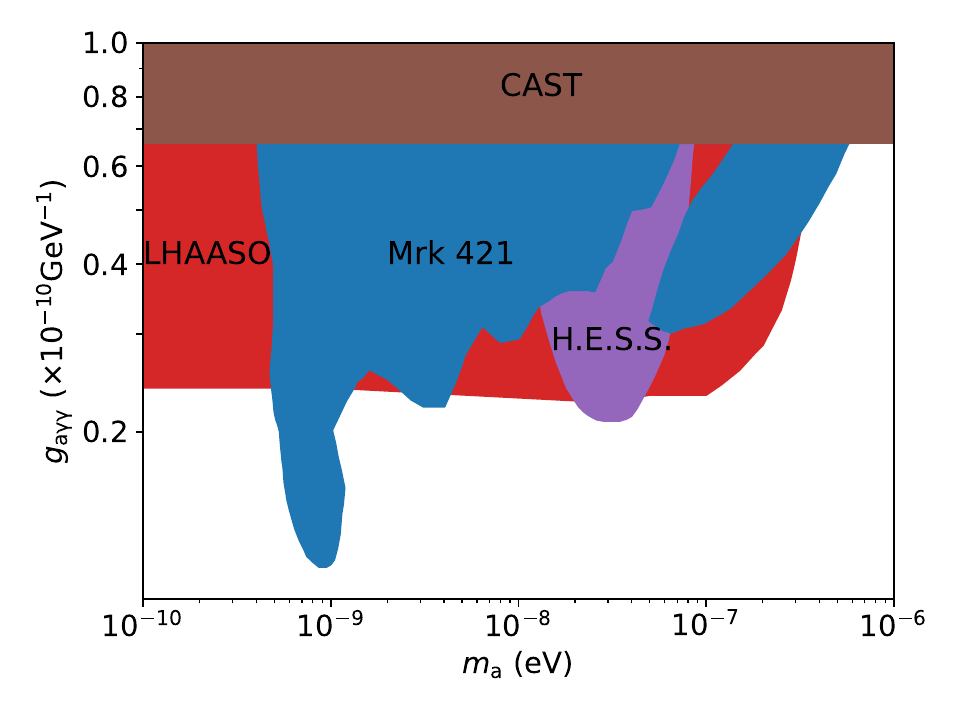}
\caption{ {\bf Constraints on the axion-gamma-ray coupling constant by LHAASO observation of the GRB 221009A. } It improves the constraint from CAST \cite{CAST:2017uph}  and is comparable with that derived from observations by HESS \cite{HESS:2013udx} and of Mrk 421 \cite{Li:2020pcn}.}
\label{axion}
\end{figure}

\begin{figure}
\centering
\includegraphics[width=15cm,height=12cm]{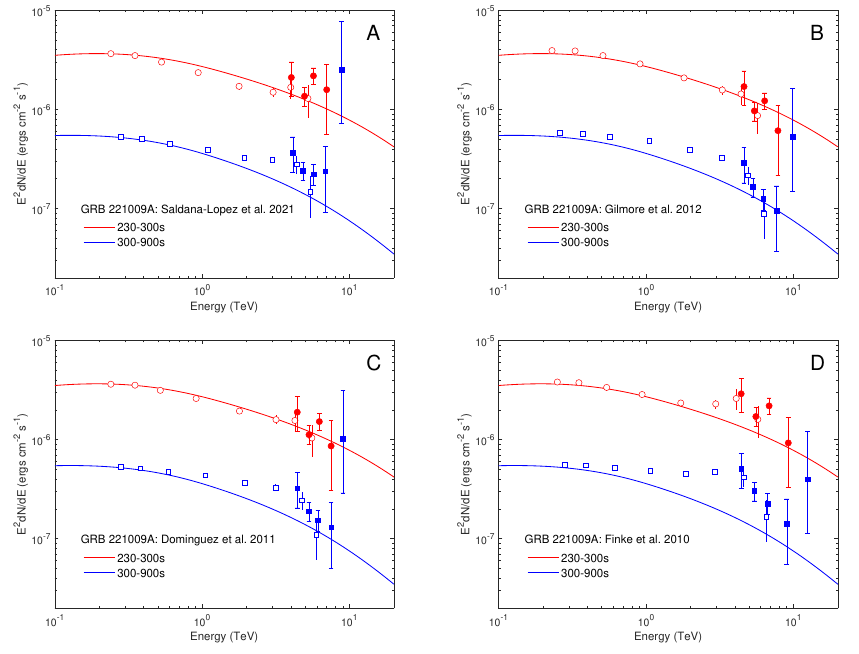}
\caption{ {\bf Comparison  between the SSC emission model and the LHAASO data for different EBL models. }  The red points are for the interval from T$_{0}$+230s to T$_{0}$+300s, while the blue points are for the interval from T$_{0}$+300s to T$_{0}$+900s. The solid lines are the fitted lines using the SSC emision model \cite{2023ScienceGRB}  for corresponding data,  respectively.
(A) The intrinsic spectrum of GRB 221009A corrected for EBL absorption using the Saldana-Lopez et al. 2021\cite{2021MNRAS.507.5144S} model.  (B) The intrinsic spectrum of GRB 221009A corrected for EBL absorption using the Gilmore et al. 2012\cite{2012MNRAS.422.3189G} model. (C) The intrinsic spectrum of GRB 221009A corrected for EBL absorption using the Dominguez et al. 2011\cite{2011MNRAS.410.2556D} model. (D) The intrinsic spectrum of GRB 221009A corrected for EBL absorption using the Finke et al. 2010\cite{2010ApJ...712..238F} model.
}
\label{fig:Comparison-KM2A}
\end{figure}

\newpage
\clearpage

\begin{table}[htb]
  \centering
  \caption{{\bf Spectral fitting results for the two time intervals from GRB 221009A.} The LP function is $\frac{dN}{dE} =J_{0}(E/1TeV)^{-a-b.log(E/1TeV)}$. The PLEC  function is $\frac{dN}{dE} =J_{0}(E/1TeV)^{-a}e^{-E/E_{cut}}$}.
  \begin{tabular}{l|ccc|cc}
    \hline
PLEC  &J$_{0}(10^{-8}$ TeV$^{-1}$ cm$^{-2}$ s$^{-1}$ ) & $a$ & E$_{cut}$ (TeV) & $\chi^{2}$/ndf  & Probability \\
    \hline
230-300s & 51.9$\pm$3.3 & 2.87$\pm$0.05   &  2.62$\pm$0.26 & 10.1/9 & 0.34 \\
300-900s & 10.5$\pm$0.6 & 2.65$\pm$0.05   &  1.99$\pm$0.15 & 18.3/10 &0.050 \\
   \hline\hline
 LP  &   &   & b  &  &  \\
    \hline
230-300s & 35.7$\pm$1.3 & 3.46$\pm$0.03   &  0.58$\pm$0.05 & 14.1/9 &  0.12 \\
300-900s & 6.37$\pm$0.16 & 3.38$\pm$0.02   &  0.73$\pm$0.05 & 37.3/10 &5.0$\times10^{-5}$ \\
\hline
\hline

  Saldana-Lopez EBL  & &   &    &   &   \\
    \hline
230-300s & 144$\pm$4 & 2.35$\pm$0.03   &  0 (fixed)  & 11.0/10 &  0.36 \\
300-900s & 24.5$\pm$0.5 & 2.26$\pm$0.02   &  0 (fixed)  & 6.6/11 &  0.83 \\
 \hline\hline
LHAASO constrained EBL  & &   &    &   &  \\
    \hline
230-300s & 214$\pm$6 & 2.12$\pm$0.03   &   0 (fixed) & 5.9/10 &  0.82 \\
300-900s & 37.8$\pm$0.9 & 2.03$\pm$0.02   &  0.15$\pm$0.06 & 5.5/10 & 0.86 \\
 \hline
  \end{tabular}
   \label{tab:fit}
 \end{table}

\newpage
\clearpage



\begin{thebibliography}{10}

\bibitem{2019Natur.575..455M}
{MAGIC Collaboration}, V.~A. {Acciari}, S.~{Ansoldi}, L.~A. {Antonelli},
  A.~{Arbet Engels}, D.~{Baack}, A.~{Babi{\'c}}, B.~{Banerjee}, U.~{Barres de
  Almeida}, J.~A. {Barrio}, J.~{Becerra Gonz{\'a}lez}, W.~{Bednarek},
  L.~{Bellizzi}, E.~{Bernardini}, A.~{Berti}, J.~{Besenrieder},
  W.~{Bhattacharyya}, C.~{Bigongiari}, A.~{Biland}, O.~{Blanch}, G.~{Bonnoli},
  {\v{Z}}.~{Bo{\v{s}}njak}, G.~{Busetto}, A.~{Carosi}, R.~{Carosi},
  G.~{Ceribella}, Y.~{Chai}, A.~{Chilingaryan}, S.~{Cikota}, S.~M. {Colak},
  U.~{Colin}, E.~{Colombo}, J.~L. {Contreras}, J.~{Cortina}, S.~{Covino},
  G.~{D'Amico}, V.~{D'Elia}, P.~{da Vela}, F.~{Dazzi}, A.~{de Angelis}, B.~{de
  Lotto}, M.~{Delfino}, J.~{Delgado}, D.~{Depaoli}, F.~{di Pierro}, L.~{di
  Venere}, E.~{Do Souto Espi{\~n}eira}, D.~{Dominis Prester}, A.~{Donini},
  D.~{Dorner}, M.~{Doro}, D.~{Elsaesser}, V.~{Fallah Ramazani}, A.~{Fattorini},
  A.~{Fern{\'a}ndez-Barral}, G.~{Ferrara}, D.~{Fidalgo}, L.~{Foffano}, M.~V.
  {Fonseca}, L.~{Font}, C.~{Fruck}, S.~{Fukami}, S.~{Gallozzi}, R.~J.
  {Garc{\'\i}a L{\'o}pez}, M.~{Garczarczyk}, S.~{Gasparyan}, M.~{Gaug},
  N.~{Giglietto}, F.~{Giordano}, N.~{Godinovi{\'c}}, D.~{Green}, D.~{Guberman},
  D.~{Hadasch}, A.~{Hahn}, J.~{Herrera}, J.~{Hoang}, D.~{Hrupec},
  M.~{H{\"u}tten}, T.~{Inada}, S.~{Inoue}, K.~{Ishio}, Y.~{Iwamura},
  L.~{Jouvin}, D.~{Kerszberg}, H.~{Kubo}, J.~{Kushida}, A.~{Lamastra},
  D.~{Lelas}, F.~{Leone}, E.~{Lindfors}, S.~{Lombardi}, F.~{Longo},
  M.~{L{\'o}pez}, R.~{L{\'o}pez-Coto}, A.~{L{\'o}pez-Oramas}, S.~{Loporchio},
  B.~{Machado de Oliveira Fraga}, C.~{Maggio}, P.~{Majumdar}, M.~{Makariev},
  M.~{Mallamaci}, G.~{Maneva}, M.~{Manganaro}, K.~{Mannheim}, L.~{Maraschi},
  M.~{Mariotti}, M.~{Mart{\'\i}nez}, S.~{Masuda}, D.~{Mazin},
  S.~{Mi{\'c}anovi{\'c}}, D.~{Miceli}, M.~{Minev}, J.~M. {Miranda},
  R.~{Mirzoyan}, E.~{Molina}, A.~{Moralejo}, D.~{Morcuende}, V.~{Moreno},
  E.~{Moretti}, P.~{Munar-Adrover}, V.~{Neustroev}, C.~{Nigro}, K.~{Nilsson},
  D.~{Ninci}, K.~{Nishijima}, K.~{Noda}, L.~{Nogu{\'e}s}, M.~{N{\"o}the},
  S.~{Nozaki}, S.~{Paiano}, J.~{Palacio}, M.~{Palatiello}, D.~{Paneque},
  R.~{Paoletti}, J.~M. {Paredes}, P.~{Pe{\~n}il}, M.~{Peresano}, M.~{Persic},
  P.~G. {Prada Moroni}, E.~{Prandini}, I.~{Puljak}, W.~{Rhode}, M.~{Rib{\'o}},
  J.~{Rico}, C.~{Righi}, A.~{Rugliancich}, L.~{Saha}, N.~{Sahakyan},
  T.~{Saito}, S.~{Sakurai}, K.~{Satalecka}, K.~{Schmidt}, T.~{Schweizer},
  J.~{Sitarek}, I.~{{\v{S}}nidari{\'c}}, D.~{Sobczynska}, A.~{Somero},
  A.~{Stamerra}, D.~{Strom}, M.~{Strzys}, Y.~{Suda}, T.~{Suri{\'c}},
  M.~{Takahashi}, F.~{Tavecchio}, P.~{Temnikov}, T.~{Terzi{\'c}}, M.~{Teshima},
  N.~{Torres-Alb{\`a}}, L.~{Tosti}, S.~{Tsujimoto}, V.~{Vagelli}, J.~{van
  Scherpenberg}, G.~{Vanzo}, M.~{Vazquez Acosta}, C.~F. {Vigorito},
  V.~{Vitale}, I.~{Vovk}, M.~{Will}, D.~{Zari{\'c}}, and L.~{Nava}.
\newblock {Teraelectronvolt emission from the {\ensuremath{\gamma}}-ray burst
  GRB 190114C}.
\newblock {\em \nat}, 575(7783):455--458, November 2019.

\bibitem{2019Natur.575..459M}
{MAGIC Collaboration}, V.~A. {Acciari}, S.~{Ansoldi}, L.~A. {Antonelli},
  A.~Arbet {Engels}, D.~{Baack}, A.~{Babi{\'c}}, B.~{Banerjee}, U.~{Barres de
  Almeida}, J.~A. {Barrio}, J.~{Becerra Gonz{\'a}lez}, W.~{Bednarek},
  L.~{Bellizzi}, E.~{Bernardini}, A.~{Berti}, J.~{Besenrieder},
  W.~{Bhattacharyya}, C.~{Bigongiari}, A.~{Biland}, O.~{Blanch}, G.~{Bonnoli},
  {\v{Z}}.~{Bo{\v{s}}njak}, G.~{Busetto}, R.~{Carosi}, G.~{Ceribella},
  Y.~{Chai}, A.~{Chilingaryan}, S.~{Cikota}, S.~M. {Colak}, U.~{Colin},
  E.~{Colombo}, J.~L. {Contreras}, J.~{Cortina}, S.~{Covino}, V.~{D'Elia},
  P.~{da Vela}, F.~{Dazzi}, A.~{de Angelis}, B.~{de Lotto}, M.~{Delfino},
  J.~{Delgado}, D.~{Depaoli}, F.~{di Pierro}, L.~{di Venere}, E.~{Do Souto
  Espi{\~n}eira}, D.~{Dominis Prester}, A.~{Donini}, D.~{Dorner}, M.~{Doro},
  D.~{Elsaesser}, V.~{Fallah Ramazani}, A.~{Fattorini}, G.~{Ferrara},
  D.~{Fidalgo}, L.~{Foffano}, M.~V. {Fonseca}, L.~{Font}, C.~{Fruck},
  S.~{Fukami}, R.~J. {Garc{\'\i}a L{\'o}pez}, M.~{Garczarczyk}, S.~{Gasparyan},
  M.~{Gaug}, N.~{Giglietto}, F.~{Giordano}, N.~{Godinovi{\'c}}, D.~{Green},
  D.~{Guberman}, D.~{Hadasch}, A.~{Hahn}, J.~{Herrera}, J.~{Hoang},
  D.~{Hrupec}, M.~{H{\"u}tten}, T.~{Inada}, S.~{Inoue}, K.~{Ishio},
  Y.~{Iwamura}, L.~{Jouvin}, D.~{Kerszberg}, H.~{Kubo}, J.~{Kushida},
  A.~{Lamastra}, D.~{Lelas}, F.~{Leone}, E.~{Lindfors}, S.~{Lombardi},
  F.~{Longo}, M.~{L{\'o}pez}, R.~{L{\'o}pez-Coto}, A.~{L{\'o}pez-Oramas},
  S.~{Loporchio}, B.~{Machado de Oliveira Fraga}, C.~{Maggio}, P.~{Majumdar},
  M.~{Makariev}, M.~{Mallamaci}, G.~{Maneva}, M.~{Manganaro}, K.~{Mannheim},
  L.~{Maraschi}, M.~{Mariotti}, M.~{Mart{\'\i}nez}, D.~{Mazin},
  S.~{Mi{\'c}anovi{\'c}}, D.~{Miceli}, M.~{Minev}, J.~M. {Miranda},
  R.~{Mirzoyan}, E.~{Molina}, A.~{Moralejo}, D.~{Morcuende}, V.~{Moreno},
  E.~{Moretti}, P.~{Munar-Adrover}, V.~{Neustroev}, C.~{Nigro}, K.~{Nilsson},
  D.~{Ninci}, K.~{Nishijima}, K.~{Noda}, L.~{Nogu{\'e}s}, S.~{Nozaki},
  S.~{Paiano}, M.~{Palatiello}, D.~{Paneque}, R.~{Paoletti}, J.~M. {Paredes},
  P.~{Pe{\~n}il}, M.~{Peresano}, M.~{Persic}, P.~G.~Prada {Moroni},
  E.~{Prandini}, I.~{Puljak}, W.~{Rhode}, M.~{Rib{\'o}}, J.~{Rico}, C.~{Righi},
  A.~{Rugliancich}, L.~{Saha}, N.~{Sahakyan}, T.~{Saito}, S.~{Sakurai},
  K.~{Satalecka}, K.~{Schmidt}, T.~{Schweizer}, J.~{Sitarek},
  I.~{{\v{S}}nidari{\'c}}, D.~{Sobczynska}, A.~{Somero}, A.~{Stamerra},
  D.~{Strom}, M.~{Strzys}, Y.~{Suda}, T.~{Suri{\'c}}, M.~{Takahashi},
  F.~{Tavecchio}, P.~{Temnikov}, T.~{Terzi{\'c}}, M.~{Teshima},
  N.~{Torres-Alb{\`a}}, L.~{Tosti}, V.~{Vagelli}, J.~{van Scherpenberg},
  G.~{Vanzo}, M.~{Vazquez Acosta}, C.~F. {Vigorito}, V.~{Vitale}, I.~{Vovk},
  M.~{Will}, D.~{Zari{\'c}}, L.~{Nava}, P.~{Veres}, P.~N. {Bhat}, M.~S.
  {Briggs}, W.~H. {Cleveland}, R.~{Hamburg}, C.~M. {Hui}, B.~{Mailyan}, R.~D.
  {Preece}, O.~J. {Roberts}, A.~{von Kienlin}, C.~A. {Wilson-Hodge},
  D.~{Kocevski}, M.~{Arimoto}, D.~{Tak}, K.~{Asano}, M.~{Axelsson},
  G.~{Barbiellini}, E.~{Bissaldi}, F.~Fana {Dirirsa}, R.~{Gill}, J.~{Granot},
  J.~{McEnery}, N.~{Omodei}, S.~{Razzaque}, F.~{Piron}, J.~L. {Racusin}, D.~J.
  {Thompson}, S.~{Campana}, M.~G. {Bernardini}, N.~P.~M. {Kuin}, M.~H.
  {Siegel}, S.~B. {Cenko}, P.~{O'Brien}, M.~{Capalbi}, A.~{Da{\i}}, M.~{de
  Pasquale}, J.~{Gropp}, N.~{Klingler}, J.~P. {Osborne}, M.~{Perri}, R.~L.~C.
  {Starling}, G.~{Tagliaferri}, A.~{Tohuvavohu}, A.~{Ursi}, M.~{Tavani},
  M.~{Cardillo}, C.~{Casentini}, G.~{Piano}, Y.~{Evangelista}, F.~{Verrecchia},
  C.~{Pittori}, F.~{Lucarelli}, A.~{Bulgarelli}, N.~{Parmiggiani}, G.~E.
  {Anderson}, J.~P. {Anderson}, G.~{Bernardi}, J.~{Bolmer}, M.~D.
  {Caballero-Garc{\'\i}a}, I.~M. {Carrasco}, A.~{Castell{\'o}n}, N.~{Castro
  Segura}, A.~J. {Castro-Tirado}, S.~V. {Cherukuri}, A.~M. {Cockeram},
  P.~{D'Avanzo}, A.~{di Dato}, R.~{Diretse}, R.~P. {Fender},
  E.~{Fern{\'a}ndez-Garc{\'\i}a}, J.~P.~U. {Fynbo}, A.~S. {Fruchter},
  J.~{Greiner}, M.~{Gromadzki}, K.~E. {Heintz}, I.~{Heywood}, A.~J. {van der
  Horst}, Y.~D. {Hu}, C.~{Inserra}, L.~{Izzo}, V.~{Jaiswal}, P.~{Jakobsson},
  J.~{Japelj}, E.~{Kankare}, D.~A. {Kann}, C.~{Kouveliotou}, S.~{Klose}, A.~J.
  {Levan}, X.~Y. {Li}, S.~{Lotti}, K.~{Maguire}, D.~B. {Malesani},
  I.~{Manulis}, M.~{Marongiu}, S.~{Martin}, A.~{Melandri}, M.~J.
  {Micha{\l}owski}, J.~C.~A. {Miller-Jones}, K.~{Misra}, A.~{Moin}, K.~P.
  {Mooley}, S.~{Nasri}, M.~{Nicholl}, A.~{Noschese}, G.~{Novara}, S.~B.
  {Pandey}, E.~{Peretti}, C.~J. {P{\'e}rez Del Pulgar}, M.~A.
  {P{\'e}rez-Torres}, D.~A. {Perley}, L.~{Piro}, F.~{Ragosta}, L.~{Resmi},
  R.~{Ricci}, A.~{Rossi}, R.~{S{\'a}nchez-Ram{\'\i}rez}, J.~{Selsing},
  S.~{Schulze}, S.~J. {Smartt}, I.~A. {Smith}, V.~V. {Sokolov}, J.~{Stevens},
  N.~R. {Tanvir}, C.~C. {Th{\"o}ne}, A.~{Tiengo}, E.~{Tremou}, E.~{Troja},
  A.~{de Ugarte Postigo}, A.~F. {Valeev}, S.~D. {Vergani}, M.~{Wieringa}, P.~A.
  {Woudt}, D.~{Xu}, O.~{Yaron}, and D.~R. {Young}.
\newblock {Observation of inverse Compton emission from a long
  {\ensuremath{\gamma}}-ray burst}.
\newblock {\em \nat}, 575(7783):459--463, November 2019.

\bibitem{2019Natur.575..464A}
H.~{Abdalla}, R.~{Adam}, F.~{Aharonian}, F.~{Ait Benkhali}, E.~O.
  {Ang{\"u}ner}, M.~{Arakawa}, C.~{Arcaro}, C.~{Armand}, H.~{Ashkar},
  M.~{Backes}, V.~{Barbosa Martins}, M.~{Barnard}, Y.~{Becherini}, D.~{Berge},
  K.~{Bernl{\"o}hr}, E.~{Bissaldi}, R.~{Blackwell}, M.~{B{\"o}ttcher},
  C.~{Boisson}, J.~{Bolmont}, S.~{Bonnefoy}, J.~{Bregeon}, M.~{Breuhaus},
  F.~{Brun}, P.~{Brun}, M.~{Bryan}, M.~{B{\"u}chele}, T.~{Bulik}, T.~{Bylund},
  M.~{Capasso}, S.~{Caroff}, A.~{Carosi}, S.~{Casanova}, M.~{Cerruti},
  T.~{Chand}, S.~{Chandra}, A.~{Chen}, S.~{Colafrancesco}, M.~{Cury{\l}o},
  I.~D. {Davids}, C.~{Deil}, J.~{Devin}, P.~{deWilt}, L.~{Dirson},
  A.~{Djannati-Ata{\"\i}}, A.~{Dmytriiev}, A.~{Donath}, V.~{Doroshenko},
  J.~{Dyks}, K.~{Egberts}, G.~{Emery}, J.~P. {Ernenwein}, S.~{Eschbach},
  K.~{Feijen}, S.~{Fegan}, A.~{Fiasson}, G.~{Fontaine}, S.~{Funk},
  M.~{F{\"u}{\ss}ling}, S.~{Gabici}, Y.~A. {Gallant}, F.~{Gat{\'e}},
  G.~{Giavitto}, L.~{Giunti}, D.~{Glawion}, J.~F. {Glicenstein},
  D.~{Gottschall}, M.~H. {Grondin}, J.~{Hahn}, M.~{Haupt}, G.~{Heinzelmann},
  G.~{Henri}, G.~{Hermann}, J.~A. {Hinton}, W.~{Hofmann}, C.~{Hoischen}, T.~L.
  {Holch}, M.~{Holler}, D.~{Horns}, D.~{Huber}, H.~{Iwasaki}, M.~{Jamrozy},
  D.~{Jankowsky}, F.~{Jankowsky}, A.~{Jardin-Blicq}, I.~{Jung-Richardt}, M.~A.
  {Kastendieck}, K.~{Katarzy{\'n}ski}, M.~{Katsuragawa}, U.~{Katz},
  D.~{Khangulyan}, B.~{Kh{\'e}lifi}, J.~{King}, S.~{Klepser},
  W.~{Klu{\'z}niak}, Nu. {Komin}, K.~{Kosack}, D.~{Kostunin}, M.~{Kreter},
  G.~{Lamanna}, A.~{Lemi{\`e}re}, M.~{Lemoine-Goumard}, J.~P. {Lenain},
  E.~{Leser}, C.~{Levy}, T.~{Lohse}, I.~{Lypova}, J.~{Mackey}, J.~{Majumdar},
  D.~{Malyshev}, V.~{Marandon}, A.~{Marcowith}, A.~{Mares}, C.~{Mariaud},
  G.~{Mart{\'\i}-Devesa}, R.~{Marx}, G.~{Maurin}, P.~J. {Meintjes}, A.~M.~W.
  {Mitchell}, R.~{Moderski}, M.~{Mohamed}, L.~{Mohrmann}, C.~{Moore},
  E.~{Moulin}, J.~{Muller}, T.~{Murach}, S.~{Nakashima}, M.~{de Naurois},
  H.~{Ndiyavala}, F.~{Niederwanger}, J.~{Niemiec}, L.~{Oakes}, P.~{O'Brien},
  H.~{Odaka}, S.~{Ohm}, E.~{de Ona Wilhelmi}, M.~{Ostrowski}, I.~{Oya},
  M.~{Panter}, R.~D. {Parsons}, C.~{Perennes}, P.~O. {Petrucci}, B.~{Peyaud},
  Q.~{Piel}, S.~{Pita}, V.~{Poireau}, A.~{Priyana Noel}, D.~A. {Prokhorov},
  H.~{Prokoph}, G.~{P{\"u}hlhofer}, M.~{Punch}, A.~{Quirrenbach}, S.~{Raab},
  R.~{Rauth}, A.~{Reimer}, O.~{Reimer}, Q.~{Remy}, M.~{Renaud}, F.~{Rieger},
  L.~{Rinchiuso}, C.~{Romoli}, G.~{Rowell}, B.~{Rudak}, E.~{Ruiz-Velasco},
  V.~{Sahakian}, S.~{Sailer}, S.~{Saito}, D.~A. {Sanchez}, A.~{Santangelo},
  M.~{Sasaki}, R.~{Schlickeiser}, F.~{Sch{\"u}ssler}, A.~{Schulz}, H.~M.
  {Schutte}, U.~{Schwanke}, S.~{Schwemmer}, M.~{Seglar-Arroyo},
  M.~{Senniappan}, A.~S. {Seyffert}, N.~{Shafi}, K.~{Shiningayamwe},
  R.~{Simoni}, A.~{Sinha}, H.~{Sol}, A.~{Specovius}, M.~{Spir-Jacob},
  {\L}.~{Stawarz}, R.~{Steenkamp}, C.~{Stegmann}, C.~{Steppa}, T.~{Takahashi},
  T.~{Tavernier}, A.~M. {Taylor}, R.~{Terrier}, D.~{Tiziani}, M.~{Tluczykont},
  C.~{Trichard}, M.~{Tsirou}, N.~{Tsuji}, R.~{Tuffs}, Y.~{Uchiyama}, D.~J. {van
  der Walt}, C.~{van Eldik}, C.~{van Rensburg}, B.~{van Soelen},
  G.~{Vasileiadis}, J.~{Veh}, C.~{Venter}, P.~{Vincent}, J.~{Vink}, H.~J.
  {V{\"o}lk}, T.~{Vuillaume}, Z.~{Wadiasingh}, S.~J. {Wagner}, R.~{White},
  A.~{Wierzcholska}, R.~{Yang}, H.~{Yoneda}, M.~{Zacharias}, R.~{Zanin}, A.~A.
  {Zdziarski}, A.~{Zech}, A.~{Ziegler}, J.~{Zorn}, N.~{{\.Z}ywucka}, F.~{de
  Palma}, M.~{Axelsson}, and O.~J. {Roberts}.
\newblock {A very-high-energy component deep in the {\ensuremath{\gamma}}-ray
  burst afterglow}.
\newblock {\em \nat}, 575(7783):464--467, November 2019.

\bibitem{2021Sci...372.1081H}
{H.~E.~S.~S. Collaboration}, H.~{Abdalla}, F.~{Aharonian}, F.~{Ait Benkhali},
  E.~O. {Ang{\"u}ner}, C.~{Arcaro}, C.~{Armand}, T.~{Armstrong}, H.~{Ashkar},
  M.~{Backes}, V.~{Baghmanyan}, V.~{Barbosa Martins}, A.~{Barnacka},
  M.~{Barnard}, Y.~{Becherini}, D.~{Berge}, K.~{Bernl{\"o}hr}, B.~{Bi},
  E.~{Bissaldi}, M.~{B{\"o}ttcher}, C.~{Boisson}, J.~{Bolmont}, M.~{de Bony de
  Lavergne}, M.~{Breuhaus}, F.~{Brun}, P.~{Brun}, M.~{Bryan}, M.~{B{\"u}chele},
  T.~{Bulik}, T.~{Bylund}, S.~{Caroff}, A.~{Carosi}, S.~{Casanova}, T.~{Chand},
  S.~{Chandra}, A.~{Chen}, G.~{Cotter}, M.~{Cury{\l}o}, J.~{Damascene
  Mbarubucyeye}, I.~D. {Davids}, J.~{Davies}, C.~{Deil}, J.~{Devin},
  L.~{Dirson}, A.~{Djannati-Ata{\"\i}}, A.~{Dmytriiev}, A.~{Donath},
  V.~{Doroshenko}, L.~{Dreyer}, C.~{Duffy}, J.~{Dyks}, K.~{Egberts},
  F.~{Eichhorn}, S.~{Einecke}, G.~{Emery}, J.~P. {Ernenwein}, K.~{Feijen},
  S.~{Fegan}, A.~{Fiasson}, G.~{Fichet de Clairfontaine}, G.~{Fontaine},
  S.~{Funk}, M.~{F{\"u}{\ss}ling}, S.~{Gabici}, Y.~A. {Gallant}, G.~{Giavitto},
  L.~{Giunti}, D.~{Glawion}, J.~F. {Glicenstein}, M.~H. {Grondin}, J.~{Hahn},
  M.~{Haupt}, G.~{Hermann}, J.~A. {Hinton}, W.~{Hofmann}, C.~{Hoischen}, T.~L.
  {Holch}, M.~{Holler}, M.~{H{\"o}rbe}, D.~{Horns}, D.~{Huber}, M.~{Jamrozy},
  D.~{Jankowsky}, F.~{Jankowsky}, A.~{Jardin-Blicq}, V.~{Joshi},
  I.~{Jung-Richardt}, E.~{Kasai}, M.~A. {Kastendieck}, K.~{Katarzy{\'n}ski},
  U.~{Katz}, D.~{Khangulyan}, B.~{Kh{\'e}lifi}, S.~{Klepser},
  W.~{Klu{\'z}niak}, Nu. {Komin}, R.~{Konno}, K.~{Kosack}, D.~{Kostunin},
  M.~{Kreter}, G.~{Lamanna}, A.~{Lemi{\`e}re}, M.~{Lemoine-Goumard}, J.~P.
  {Lenain}, F.~{Leuschner}, C.~{Levy}, T.~{Lohse}, I.~{Lypova}, J.~{Mackey},
  J.~{Majumdar}, D.~{Malyshev}, D.~{Malyshev}, V.~{Marandon}, P.~{Marchegiani},
  A.~{Marcowith}, A.~{Mares}, G.~{Mart{\'\i}-Devesa}, R.~{Marx}, G.~{Maurin},
  P.~J. {Meintjes}, M.~{Meyer}, A.~{Mitchell}, R.~{Moderski}, L.~{Mohrmann},
  A.~{Montanari}, C.~{Moore}, P.~{Morris}, E.~{Moulin}, J.~{Muller},
  T.~{Murach}, K.~{Nakashima}, A.~{Nayerhoda}, M.~{de Naurois}, H.~{Ndiyavala},
  J.~{Niemiec}, L.~{Oakes}, P.~{O'Brien}, H.~{Odaka}, S.~{Ohm},
  L.~{Olivera-Nieto}, E.~{de Ona Wilhelmi}, M.~{Ostrowski}, S.~{Panny},
  M.~{Panter}, R.~D. {Parsons}, G.~{Peron}, B.~{Peyaud}, Q.~{Piel}, S.~{Pita},
  V.~{Poireau}, A.~{Priyana Noel}, D.~A. {Prokhorov}, H.~{Prokoph},
  G.~{P{\"u}hlhofer}, M.~{Punch}, A.~{Quirrenbach}, S.~{Raab}, R.~{Rauth},
  P.~{Reichherzer}, A.~{Reimer}, O.~{Reimer}, Q.~{Remy}, M.~{Renaud},
  F.~{Rieger}, L.~{Rinchiuso}, C.~{Romoli}, G.~{Rowell}, B.~{Rudak},
  E.~{Ruiz-Velasco}, V.~{Sahakian}, S.~{Sailer}, H.~{Salzmann}, D.~A.
  {Sanchez}, A.~{Santangelo}, M.~{Sasaki}, M.~{Scalici}, J.~{Sch{\"a}fer},
  F.~{Sch{\"u}ssler}, H.~M. {Schutte}, U.~{Schwanke}, M.~{Seglar-Arroyo},
  M.~{Senniappan}, A.~S. {Seyffert}, N.~{Shafi}, J.~N.~S. {Shapopi},
  K.~{Shiningayamwe}, R.~{Simoni}, A.~{Sinha}, H.~{Sol}, A.~{Specovius},
  S.~{Spencer}, M.~{Spir-Jacob}, {\L}.~{Stawarz}, L.~{Sun}, R.~{Steenkamp},
  C.~{Stegmann}, S.~{Steinmassl}, C.~{Steppa}, T.~{Takahashi}, T.~{Tam},
  T.~{Tavernier}, A.~M. {Taylor}, R.~{Terrier}, J.~H.~E. {Thiersen},
  D.~{Tiziani}, M.~{Tluczykont}, L.~{Tomankova}, M.~{Tsirou}, R.~{Tuffs},
  Y.~{Uchiyama}, D.~J. {van der Walt}, C.~{van Eldik}, C.~{van Rensburg},
  B.~{van Soelen}, G.~{Vasileiadis}, J.~{Veh}, C.~{Venter}, P.~{Vincent},
  J.~{Vink}, H.~J. {V{\"o}lk}, Z.~{Wadiasingh}, S.~J. {Wagner}, J.~{Watson},
  F.~{Werner}, R.~{White}, A.~{Wierzcholska}, Yu~Wun {Wong}, A.~{Yusafzai},
  M.~{Zacharias}, R.~{Zanin}, D.~{Zargaryan}, A.~A. {Zdziarski}, A.~{Zech},
  S.~J. {Zhu}, J.~{Zorn}, S.~{Zouari}, N.~{{\.Z}ywucka}, P.~{Evans}, and
  K.~{Page}.
\newblock {Revealing x-ray and gamma ray temporal and spectral similarities in
  the GRB 190829A afterglow}.
\newblock {\em Science}, 372(6546):1081--1085, June 2021.

\bibitem{2019ApJ...880L..27D}
Evgeny {Derishev} and Tsvi {Piran}.
\newblock {The Physical Conditions of the Afterglow Implied by
  MAGIC{\textquoteright}s Sub-TeV Observations of GRB 190114C}.
\newblock {\em \apjl}, 880(2):L27, August 2019.

\bibitem{2019ApJ...884..117W}
Xiang-Yu {Wang}, Ruo-Yu {Liu}, Hai-Ming {Zhang}, Shao-Qiang {Xi}, and Bing
  {Zhang}.
\newblock {Synchrotron Self-Compton Emission from External Shocks as the Origin
  of the Sub-TeV Emission in GRB 180720B and GRB 190114C}.
\newblock {\em \apj}, 884(2):117, October 2019.

\bibitem{2023arXiv230314172L}
S.~{Lesage}, P.~{Veres}, M.~S. {Briggs}, A.~{Goldstein}, D.~{Kocevski},
  E.~{Burns}, C.~A. {Wilson-Hodge}, P.~N. {Bhat}, D.~{Huppenkothen}, C.~L.
  {Fryer}, R.~{Hamburg}, J.~{Racusin}, E.~{Bissaldi}, W.~H. {Cleveland},
  S.~{Dalessi}, C.~{Fletcher}, M.~M. {Giles}, B.~A. {Hristov}, C.~M. {Hui},
  B.~{Mailyan}, S.~{Poolakkil}, O.~J. {Roberts}, A.~{von Kienlin}, J.~{Wood},
  M.~{Ajello}, M.~{Arimoto}, L.~{Baldini}, J.~{Ballet}, M.~G. {Baring},
  D.~{Bastieri}, J.~{Becerra Gonzalez}, R.~{Bellazzini}, E.~{Bissaldi}, R.~D.
  {Blandford}, R.~{Bonino}, P.~{Bruel}, S.~{Buson}, R.~A. {Cameron},
  R.~{Caputo}, P.~A. {Caraveo}, E.~{Cavazzuti}, G.~{Chiaro}, N.~{Cibrario},
  S.~{Ciprini}, P.~{Cristarella Orestano}, M.~{Crnogorcevic}, A.~{Cuoco},
  S.~{Cutini}, F.~{DAmmando}, S.~{De Gaetano}, N.~{Di Lalla}, L.~{Di Venere},
  A.~{Dominguez}, S.~J. {Fegan}, E.~C. {Ferrara}, H.~{Fleischhack},
  Y.~{Fukazawa}, S.~{Funk}, P.~{Fusco}, G.~{Galanti}, V.~{Gammaldi},
  F.~{Gargano}, C.~{Gasbarra}, D.~{Gasparrini}, S.~{Germani}, F.~{Giacchino},
  N.~{Giglietto}, R.~{Gill}, M.~{Giroletti}, J.~{Granot}, D.~{Green}, I.~A.
  {Grenier}, S.~{Guiriec}, M.~{Gustafsson}, E.~{Hays}, J.~W. {Hewitt},
  D.~{Horan}, X.~{Hou}, M.~{Kuss}, L.~{Latronico}, A.~{Laviron},
  M.~{Lemoine-Goumard}, J.~{Li}, I.~{Liodakis}, F.~{Longo}, F.~{Loparco},
  L.~{Lorusso}, M.~N. {Lovellette}, P.~{Lubrano}, S.~{Maldera}, A.~{Manfreda},
  G.~{Marti-Devesa}, M.~N. {Mazziotta}, J.~E. {McEnery}, I.~{Mereu},
  M.~{Meyer}, P.~F. {Michelson}, T.~{Mizuno}, M.~E. {Monzani}, A.~{Morselli},
  I.~V. {Moskalenko}, M.~{Negro}, E.~{Nuss}, N.~{Omodei}, E.~{Orlando}, J.~F.
  {Ormes}, D.~{Paneque}, G.~{Panzarini}, M.~{Persic}, M.~{Pesce-Rollins},
  R.~{Pillera}, F.~{Piron}, H.~{Poon}, T.~A. {Porter}, G.~{Principe},
  S.~{Raino}, R.~{Rando}, B.~{Rani}, M.~{Razzano}, S.~{Razzaque}, A.~{Reimer},
  O.~{Reimer}, F.~{Ryde}, M.~{Sanchez-Conde}, P.~M. {Saz Parkinson},
  L.~{Scotton}, D.~{Serini}, C.~{Sgro}, V.~{Sharma}, E.~J. {Siskind},
  G.~{Spandre}, P.~{Spinelli}, H.~{Tajima}, D.~F. {Torres}, J.~{Valverde},
  T.~{Venters}, Z.~{Wadiasingh}, K.~{Wood}, and G.~{Zaharijas}.
\newblock {Fermi-GBM Discovery of GRB 221009A: An Extraordinarily Bright GRB
  from Onset to Afterglow}.
\newblock {\em arXiv e-prints}, page arXiv:2303.14172, March 2023.

\bibitem{2022GCN.32635....1K}
J.~A. {Kennea}, M.~{Williams}, and {Swift Team}.
\newblock {GRB 221009A: Swift detected transient may be GRB}.
\newblock {\em GRB Coordinates Network}, 32635:1, October 2022.

\bibitem{2022GCN.32658....1P}
R.~{Pillera}, E.~{Bissaldi}, N.~{Omodei}, G.~{La Mura}, F.~{Longo}, and
  {Fermi-LAT team}.
\newblock {GRB 221009A: Fermi-LAT refined analysis}.
\newblock {\em GRB Coordinates Network}, 32658:1, October 2022.

\bibitem{2023arXiv230301203A}
Zheng-Hua {An}, S.~{Antier}, Xing-Zi {Bi}, Qing-Cui {Bu}, Ce~{Cai}, Xue-Lei
  {Cao}, Anna-Elisa {Camisasca}, Zhi {Chang}, Gang {Chen}, Li~{Chen},
  Tian-Xiang {Chen}, Wen {Chen}, Yi-Bao {Chen}, Yong {Chen}, Yu-Peng {Chen},
  Michael~W. {Coughlin}, Wei-Wei {Cui}, Zi-Gao {Dai}, T.~{Hussenot-Desenonges},
  Yan-Qi {Du}, Yuan-Yuan {Du}, Yun-Fei {Du}, Cheng-Cheng {Fan}, Filippo
  {Frontera}, He~{Gao}, Min {Gao}, Ming-Yu {Ge}, Ke~{Gong}, Yu-Dong {Gu},
  Ju~{Guan}, Dong-Ya {Guo}, Zhi-Wei {Guo}, Cristiano {Guidorzi}, Da-Wei {Han},
  Jian-Jian {He}, Jun-Wang {He}, Dong-Jie {Hou}, Yue {Huang}, Jia {Huo}, Zhen
  {Ji}, Shu-Mei {Jia}, Wei-Chun {Jiang}, David~Alexander {Kann}, A.~{Klotz},
  Ling-Da {Kong}, Lin {Lan}, An~{Li}, Bing {Li}, Chao-Yang {Li}, Cheng-Kui
  {Li}, Gang {Li}, Mao-Shun {Li}, Ti-Pei {Li}, Wei {Li}, Xiao-Bo {Li}, Xin-Qiao
  {Li}, Xu-Fang {Li}, Yan-Guo {Li}, Zheng-Wei {Li}, Jing {Liang}, Xiao-Hua
  {Liang}, Jin-Yuan {Liao}, Lin {Lin}, Cong-Zhan {Liu}, He-Xin {Liu}, Hong-Wei
  {Liu}, Jia-Cong {Liu}, Xiao-Jing {Liu}, Ya-Qing {Liu}, Yu-Rong {Liu},
  Fang-Jun {Lu}, Hong {Lu}, Xue-Feng {Lu}, Qi~{Luo}, Tao {Luo}, Bin-Yuan {Ma},
  Fu-Li {Ma}, Rui-Can {Ma}, Xiang {Ma}, Romain {Maccary}, Ji-Rong {Mao}, Bin
  {Meng}, Jian-Yin {Nie}, Mauro {Orlandini}, Ge~{Ou}, Jing-Qiang {Peng}, Wen-Xi
  {Peng}, Rui {Qiao}, Jin-Lu {Qu}, Xiao-Qin {Ren}, Jing-Yan {Shi}, Qi~{Shi},
  Li-Ming {Song}, Xin-Ying {Song}, Ju~{Su}, Gong-Xing {Sun}, Liang {Sun},
  Xi-Lei {Sun}, Wen-Jun {Tan}, Ying {Tan}, Lian {Tao}, You-Li {Tuo}, Damien
  {Turpin}, Jin-Zhou {Wang}, Chen {Wang}, Chen-Wei {Wang}, Hong-Jun {Wang}, Hui
  {Wang}, Jin {Wang}, Ling-Jun {Wang}, Peng-Ju {Wang}, Ping {Wang}, Wen-Shuai
  {Wang}, Xiang-Yu {Wang}, Xi-Lu {Wang}, Yu-Sa {Wang}, Yue {Wang}, Xiang-Yang
  {Wen}, Bo-Bing {Wu}, Bai-Yang {Wu}, Hong {Wu}, Sheng-Hui {Xiao}, Shuo {Xiao},
  Yun-Xiang {Xiao}, Sheng-Lun {Xie}, Shao-Lin {Xiong}, Sen-Lin {Xiong}, Dong
  {Xu}, He~{Xu}, Yan-Jun {Xu}, Yan-Bing {Xu}, Ying-Chen {Xu}, Yu-Peng {Xu},
  Wang-Chen {Xue}, Sheng {Yang}, Yan-Ji {Yang}, Zi-Xu {Yang}, Wen-Tao {Ye},
  Qi-Bin {Yi}, Shu-Xu {Yi}, Qian-Qing {Yin}, Yuan {You}, Yun-Wei {Yu}, Wei
  {Yu}, Wen-Hui {Yu}, Ming {Zeng}, Bing {Zhang}, Bin-Bin {Zhang}, Da-Li
  {Zhang}, Fan {Zhang}, Hong-Mei {Zhang}, Juan {Zhang}, Liang {Zhang}, Peng
  {Zhang}, Peng {Zhang}, Shu {Zhang}, Shuang-Nan {Zhang}, Wan-Chang {Zhang},
  Xiao-Feng {Zhang}, Xiao-Lu {Zhang}, Yan-Qiu {Zhang}, Yan-Ting {Zhang}, Yi-Fei
  {Zhang}, Yuan-Hang {Zhang}, Zhen {Zhang}, Guo-Ying {Zhao}, Hai-Sheng {Zhao},
  Hong-Yu {Zhao}, Qing-Xia {Zhao}, Shu-Jie {Zhao}, Xiao-Yun {Zhao}, Xiao-Fan
  {Zhao}, Yi~{Zhao}, Chao {Zheng}, Shi-Jie {Zheng}, Deng-Ke {Zhou}, Xing
  {Zhou}, and Xiao-Cheng {Zhu}.
\newblock {Insight-HXMT and GECAM-C observations of the brightest-of-all-time
  GRB 221009A}.
\newblock {\em arXiv e-prints}, page arXiv:2303.01203, March 2023.

\bibitem{2023arXiv230213383F}
D.~{Frederiks}, D.~{Svinkin}, A.~L. {Lysenko}, S.~{Molkov}, A.~{Tsvetkova},
  M.~{Ulanov}, A.~{Ridnaia}, A.~A. {Lutovinov}, I.~{Lapshov}, A.~{Tkachenko},
  and V.~{Levin}.
\newblock {Properties of the extremely energetic GRB 221009A from Konus-WIND
  and SRG/ART-XC observations}.
\newblock {\em arXiv e-prints}, page arXiv:2302.13383, February 2023.

\bibitem{2022GCN.32648....1D}
A.~{de Ugarte Postigo}, L.~{Izzo}, G.~{Pugliese}, D.~{Xu}, B.~{Schneider},
  J.~P.~U. {Fynbo}, N.~R. {Tanvir}, D.~B. {Malesani}, A.~{Saccardi}, D.~A.
  {Kann}, K.~{Wiersema}, B.~P. {Gompertz}, C.~C. {Thoene}, A.~J. {Levan}, and
  {Stargate Collaboration}.
\newblock {GRB 221009A: Redshift from X-shooter/VLT}.
\newblock {\em GRB Coordinates Network}, 32648:1, October 2022.

\bibitem{2022ChPhC..46c0001M}
Xin-Hua {Ma}, Yu-Jiang {Bi}, Zhen {Cao}, Ming-Jun {Chen}, Song-Zhan {Chen},
  Yao-Dong {Cheng}, Guang-Hua {Gong}, Min-Hao {Gu}, Hui-Hai {He}, Chao {Hou},
  Wen-Hao {Huang}, Xing-Tao {Huang}, Cheng {Liu}, Oleg {Shchegolev}, Xiang-Dong
  {Sheng}, Yuri {Stenkin}, Chao-Yong {Wu}, Han-Rong {Wu}, Sha {Wu}, Gang
  {Xiao}, Zhi-Guo {Yao}, Shou-Shan {Zhang}, Yi~{Zhang}, and Xiong {Zuo}.
\newblock {Chapter 1 LHAASO Instruments and Detector technology}.
\newblock {\em Chinese Physics C}, 46(3):030001, March 2022.

\bibitem{2022GCN.32677....1H}
Yong {Huang}, Shicong {Hu}, Songzhan {Chen}, Min {Zha}, Cheng {Liu}, Zhiguo
  {Yao}, Zhen {Cao}, and The~Lhaaso {Experiment}.
\newblock {LHAASO observed GRB 221009A with more than 5000 VHE photons up to
  around 18 TeV}.
\newblock {\em GRB Coordinates Network}, 32677:1, October 2022.

\bibitem{2023ScienceGRB}
{LHAASO Collaboration}, Z.~{Cao}, F.~{Aharonian}, Q.~{An}, A.~{Axikegu}, L.~X.
  {Bai}, Y.~X. {Bai}, Y.~W. {Bao}, D.~{Bastieri}, X.~J. {Bi}, Y.~J. {Bi}, J.~T.
  {Cai}, Q.~{Cao}, W.~Y. {Cao}, Z.~{Cao}, J.~{Chang}, J.~F. {Chang}, E.~S.
  {Chen}, L.~{Chen}, L.~{Chen}, L.~{Chen}, M.~J. {Chen}, M.~L. {Chen}, Q.~H.
  {Chen}, S.~H. {Chen}, S.~Z. {Chen}, T.~L. {Chen}, Y.~{Chen}, H.~L. {Cheng},
  N.~{Cheng}, Y.~D. {Cheng}, S.~W. {Cui}, X.~H. {Cui}, Y.~D. {Cui}, B.~Z.
  {Dai}, H.~L. {Dai}, D.~{Danzengluobu}, D.~{Della Volpe}, X.~Q. {Dong}, K.~K.
  {Duan}, J.~H. {Fan}, Y.~Z. {Fan}, J.~{Fang}, K.~{Fang}, C.~F. {Feng},
  L.~{Feng}, S.~H. {Feng}, X.~T. {Feng}, Y.~L. {Feng}, B.~{Gao}, C.~D. {Gao},
  L.~Q. {Gao}, Q.~{Gao}, W.~{Gao}, W.~K. {Gao}, M.~M. {Ge}, L.~S. {Geng}, G.~H.
  {Gong}, Q.~B. {Gou}, M.~H. {Gu}, F.~L. {Guo}, X.~L. {Guo}, Y.~Q. {Guo}, Y.~Y.
  {Guo}, Y.~A. {Han}, H.~H. {He}, H.~N. {He}, J.~Y. {He}, X.~B. {He}, Y.~{He},
  M.~{Heller}, Y.~K. {Hor}, B.~W. {Hou}, C.~{Hou}, X.~{Hou}, H.~B. {Hu},
  Q.~{Hu}, S.~C. {Hu}, D.~H. {Huang}, T.~Q. {Huang}, W.~J. {Huang}, X.~T.
  {Huang}, Z.~C. {Huang}, X.~L. {Ji}, H.~Y. {Jia}, K.~{Jia}, K.~{Jiang}, X.~W.
  {Jiang}, Z.~J. {Jiang}, M.~{Jin}, M.~M. {Kang}, T.~{Ke}, D.~{Kuleshov},
  K.~{Kurinov}, B.~B. {Li}, C.~{Li}, C.~{Li}, D.~{Li}, F.~{Li}, H.~B. {Li},
  H.~C. {Li}, H.~Y. {Li}, J.~{Li}, J.~{Li}, J.~{Li}, K.~{Li}, W.~L. {Li}, W.~L.
  {Li}, X.~R. {Li}, X.~{Li}, Y.~Z. {Li}, Z.~{Li}, Z.~{Li}, E.~W. {Liang}, Y.~F.
  {Liang}, S.~J. {Lin}, B.~{Liu}, C.~{Liu}, D.~{Liu}, H.~{Liu}, H.~D. {Liu},
  J.~{Liu}, J.~L. {Liu}, J.~L. {Liu}, J.~S. {Liu}, J.~Y. {Liu}, M.~Y. {Liu},
  R.~Y. {Liu}, S.~M. {Liu}, W.~{Liu}, Y.~{Liu}, Y.~N. {Liu}, W.~J. {Long},
  R.~{Lu}, Q.~{Luo}, H.~K. {Lv}, B.~Q. {Ma}, L.~L. {Ma}, X.~H. {Ma}, J.~R.
  {Mao}, Z.~{Min}, W.~{Mitthumsiri}, Y.~C. {Nan}, Z.~W. {Ou}, B.~Y. {Pang},
  P.~{Pattarakijwanich}, Z.~Y. {Pei}, M.~Y. {Qi}, Y.~Q. {Qi}, B.~Q. {Qiao},
  J.~J. {Qin}, D.~{Ruffolo}, A.~{Saiz}, C.~Y. {Shao}, L.~{Shao},
  O.~{Shchegolev}, X.~D. {Sheng}, H.~C. {Song}, Y.~V. {Stenkin}, V.~{Stepanov},
  Y.~{Su}, Q.~N. {Sun}, X.~N. {Sun}, Z.~B. {Sun}, P.~H.~T. {Tam}, Z.~B. {Tang},
  W.~W. {Tian}, C.~{Wang}, C.~B. {Wang}, G.~W. {Wang}, H.~G. {Wang}, H.~H.
  {Wang}, J.~C. {Wang}, J.~S. {Wang}, K.~{Wang}, L.~P. {Wang}, L.~Y. {Wang},
  P.~H. {Wang}, R.~{Wang}, W.~{Wang}, X.~G. {Wang}, Y.~D. {Wang}, Y.~J. {Wang},
  Z.~H. {Wang}, Z.~X. {Wang}, Z.~{Wang}, D.~M. {Wei}, J.~J. {Wei}, Y.~J. {Wei},
  T.~{Wen}, C.~Y. {Wu}, H.~R. {Wu}, S.~{Wu}, X.~F. {Wu}, Y.~S. {Wu}, S.~Q.
  {Xi}, J.~{Xia}, J.~J. {Xia}, G.~M. {Xiang}, D.~X. {Xiao}, G.~{Xiao}, G.~G.
  {Xin}, Y.~L. {Xin}, Y.~{Xing}, Z.~{Xiong}, D.~L. {Xu}, R.~F. {Xu}, R.~X.
  {Xu}, L.~{Xue}, D.~H. {Yan}, J.~Z. {Yan}, T.~{Yan}, C.~W. {Yang}, F.~{Yang},
  F.~F. {Yang}, H.~W. {Yang}, J.~Y. {Yang}, L.~L. {Yang}, M.~J. {Yang}, R.~Z.
  {Yang}, S.~B. {Yang}, Y.~H. {Yao}, Y.~M. {Ye}, L.~Q. {Yin}, N.~{Yin}, X.~H.
  {You}, Z.~Y. {You}, Y.~H. {Yu}, Q.~{Yuan}, H.~{Yue}, H.~D. {Zeng}, T.~X.
  {Zeng}, W.~{Zeng}, Z.~K. {Zeng}, B.~{Zhang}, B.~B. {Zhang}, F.~{Zhang}, H.~M.
  {Zhang}, H.~Y. {Zhang}, J.~L. {Zhang}, L.~X. {Zhang}, L.~{Zhang}, P.~F.
  {Zhang}, P.~P. {Zhang}, R.~{Zhang}, S.~B. {Zhang}, S.~R. {Zhang}, S.~S.
  {Zhang}, X.~{Zhang}, X.~P. {Zhang}, Y.~F. {Zhang}, Y.~{Zhang}, Y.~{Zhang},
  B.~{Zhao}, J.~{Zhao}, L.~{Zhao}, L.~Z. {Zhao}, S.~P. {Zhao}, F.~{Zheng},
  B.~{Zhou}, H.~{Zhou}, J.~N. {Zhou}, P.~{Zhou}, R.~{Zhou}, X.~X. {Zhou}, C.~G.
  {Zhu}, F.~R. {Zhu}, H.~{Zhu}, K.~J. {Zhu}, and X.~{Zuo}.
\newblock {A tera-electron volt afterglow from a narrow jet in an extremely
  bright gamma-ray burst.}
\newblock {\em Science}, 380(6652):1390--1396, June 2023.

\bibitem{2021ChPhC..45b5002A}
F.~{Aharonian}, Q.~{An}, {Axikegu}, L.~X. {Bai}, Y.~X. {Bai}, Y.~W. {Bao},
  D.~{Bastieri}, X.~J. {Bi}, Y.~J. {Bi}, H.~{Cai}, J.~T. {Cai}, Z.~{Cao},
  Z.~{Cao}, J.~{Chang}, J.~F. {Chang}, X.~C. {Chang}, B.~M. {Chen}, J.~{Chen},
  L.~{Chen}, L.~{Chen}, L.~{Chen}, M.~J. {Chen}, M.~L. {Chen}, Q.~H. {Chen},
  S.~H. {Chen}, S.~Z. {Chen}, T.~L. {Chen}, X.~L. {Chen}, Y.~{Chen},
  N.~{Cheng}, Y.~D. {Cheng}, S.~W. {Cui}, X.~H. {Cui}, Y.~D. {Cui}, B.~Z.
  {Dai}, H.~L. {Dai}, Z.~G. {Dai}, {Danzengluobu}, D.~{Della Volpe},
  B.~D'ettorre {Piazzoli}, X.~J. {Dong}, J.~H. {Fan}, Y.~Z. {Fan}, Z.~X. {Fan},
  J.~{Fang}, K.~{Fang}, C.~F. {Feng}, L.~{Feng}, S.~H. {Feng}, Y.~L. {Feng},
  B.~{Gao}, C.~D. {Gao}, Q.~{Gao}, W.~{Gao}, M.~M. {Ge}, L.~S. {Geng}, G.~H.
  {Gong}, Q.~B. {Gou}, M.~H. {Gu}, J.~G. {Guo}, X.~L. {Guo}, Y.~Q. {Guo}, Y.~Y.
  {Guo}, Y.~A. {Han}, H.~H. {He}, H.~N. {He}, J.~C. {He}, S.~L. {He}, X.~B.
  {He}, Y.~{He}, M.~{Heller}, Y.~K. {Hor}, C.~{Hou}, X.~{Hou}, H.~B. {Hu},
  S.~{Hu}, S.~C. {Hu}, X.~J. {Hu}, D.~H. {Huang}, Q.~L. {Huang}, W.~H. {Huang},
  X.~T. {Huang}, Z.~C. {Huang}, F.~{Ji}, X.~L. {Ji}, H.~Y. {Jia}, K.~{Jiang},
  Z.~J. {Jiang}, C.~{Jin}, D.~{Kuleshov}, K.~{Levochkin}, B.~B. {Li}, C.~{Li},
  C.~{Li}, F.~{Li}, H.~B. {Li}, H.~C. {Li}, H.~Y. {Li}, J.~{Li}, K.~{Li}, W.~L.
  {Li}, X.~{Li}, X.~{Li}, X.~R. {Li}, Y.~{Li}, Y.~Z. {Li}, Z.~{Li}, Z.~{Li},
  E.~W. {Liang}, Y.~F. {Liang}, S.~J. {Lin}, B.~{Liu}, C.~{Liu}, D.~{Liu},
  H.~{Liu}, H.~D. {Liu}, J.~{Liu}, J.~L. {Liu}, J.~S. {Liu}, J.~Y. {Liu}, M.~Y.
  {Liu}, R.~Y. {Liu}, S.~M. {Liu}, W.~{Liu}, Y.~N. {Liu}, Z.~X. {Liu}, W.~J.
  {Long}, R.~{Lu}, H.~K. {Lv}, B.~Q. {Ma}, L.~L. {Ma}, X.~H. {Ma}, J.~R. {Mao},
  A.~{Masood}, W.~{Mitthumsiri}, T.~{Montaruli}, Y.~C. {Nan}, B.~Y. {Pang},
  P.~{Pattarakijwanich}, Z.~Y. {Pei}, M.~Y. {Qi}, D.~{Ruffolo}, V.~{Rulev},
  A.~{S{\'a}iz}, L.~{Shao}, O.~{Shchegolev}, X.~D. {Sheng}, J.~R. {Shi}, H.~C.
  {Song}, Yu.~V. {Stenkin}, V.~{Stepanov}, Q.~N. {Sun}, X.~N. {Sun}, Z.~B.
  {Sun}, P.~H.~T. {Tam}, Z.~B. {Tang}, W.~W. {Tian}, B.~D. {Wang}, C.~{Wang},
  H.~{Wang}, H.~G. {Wang}, J.~C. {Wang}, J.~S. {Wang}, L.~P. {Wang}, L.~Y.
  {Wang}, R.~N. {Wang}, W.~{Wang}, W.~{Wang}, X.~G. {Wang}, X.~J. {Wang}, X.~Y.
  {Wang}, Y.~D. {Wang}, Y.~J. {Wang}, Y.~P. {Wang}, Z.~{Wang}, Z.~{Wang}, Z.~H.
  {Wang}, Z.~X. {Wang}, D.~M. {Wei}, J.~J. {Wei}, Y.~J. {Wei}, T.~{Wen}, C.~Y.
  {Wu}, H.~R. {Wu}, S.~{Wu}, W.~X. {Wu}, X.~F. {Wu}, S.~Q. {Xi}, J.~{Xia},
  J.~J. {Xia}, G.~M. {Xiang}, G.~{Xiao}, H.~B. {Xiao}, G.~G. {Xin}, Y.~L.
  {Xin}, Y.~{Xing}, D.~L. {Xu}, R.~X. {Xu}, L.~{Xue}, D.~H. {Yan}, C.~W.
  {Yang}, F.~F. {Yang}, J.~Y. {Yang}, L.~L. {Yang}, M.~J. {Yang}, R.~Z. {Yang},
  S.~B. {Yang}, Y.~H. {Yao}, Z.~G. {Yao}, Y.~M. {Ye}, L.~Q. {Yin}, N.~{Yin},
  X.~H. {You}, Z.~Y. {You}, Y.~H. {Yu}, Q.~{Yuan}, H.~D. {Zeng}, T.~X. {Zeng},
  W.~{Zeng}, Z.~K. {Zeng}, M.~{Zha}, X.~X. {Zhai}, B.~B. {Zhang}, H.~M.
  {Zhang}, H.~Y. {Zhang}, J.~L. {Zhang}, J.~W. {Zhang}, L.~{Zhang}, L.~{Zhang},
  L.~X. {Zhang}, P.~F. {Zhang}, P.~P. {Zhang}, R.~{Zhang}, S.~R. {Zhang}, S.~S.
  {Zhang}, X.~{Zhang}, X.~P. {Zhang}, Y.~{Zhang}, Y.~{Zhang}, Y.~F. {Zhang},
  Y.~L. {Zhang}, B.~{Zhao}, J.~{Zhao}, L.~{Zhao}, L.~Z. {Zhao}, S.~P. {Zhao},
  F.~{Zheng}, Y.~{Zheng}, B.~{Zhou}, H.~{Zhou}, J.~N. {Zhou}, P.~{Zhou},
  R.~{Zhou}, X.~X. {Zhou}, C.~G. {Zhu}, F.~R. {Zhu}, H.~{Zhu}, K.~J. {Zhu},
  X.~{Zuo}, and {(Lhaaso Collaboration)}.
\newblock {Observation of the Crab Nebula with LHAASO-KM2A - a performance
  study}.
\newblock {\em Chinese Physics C}, 45(2):025002, February 2021.

\bibitem{2021MNRAS.507.5144S}
Alberto {Saldana-Lopez}, Alberto {Dom{\'\i}nguez}, Pablo~G.
  {P{\'e}rez-Gonz{\'a}lez}, Justin {Finke}, Marco {Ajello}, Joel~R. {Primack},
  Vaidehi~S. {Paliya}, and Abhishek {Desai}.
\newblock {An observational determination of the evolving extragalactic
  background light from the multiwavelength HST/CANDELS survey in the Fermi and
  CTA era}.
\newblock {\em \mnras}, 507(4):5144--5160, November 2021.

\bibitem{2012MNRAS.422.3189G}
Rudy~C. {Gilmore}, Rachel~S. {Somerville}, Joel~R. {Primack}, and Alberto
  {Dom{\'\i}nguez}.
\newblock {Semi-analytic modelling of the extragalactic background light and
  consequences for extragalactic gamma-ray spectra}.
\newblock {\em \mnras}, 422(4):3189--3207, June 2012.

\bibitem{2011MNRAS.410.2556D}
A.~{Dom{\'\i}nguez}, J.~R. {Primack}, D.~J. {Rosario}, F.~{Prada}, R.~C.
  {Gilmore}, S.~M. {Faber}, D.~C. {Koo}, R.~S. {Somerville}, M.~A.
  {P{\'e}rez-Torres}, P.~{P{\'e}rez-Gonz{\'a}lez}, J.~S. {Huang}, M.~{Davis},
  P.~{Guhathakurta}, P.~{Barmby}, C.~J. {Conselice}, M.~{Lozano}, J.~A.
  {Newman}, and M.~C. {Cooper}.
\newblock {Extragalactic background light inferred from AEGIS galaxy-SED-type
  fractions}.
\newblock {\em \mnras}, 410(4):2556--2578, February 2011.

\bibitem{2010ApJ...712..238F}
Justin~D. {Finke}, Soebur {Razzaque}, and Charles~D. {Dermer}.
\newblock {Modeling the Extragalactic Background Light from Stars and Dust}.
\newblock {\em \apj}, 712(1):238--249, March 2010.

\bibitem{1999A&A...349...11A}
F.~A. {Aharonian}, A.~G. {Akhperjanian}, J.~A. {Barrio}, K.~{Bernl{\"o}hr},
  H.~{Bojahr}, I.~{Calle}, J.~L. {Contreras}, J.~{Cortina}, A.~{Daum},
  T.~{Deckers}, S.~{Denninghoff}, V.~{Fonseca}, J.~C. {Gonzalez},
  G.~{Heinzelmann}, M.~{Hemberger}, G.~{Hermann}, M.~{He{\ss}}, A.~{Heusler},
  W.~{Hofmann}, H.~{Hohl}, D.~{Horns}, A.~{Ibarra}, R.~{Kankanyan},
  J.~{Kettler}, C.~{K{\"o}hler}, A.~{Konopelko}, H.~{Kornmeyer}, M.~{Kestel},
  D.~{Kranich}, H.~{Krawczynski}, H.~{Lampeitl}, A.~{Lindner}, E.~{Lorenz},
  N.~{Magnussen}, H.~{Meyer}, R.~{Mirzoyan}, A.~{Moralejo}, L.~{Padilla},
  M.~{Panter}, D.~{Petry}, R.~{Plaga}, A.~{Plyasheshnikov}, J.~{Prahl},
  G.~{P{\"u}hlhofer}, G.~{Rauterberg}, C.~{Renault}, W.~{Rhode},
  A.~{R{\"o}hring}, V.~{Sahakian}, M.~{Samorski}, D.~{Schmele},
  F.~{Schr{\"o}der}, W.~{Stamm}, H.~J. {V{\"o}lk}, B.~{Wiebel-Sooth},
  C.~{Wiedner}, M.~{Willmer}, and W.~{Wittek}.
\newblock {The time averaged TeV energy spectrum of MKN 501 of the
  extraordinary 1997 outburst as measured with the stereoscopic Cherenkov
  telescope system of HEGRA}.
\newblock {\em \aap}, 349:11--28, September 1999.

\bibitem{2002A&A...384..834A}
F.~A. {Aharonian}, A.~N. {Timokhin}, and A.~V. {Plyasheshnikov}.
\newblock {On the origin of highest energy gamma-rays from Mkn 501}.
\newblock {\em \aap}, 384:834--847, March 2002.

\bibitem{2001ICRC...27I.250A}
F.~A. {Aharonian}.
\newblock {TeV blazars and cosmic infrared background radiation}.
\newblock In {\em 27th International Cosmic Ray Conference (ICRC27)}, volume~27
  of {\em International Cosmic Ray Conference}, page 250, January 2001.

\bibitem{Berta:2010rc}
S.~Berta et~al.
\newblock {Dissecting the cosmic infra-red background with Herschel/PEP}.
\newblock {\em Astron. Astrophys.}, 518:L30, 2010.

\bibitem{CAST:2017uph}
V.~Anastassopoulos et~al.
\newblock {New CAST Limit on the Axion-Photon Interaction}.
\newblock {\em Nature Phys.}, 13:584--590, 2017.

\bibitem{HESS:2013udx}
A.~Abramowski et~al.
\newblock {Constraints on axionlike particles with H.E.S.S. from the
  irregularity of the PKS 2155-304 energy spectrum}.
\newblock {\em Phys. Rev. D}, 88(10):102003, 2013.

\bibitem{Li:2020pcn}
Hai-Jun Li, Jun-Guang Guo, Xiao-Jun Bi, Su-Jie Lin, and Peng-Fei Yin.
\newblock {Limits on axion-like particles from Mrk 421 with 4.5-year period
  observations by ARGO-YBJ and Fermi-LAT}.
\newblock {\em Phys. Rev. D}, 103(8):083003, 2021.

\bibitem{LHAASO:2021opi}
Zhen Cao et~al.
\newblock {Exploring Lorentz Invariance Violation from Ultrahigh-Energy
  \ensuremath{\gamma} Rays Observed by LHAASO}.
\newblock {\em Phys. Rev. Lett.}, 128(5):051102, 2022.

\bibitem{FermiGBMLAT:2009nfe}
M.~Ackermann et~al.
\newblock {A limit on the variation of the speed of light arising from quantum
  gravity effects}.
\newblock {\em Nature}, 462:331--334, 2009.

\bibitem{2009ApJ...703..675N}
Ehud {Nakar}, Shin'ichiro {Ando}, and Re'em {Sari}.
\newblock {Klein-Nishina Effects on Optically Thin Synchrotron and Synchrotron
  Self-Compton Spectrum}.
\newblock {\em \apj}, 703(1):675--691, September 2009.

\bibitem{2023ApJ...947L..14Z}
B.~Theodore {Zhang}, Kohta {Murase}, Kunihito {Ioka}, Deheng {Song}, Chengchao
  {Yuan}, and P{\'e}ter {M{\'e}sz{\'a}ros}.
\newblock {External Inverse-compton and Proton Synchrotron Emission from the
  Reverse Shock as the Origin of VHE Gamma Rays from the Hyper-bright GRB
  221009A}.
\newblock {\em \apjl}, 947(1):L14, April 2023.

\bibitem{2023A&A...670L..12D}
Saikat {Das} and Soebur {Razzaque}.
\newblock {Ultrahigh-energy cosmic-ray signature in GRB 221009A}.
\newblock {\em \aap}, 670:L12, February 2023.

\bibitem{2023ApJ...947...87K}
Dmitry {Khangulyan}, Andrew~M. {Taylor}, and Felix {Aharonian}.
\newblock {The Formation of Hard Very High Energy Spectra from Gamma-ray Burst
  Afterglows via Two-zone Synchrotron Self-Compton Emission}.
\newblock {\em \apj}, 947(2):87, April 2023.

\bibitem{2016PhRvD..94b3005A}
Katsuaki {Asano} and Peter {M{\'e}sz{\'a}ros}.
\newblock {Ultrahigh-energy cosmic ray production by turbulence in gamma-ray
  burst jets and cosmogenic neutrinos}.
\newblock {\em \prd}, 94(2):023005, July 2016.

\bibitem{Hauser:1998ri}
M.~G. Hauser et~al.
\newblock {The COBE diffuse infrared background experiment search for the
  cosmic infrared background. 1. Limits and detections}.
\newblock {\em Astrophys. J.}, 508:25, 1998.

\bibitem{Lagache:1999ji}
G.~Lagache, L.~M. Haffner, R.~J. Reynolds, and S.~L. Tufte.
\newblock {Evidence for dust emission in the warm ionised medium using wham
  data}.
\newblock {\em Astron. Astrophys.}, 354:247, 2000.

\bibitem{Gardner:1999jy}
J.~P. Gardner et~al.
\newblock {The hubble deep field south - stis imaging}.
\newblock {\em Astron. J.}, 119:486, 2000.

\bibitem{Elbaz:2002vd}
D.~Elbaz, C.~Cesarsky, P.~Chanial, H.~Aussel, A.~Franceschini, D.~Fadda, and
  R.~Chary.
\newblock {The Bulk of the cosmic infrared background resolved by ISOCAM}.
\newblock {\em Astron. Astrophys.}, 384:848--865, 2002.

\bibitem{Fazio:2004kx}
G.~G. Fazio et~al.
\newblock {Number counts at 3 \ensuremath{<} lambda \ensuremath{<} 10 um from
  the Spitzer Space Telescope}.
\newblock {\em Astrophys. J. Suppl.}, 154:39--43, 2004.

\bibitem{Xu:2004zg}
C.~Kevin Xu et~al.
\newblock {Number counts of GALEX sources in FUV (1530A) and NUV (2310A)
  bands}.
\newblock {\em Astrophys. J. Lett.}, 619:L11--L14, 2005.

\bibitem{Bethermin:2010jb}
Matthieu Bethermin, Herve Dole, Alexandre Beelen, and Herve Aussel.
\newblock {Spitzer Deep and Wide Legacy Mid- and Far-Infrared Number Counts and
  Lower Limits of Cosmic Infrared Background}.
\newblock {\em Astron. Astrophys.}, 512:A78, 2010.

\bibitem{Matsuura:2010rb}
S.~Matsuura et~al.
\newblock {Detection of the Cosmic Far-Infrared Background in the AKARI Deep
  Field South}.
\newblock {\em Astrophys. J.}, 737:2, 2011.

\bibitem{Voyer:2011mx}
Elysse~N. Voyer, Jonathan~P. Gardner, Harry~I. Teplitz, Brian~D. Siana, and
  Duilia~F. de~Mello.
\newblock {Far-Ultraviolet Number Counts of Field Galaxies}.
\newblock {\em Astrophys. J.}, 736:80, 2011.

\bibitem{Zemcov:2014eca}
Michael Zemcov et~al.
\newblock {On the Origin of Near-Infrared Extragalactic Background Light
  Anisotropy}.
\newblock {\em Science}, 346:732, 2014.

\bibitem{Driver:2016krv}
Simon~P. Driver, Stephen~K. Andrews, Luke~J. Davies, Aaron S.~G. Robotham,
  Angus~H. Wright, Rogier~A. Windhorst, Seth Cohen, Kim Emig, Rolf~A. Jansen,
  and Loretta Dunne.
\newblock {Measurements of Extragalactic Background Light From the far UV to
  the far IR From Deep Ground- and Space-based Galaxy Counts}.
\newblock {\em Astrophys. J.}, 827(2):108, 2016.

\bibitem{Finkbeiner:2000vr}
Douglas~P. Finkbeiner, Marc Davis, and David~J. Schlegel.
\newblock {Detection of a far ir excess with dirbe at 60 and 100 microns}.
\newblock {\em Astrophys. J.}, 544:81--97, 2000.

\bibitem{Madau:1999yh}
Piero Madau and Lucia Pozzetti.
\newblock {Deep galaxy counts, extragalactic background light, and the stellar
  baryon budget}.
\newblock {\em Mon. Not. Roy. Astron. Soc.}, 312:L9, 2000.

\bibitem{Metcalfe:2003zi}
L.~Metcalfe et~al.
\newblock {An ISOCAM survey through gravitationally lensing galaxy clusters. 1.
  Source lists and source counts for A370, A2218 and A2390}.
\newblock {\em Astron. Astrophys.}, 407:791--822, 2003.

\bibitem{Papovich:2004vh}
Casey Papovich et~al.
\newblock {The 24 micron source counts in deep spitzer surveys}.
\newblock {\em Astrophys. J. Suppl.}, 154:70--74, 2004.

\bibitem{Frayer:2006qq}
David~T. Frayer, M.~T. Huynh, R.~Chary, M.~Dickinson, D.~Elbaz, D.~Fadda, J.~A.
  Surace, H.~I. Teplitz, L.~Yan, and B.~Mobasher.
\newblock {Spitzer 70-micron Source Counts in GOODS-North}.
\newblock {\em Astrophys. J. Lett.}, 647:L9--L12, 2006.

\bibitem{Keenan:2010na}
Ryan~C. Keenan, Amy~J. Barger, Lennox~L. Cowie, and Wei-Hao Wang.
\newblock {The Resolved Near-Infrared Extragalactic Background}.
\newblock {\em Astrophys. J.}, 723:40--46, 2010.

\bibitem{Matsuoka:2011hb}
Y.~Matsuoka, N.~Ienaka, K.~Kawara, and S.~Oyabu.
\newblock {Cosmic Optical Background: the View from Pioneer 10/11}.
\newblock {\em Astrophys. J.}, 736:119, 2011.

\bibitem{Penin:2011nq}
Aurelie Penin, Guilaine Lagache, Alberto Noriega-Crepo, Julien Grain,
  Marc-Antoine Miville-Deschenes, Nicolas Ponthieu, Peter Martin, Kevin
  Blagrave, and Felix~J. Lockman.
\newblock {An accurate measurement of the anisotropies and mean level of the
  Cosmic Infrared Background at 100 and 160 um}.
\newblock {\em Astron. Astrophys.}, 543:A123, 2012.

\bibitem{10.1093/mnras/stx1296}
K.~Mattila, P.~Väisänen, K.~Lehtinen, G.~von Appen-Schnur, and Ch. Leinert.
\newblock {Extragalactic background light: a measurement at 400 nm using dark
  cloud shadow – II. Spectroscopic separation of the dark cloud’s light,
  and results}.
\newblock {\em Monthly Notices of the Royal Astronomical Society},
  470(2):2152--2169, 05 2017.

\bibitem{2021ChPhC..45h5002A}
F.~{Aharonian}, Q.~{An}, {Axikegu}, L.~X. {Bai}, Y.~X. {Bai}, Y.~W. {Bao},
  D.~{Bastieri}, X.~J. {Bi}, Y.~J. {Bi}, H.~{Cai}, J.~T. {Cai}, Z.~{Cao},
  Z.~{Cao}, J.~{Chang}, J.~F. {Chang}, X.~C. {Chang}, B.~M. {Chen}, J.~{Chen},
  L.~{Chen}, L.~{Chen}, L.~{Chen}, M.~J. {Chen}, M.~L. {Chen}, Q.~H. {Chen},
  S.~H. {Chen}, S.~Z. {Chen}, T.~L. {Chen}, X.~L. {Chen}, Y.~{Chen},
  N.~{Cheng}, Y.~D. {Cheng}, S.~W. {Cui}, X.~H. {Cui}, Y.~D. {Cui}, B.~Z.
  {Dai}, H.~L. {Dai}, Z.~G. {Dai}, {Danzengluobu}, D.~{Della Volpe},
  B.~D'ettorre {Piazzoli}, X.~J. {Dong}, J.~H. {Fan}, Y.~Z. {Fan}, Z.~X. {Fan},
  J.~{Fang}, K.~{Fang}, C.~F. {Feng}, L.~{Feng}, S.~H. {Feng}, Y.~L. {Feng},
  B.~{Gao}, C.~D. {Gao}, Q.~{Gao}, W.~{Gao}, M.~M. {Ge}, L.~S. {Geng}, G.~H.
  {Gong}, Q.~B. {Gou}, M.~H. {Gu}, J.~G. {Guo}, X.~L. {Guo}, Y.~Q. {Guo}, Y.~Y.
  {Guo}, Y.~A. {Han}, H.~H. {He}, H.~N. {He}, J.~C. {He}, S.~L. {He}, X.~B.
  {He}, Y.~{He}, M.~{Heller}, Y.~K. {Hor}, C.~{Hou}, X.~{Hou}, H.~B. {Hu},
  S.~{Hu}, S.~C. {Hu}, X.~J. {Hu}, D.~H. {Huang}, Q.~L. {Huang}, W.~H. {Huang},
  X.~T. {Huang}, Z.~C. {Huang}, F.~{Ji}, X.~L. {Ji}, H.~Y. {Jia}, K.~{Jiang},
  Z.~J. {Jiang}, C.~{Jin}, D.~{Kuleshov}, K.~{Levochkin}, B.~B. {Li}, C.~{Li},
  C.~{Li}, F.~{Li}, H.~B. {Li}, H.~C. {Li}, H.~Y. {Li}, J.~{Li}, K.~{Li}, W.~L.
  {Li}, X.~{Li}, X.~{Li}, X.~R. {Li}, Y.~{Li}, Y.~Z. {Li}, Z.~{Li}, Z.~{Li},
  E.~W. {Liang}, Y.~F. {Liang}, S.~J. {Lin}, B.~{Liu}, C.~{Liu}, D.~{Liu},
  H.~{Liu}, H.~D. {Liu}, J.~{Liu}, J.~L. {Liu}, J.~S. {Liu}, J.~Y. {Liu}, M.~Y.
  {Liu}, R.~Y. {Liu}, S.~M. {Liu}, W.~{Liu}, Y.~N. {Liu}, Z.~X. {Liu}, W.~J.
  {Long}, R.~{Lu}, H.~K. {Lv}, B.~Q. {Ma}, L.~L. {Ma}, X.~H. {Ma}, J.~R. {Mao},
  A.~{Masood}, W.~{Mitthumsiri}, T.~{Montaruli}, Y.~C. {Nan}, B.~Y. {Pang},
  P.~{Pattarakijwanich}, Z.~Y. {Pei}, M.~Y. {Qi}, B.~Q. {Qiao}, D.~{Ruffolo},
  V.~{Rulev}, A.~{S{\'a}iz}, L.~{Shao}, O.~{Shchegolev}, X.~D. {Sheng}, J.~R.
  {Shi}, H.~C. {Song}, Yu.~V. {Stenkin}, V.~{Stepanov}, Q.~N. {Sun}, X.~N.
  {Sun}, Z.~B. {Sun}, P.~H.~T. {Tam}, Z.~B. {Tang}, W.~W. {Tian}, B.~D. {Wang},
  C.~{Wang}, H.~{Wang}, H.~G. {Wang}, J.~C. {Wang}, J.~S. {Wang}, L.~P. {Wang},
  L.~Y. {Wang}, R.~N. {Wang}, W.~{Wang}, W.~{Wang}, X.~G. {Wang}, X.~J. {Wang},
  X.~Y. {Wang}, Y.~D. {Wang}, Y.~J. {Wang}, Y.~P. {Wang}, Z.~{Wang}, Z.~{Wang},
  Z.~H. {Wang}, Z.~X. {Wang}, D.~M. {Wei}, J.~J. {Wei}, Y.~J. {Wei}, T.~{Wen},
  C.~Y. {Wu}, H.~R. {Wu}, S.~{Wu}, W.~X. {Wu}, X.~F. {Wu}, S.~Q. {Xi},
  J.~{Xia}, J.~J. {Xia}, G.~M. {Xiang}, G.~{Xiao}, H.~B. {Xiao}, G.~G. {Xin},
  Y.~L. {Xin}, Y.~{Xing}, D.~L. {Xu}, R.~X. {Xu}, L.~{Xue}, D.~H. {Yan}, C.~W.
  {Yang}, F.~F. {Yang}, J.~Y. {Yang}, L.~L. {Yang}, M.~J. {Yang}, R.~Z. {Yang},
  S.~B. {Yang}, Y.~H. {Yao}, Z.~G. {Yao}, Y.~M. {Ye}, L.~Q. {Yin}, N.~{Yin},
  X.~H. {You}, Z.~Y. {You}, Y.~H. {Yu}, Q.~{Yuan}, H.~D. {Zeng}, T.~X. {Zeng},
  W.~{Zeng}, Z.~K. {Zeng}, M.~{Zha}, X.~X. {Zhai}, B.~B. {Zhang}, H.~M.
  {Zhang}, H.~Y. {Zhang}, J.~L. {Zhang}, J.~W. {Zhang}, L.~{Zhang}, L.~{Zhang},
  L.~X. {Zhang}, P.~F. {Zhang}, P.~P. {Zhang}, R.~{Zhang}, S.~R. {Zhang}, S.~S.
  {Zhang}, X.~{Zhang}, X.~P. {Zhang}, Y.~{Zhang}, Y.~{Zhang}, Y.~F. {Zhang},
  Y.~L. {Zhang}, B.~{Zhao}, J.~{Zhao}, L.~{Zhao}, L.~Z. {Zhao}, S.~P. {Zhao},
  F.~{Zheng}, Y.~{Zheng}, B.~{Zhou}, H.~{Zhou}, J.~N. {Zhou}, P.~{Zhou},
  R.~{Zhou}, X.~X. {Zhou}, C.~G. {Zhu}, F.~R. {Zhu}, H.~{Zhu}, K.~J. {Zhu},
  X.~{Zuo}, and (The~Lhaaso {Collaboration)}.
\newblock {Performance of LHAASO-WCDA and observation of the Crab Nebula as a
  standard candle}.
\newblock {\em Chinese Physics C}, 45(8):085002, August 2021.

\bibitem{2022arXiv221012855A}
Rafael {Alves Batista}.
\newblock {GRB 221009A: a potential source of ultra-high-energy cosmic rays}.
\newblock {\em arXiv e-prints}, page arXiv:2210.12855, October 2022.

\bibitem{2023ApJ...944L..34R}
Annika {Rudolph}, Maria {Petropoulou}, Walter {Winter}, and {\v{Z}}eljka
  {Bo{\v{s}}njak}.
\newblock {Multi-messenger Model for the Prompt Emission from GRB 221009A}.
\newblock {\em \apjl}, 944(2):L34, February 2023.

\bibitem{Raffelt:1987im}
Georg Raffelt and Leo Stodolsky.
\newblock {Mixing of the Photon with Low Mass Particles}.
\newblock {\em Phys. Rev. D}, 37:1237, 1988.

\bibitem{Fletcher:2011fn}
Andrew Fletcher.
\newblock {Magnetic fields in nearby galaxies}.
\newblock {\em ASP Conf. Ser.}, 438:197--210, 2011.

\bibitem{Jansson:2012pc}
Ronnie Jansson and Glennys~R. Farrar.
\newblock {A New Model of the Galactic Magnetic Field}.
\newblock {\em Astrophys. J.}, 757:14, 2012.

\bibitem{Cordes:2002wz}
James~M. Cordes and T.~J.~W. Lazio.
\newblock {NE2001. 1. A New model for the galactic distribution of free
  electrons and its fluctuations}.
\newblock 7 2002.

\bibitem{Biteau:2015xpa}
Jonathan Biteau and David~A. Williams.
\newblock {The extragalactic background light, the Hubble constant, and
  anomalies: conclusions from 20 years of TeV gamma-ray observations}.
\newblock {\em Astrophys. J.}, 812(1):60, 2015.

\end{thebibliography}


\section*{Acknowledgments}
We would like to thank all staff members who work at the LHAASO site above 4400 meters above sea level year-round to maintain the detector and keep the water recycling system, electricity power supply and other components of the experiment operating smoothly. We are grateful to Chengdu Management Committee of Tianfu New Area for the constant financial support for research with LHAASO data. We deeply appreciate the computing and data service support provided by the National High Energy Physics Data Center for the data analysis in this paper.
\noindent {\bf Funding:}
This research work is supported by the following grants: The National Key R\&D program of China No.2018YFA0404201, No.2018YFA0404202, No.2018YFA0404203, No.2018YFA0404204,  National Natural Science Foundation of China No.12022502, No.U1831208, No.12205314, No.12105301, No.12261160362, No.12105294, No.U1931201, No.12005246, No.12173039, No.12121003, No.12333006, Department of Science and Technology of Sichuan Province, China No.2021YFSY0030, Project for Young Scientists in Basic Research of Chinese Academy of Sciences No.YSBR-061,  and in Thailand by the NSRF via the Program Management Unit for Human Resources \& Institutional Development, Research and Innovation (No. B37G660015).

\noindent {\bf Author contributions:} S.Z.~Chen and X.J.~Bi led the writing of the text about the data analysis and interpretation, respectively. S.Z.~Chen performed the data analysis of KM2A and S.~Wu provided the cross-check. S.C.~Hu (supervised by M.~Zha) performed the spectrum analysis of WCDA. X.J.~Bi led the interpretation on EBL, LIV and axions.  X.Y.~Wang and J. H.~Zheng performed modelling of the SED using a GRB afterglow model.  Zhen~Cao is the spokesperson of the LHAASO Collaboration and the principal investigator of the LHAASO project and coordinated the specific working group for this paper involving all corresponding authors. F.~Aharonian provided crucial comments on drafting of the manuscript. All other authors participated in data analysis, including detector calibration, data processing, event reconstruction, data quality check, and various simulations, and provided comments on the manuscript.

\noindent {\bf Competing Interests:}
There is no conflict of interest of the collaboration members. All relevant funding grants are listed in the Acknowledgments section.

\noindent {\bf Data and materials availability:}
The data supporting the conclusions of this paper are available in the Supplementary Materials section.
The data of figures can be found at \url{https://cstr.cn/CSTR:17081.11.opendata.LHAASO.20231121112426}

\newpage

\setcounter{page}{1}
\renewcommand{\thepage}{S\arabic{page}}

\section*{}
\begin{center}
{\large Supplementary Materials for}

{\bf Very high energy gamma-ray emission beyond 10 TeV  from GRB 221009A}

The LHAASO Collaboration$^\ast$\\

$^\ast$~Corresponding authors:  S.Z.~Chen(chensz@ihep.ac.cn), X.J.~Bi (bixj@ihep.ac.cn),  S.C.~Hu(hushicong@ihep.ac.cn),  X.Y.~Wang (xywang@nju.edu.cn)\\
\end{center}

\vskip 2cm
\noindent{\bf This PDF file includes:}\\

\noindent\hspace*{1cm} LHAASO Collaboration Author List\\
\noindent\hspace*{1cm} Supplementary Materials\\
\noindent\hspace*{1cm} Figures S1 to S3\\
\noindent\hspace*{1cm} Tables S1 to S3\\
\clearpage

\section*{LHAASO Collaboration authors and affiliations}
Zhen Cao$^{1,2,3}$,
F. Aharonian$^{4,5}$,
Q. An$^{6,7}$,
Axikegu$^{8}$,
Y.X. Bai$^{1,3}$,
Y.W. Bao$^{9}$,
D. Bastieri$^{10}$,
X.J. Bi$^{1,2,3}$,
Y.J. Bi$^{1,3}$,
J.T. Cai$^{10}$
% ...
(The LHAASO Collaboration)\\
$^{1}$ Key Laboratory of Particle Astrophysics \& Experimental Physics Division \& Computing Center, Institute of High Energy Physics, Chinese Academy of Sciences, 100049 Beijing, China\\
$^{2}$ University of Chinese Academy of Sciences, 100049 Beijing, China\\
$^{3}$ TIANFU Cosmic Ray Research Center, Chengdu, Sichuan,  China\\
$^{4}$ Dublin Institute for Advanced Studies, 31 Fitzwilliam Place, 2 Dublin, Ireland \\
$^{5}$ Max-Planck-Institut for Nuclear Physics, P.O. Box 103980, 69029  Heidelberg, Germany\\
$^{6}$ State Key Laboratory of Particle Detection and Electronics, China\\
$^{7}$ University of Science and Technology of China, 230026 Hefei, Anhui, China\\
$^{8}$ School of Physical Science and Technology \&  School of Information Science and Technology, Southwest Jiaotong University, 610031 Chengdu, Sichuan, China\\
$^{9}$ School of Astronomy and Space Science, Nanjing University, 210023 Nanjing, Jiangsu, China\\
$^{10}$ Center for Astrophysics, Guangzhou University, 510006 Guangzhou, Guangdong, China\\
% ...

\setcounter{figure}{0}
\setcounter{table}{0}
\renewcommand{\figurename}{Figure}
\renewcommand{\thefigure}{S\arabic{figure}} 
\renewcommand{\thetable}{S\arabic{table}}   
\setcounter{section}{0}
\renewcommand{\thesection}{S\arabic{section}}
\setcounter{equation}{0}
\renewcommand{\theequation}{S\arabic{equation}}

\section{Supplementary Materials}
\subsection{LHAASO detector}

LHAASO\cite{2022ChPhC..46c0001M,2023ScienceGRB} consists of three detector arrays:  square kilometer array (KM2A), Water Cherenkov Detector Array (WCDA), and Wide-Field-of-view Cherenkov Telescope Array (WFCTA). WFCTA is mainly for cosmic ray physics, while the two particle detector arrays KM2A and WCDA are mainly for gamma-ray physics. When a high-energy extraterrestrial particle, a gamma-ray or cosmic ray, enters Earth's atmosphere, it initiates a cascade consisting of secondary hadrons, muons, leptons, and gamma-rays known as an air shower. The WCDA and KM2A detectors record different components of these air showers, which are used to reconstruct the type, energy, and arrival direction of the primary particles.

WCDA consists of three water ponds with a total area of 300 m $\times$ 260 m and 3120 detector units. Each detector unit is 5 m $\times$ 5 m and is separated by non-reflecting black plastic curtains and equipped with two upward-facing PMTs on the bottom at the center of the unit. Each pond is filled with purified water up to 4 m above the photo-cathodes of the PMTs. The whole LHAASO-WCDA detector has been operational since March 5th, 2021, and the duty cycle is about 98\%. A trigger algorithm was implemented to record air showers by requiring at least 30 PMTs fired among a 12 $\times$ 12 PMT array simultaneously within a window of 250 ns, and the trigger rate is around 35 kHz. The event reconstruction method and the corresponding performance of the array is described elsewhere\cite{2021ChPhC..45h5002A}.

KM2A is composed of 5216 electromagnetic particle detectors (EDs) and 1188 muon detectors (MDs), which are distributed in an area of 1.3 km$^2$. Each ED consists of a 1 m$^2$ plastic scintillator covered by a 0.5 cm thick lead plate and equipped with a 1.5-inch photomultiplier tube (PMT). Each MD consists of a cylindrical water tank, with a diameter of 6.8 m and a height of 1.2 m, and an 8-inch PMT, which is buried under 2.5 m of soil. The MDs are designed to detect the muon component of showers, which is used to discriminate between gamma-ray and hadron-induced showers. The whole KM2A detector was completed and operational on July 19th, 2021, and the duty cycle is about 99\%. A trigger is generated when 20 EDs are fired within a 400 ns window, and the trigger rate is about 2.5 kHz. The performance, including angular resolution, energy resolution, and gamma-ray/cosmic-ray discrimination power, of KM2A for gamma-rays has been thoroughly tested using the observation of the Crab Nebula\cite{2021ChPhC..45b5002A}.

\subsection{The detailed spectral information observed by KM2A}
The resulting differential flux has been shown in Figure \ref{fig:sed}. The detailed information about these results, including the number of events from the source region (N$_{on}$), the number of background events (N$_b$), and the differential flux at the median energy (E$_{LP}$ for LP spectrum and E$_{PLEC}$ for PLEC spectrum) of the bin, are listed in  Table \ref{tab:sed}.

\subsection{The highest energy events observed by KM2A}

During the period from T$_0$+230s to T$_0$+900s, nine events with reconstructed energy above 10 TeV were observed by KM2A, adopting a primary reconstruction as listed in  Table \ref{tab:sed}.
Detailed information about these events is listed in  Table \ref{tab:event}, including the number of detected secondary electromagnetic particles (N$_e$), the number of detected muons (N$_{\mu}$), the incident zenith angle ($\theta$), the distance of the shower core from the nearest edge of the active detector array (D$_{edge}$), the space angle ($\Delta \psi$) between the event and the direction of the GRB, the arrival time (T$_{event}$) since T$_0$, median energy  (E$_{LP}$) and its errors (using the LP spectrum shown in panel A of Figure \ref{fig:sed}, median energy (E$_{PLEC}$) and its errors using the PLEC spectrum shown in panel B of Figure \ref{fig:sed}, and median energy (E$_{EBL}$) and its errors using the EBL model and spectrum shown in panel B of Figure \ref{fig:intrinsic}.
The chance probability of each event due to background is estimated using the characteristics of each event. Firstly, for each event, the number of background (denoted as $b$) and signal (denoted as $s$) events, with arriving time T$_{event}$ during the period from  T$_0$+230s to T$_0$+900s, ratio less than $log((N_{\mu}+0.0001)/N_{e})$, space angle with the GRB less than ($\Delta \psi$), and reconstructed energy above E$_{rec}$ of the event, is estimated. Hereafter, the chance probability of the event is calculated using $b/(s+b)$. The chance probability (denoted as P) of each event is also listed in   Table \ref{tab:event}.

\subsection{SED fitting using different EBL models}

\begin{figure}
\centering
\includegraphics[width=0.48\textwidth,height=6.5cm]{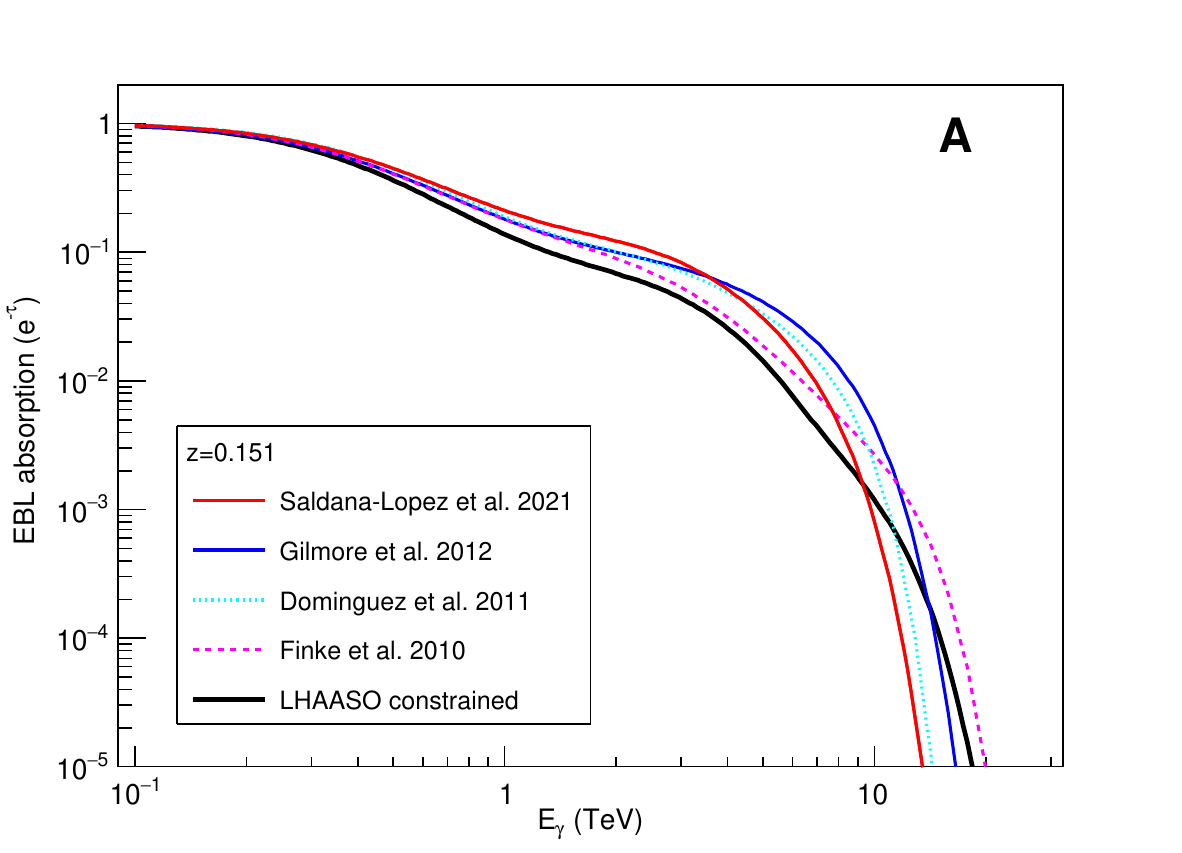}
\includegraphics[width=0.48\textwidth,height=6.5cm]{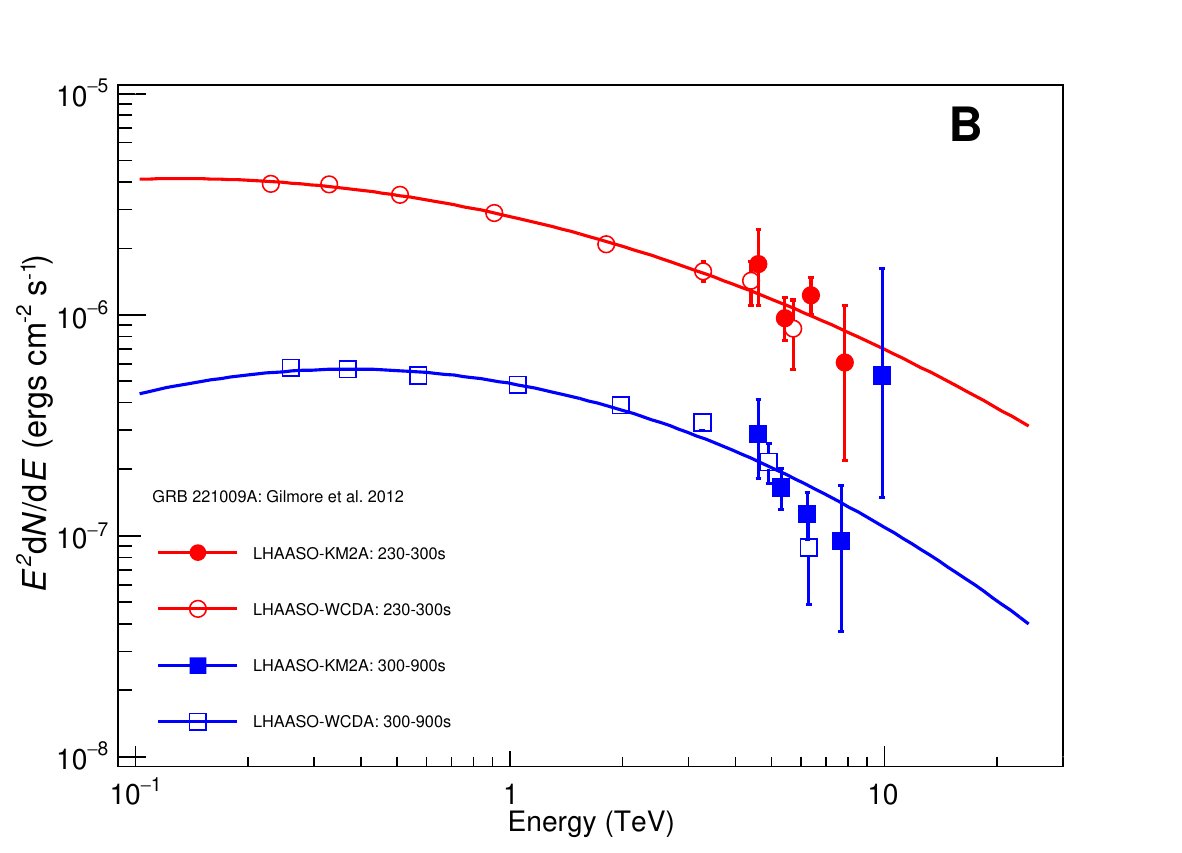}
\includegraphics[width=0.48\textwidth,height=6.5cm]{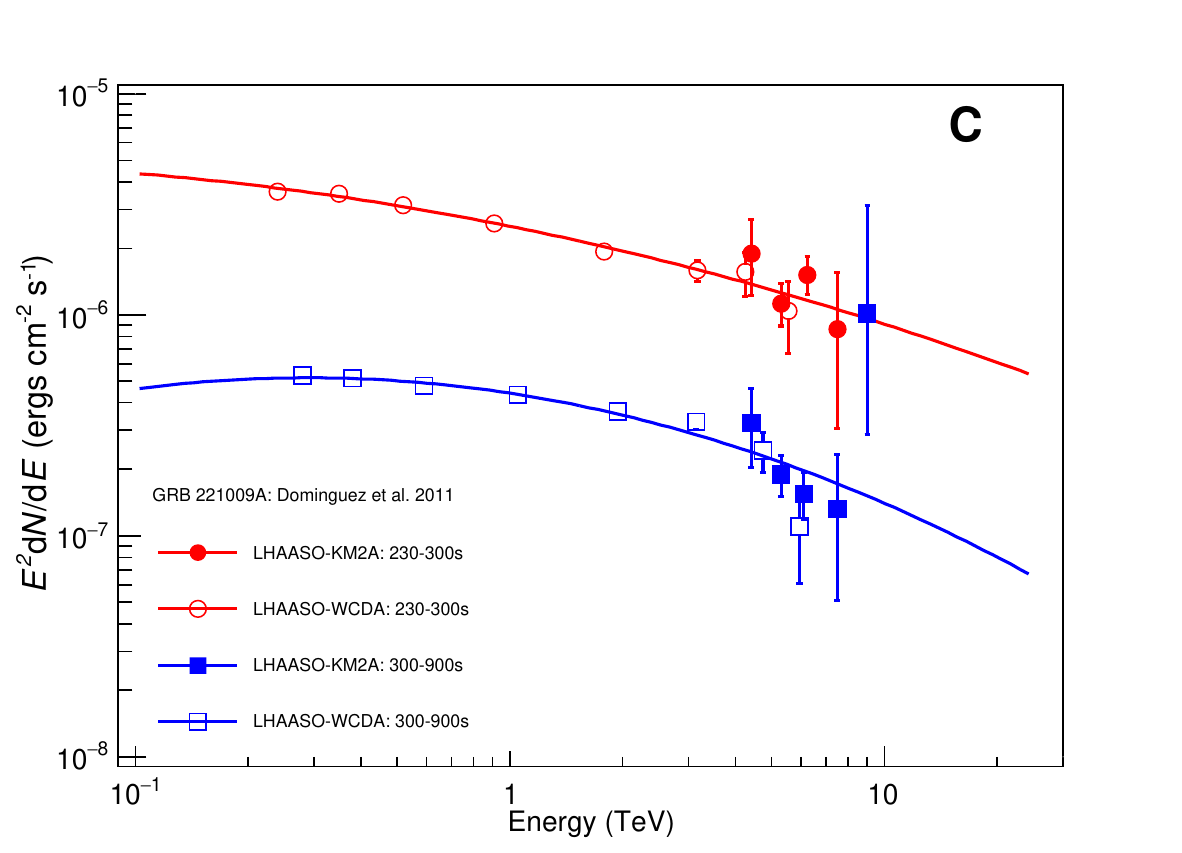}
\includegraphics[width=0.48\textwidth,height=6.5cm]{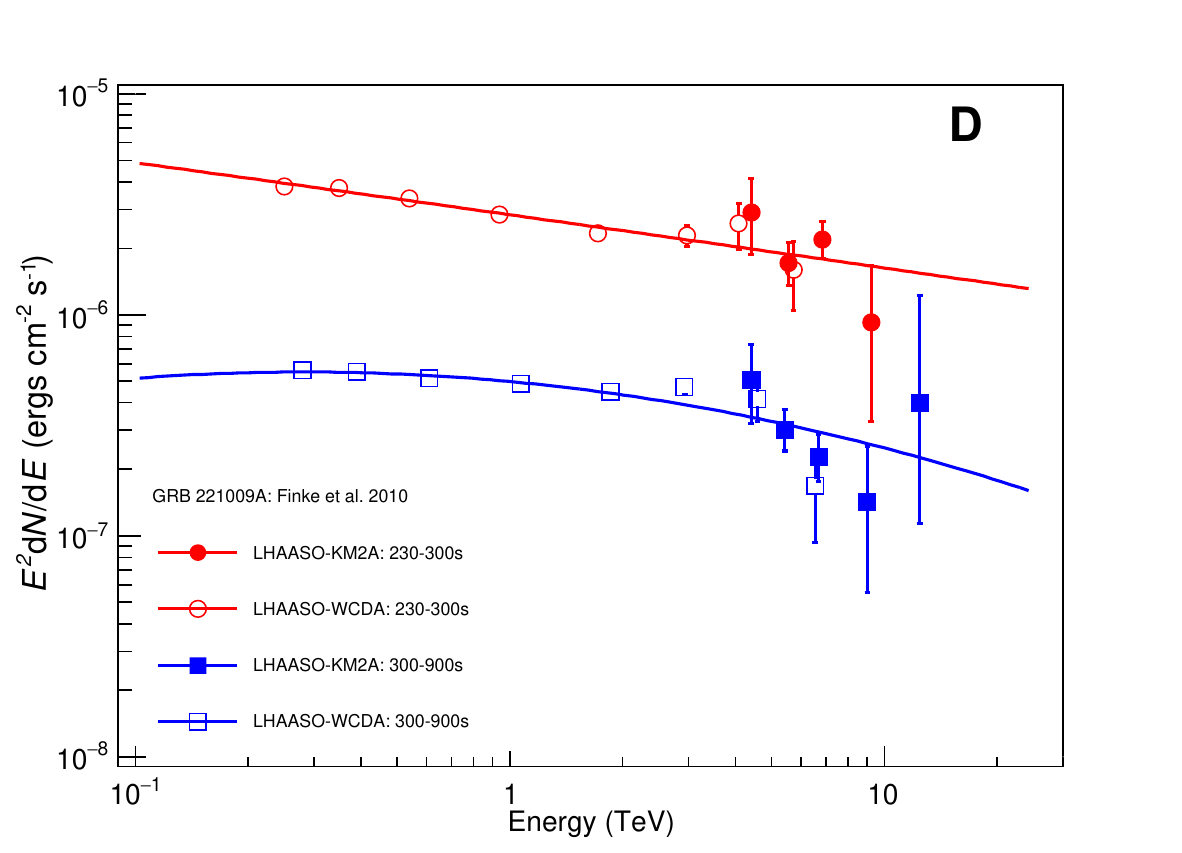}
\caption{ {\bf The EBL absorption and the intrinsic spectrum for gamma-rays from GRB 221009A using different EBL models. }
Panel (A) shows the EBL absorption models for VHE gamma-rays from a redshift of z=0.151. Different lines represent different EBL models, i.e., Saldana-Lopez et al. 2021\cite{2021MNRAS.507.5144S}, Gilmore et al. 2012\cite{2012MNRAS.422.3189G}, Dominguez et al. 2011\cite{2011MNRAS.410.2556D}, and Finke et al. 2010\cite{2010ApJ...712..238F}. Panel (B) shows the intrinsic spectrum of GRB 221009A corrected for EBL absorption using the Gilmore et al. 2012 model. The red points are for the interval from T$_{0}$+230s to T$_{0}$+300s, while the blue points are for the interval from T$_{0}$+300s to T$_{0}$+900s. The filled points are obtained using KM2A data, while the unfilled points are obtained using WCDA data. The solid lines are the fitting result using the log-parabolic function. Panels (C) and (D) show the intrinsic spectrum of GRB 221009A corrected for EBL absorption using the EBL models of Dominguez et al. 2011 and Finke et al. 2010, respectively.
}
\label{fig:ebls}
\end{figure}

The gamma-ray flux from GRB 221009A is estimated using the number of excess events and the corresponding statistical uncertainty in each energy bin. To test different EBL models, such as Saldana-Lopez et al. 2021 \cite{2021MNRAS.507.5144S}, Gilmore et al. 2012 \cite{2012MNRAS.422.3189G}, Dominguez et al. 2011 \cite{2011MNRAS.410.2556D}, and Finke et al. 2010 \cite{2010ApJ...712..238F}, we adopt a log-parabolic form to characterize the intrinsic GRB spectrum (corrected for EBL absorption) and then fit an attenuated model of the form dN/dE = J$_0$E$^{a+b.log(E)}$e$^{-\tau(E)}$ to the data. The intrinsic spectrum using different EBL models is shown in  Figure \ref{fig:ebls}. The $\chi^{2}/ndf$ values for  different EBL models are listed in  Table \ref{tab:ebls}. For the interval from T$_{0}$+230s to T$_{0}$+300s, the minimum $\chi^{2}/ndf$ is achieved using the Gilmore et al. 2012 model. However, the minimum $\chi^{2}/ndf$ is achieved using the Saldana-Lopez et al. 2021 model for the interval from T$_{0}$+300s to T$_{0}$+900s. The total $\chi^{2}/ndf$ after summing up the two intervals is comparable in the two cases.

To further test the EBL model, we divided the distribution of EBL adopted in the Saldana-Lopez et al. 2021\cite{2021MNRAS.507.5144S} model into three wavelength ranges, i.e., $<$8 $\mu$m, 8 to 28 $\mu$m, and $>$28 $\mu$m. We also adopted a log-parabolic form to characterize the intrinsic GRB spectrum and then fit an attenuated model to the spectral data. We then tuned the scaling factor of each EBL range to minimize the total $\chi^{2}/ndf$ with the sum of the two intervals. According to our fitting result, the best fit factors for the three EBL ranges are 1.30$_{-0.20}^{+0.33}$, 1.20$_{-0.20}^{+0.23}$, and 0.40$_{-0.16}^{+0.44}$, respectively. The LHAASO constrained EBL model is plotted in Fig. \ref{fig:EBL-fit}. For comparation the Saldana-Lopez EBL model and data points from different EBL measurements are also shown in the figure.
The intrinsic SEDs are presented in Figure \ref{fig:intrinsic}. The corresponding $\chi^{2}/ndf$ values using the best fitting factors for the two intervals are also listed in Table \ref{tab:ebls}.

\subsection{Origin of the $\sim 10$ TeV gamma-rays}

In the work on the WCDA result of GRB 221009A\cite{2023ScienceGRB}, the multi-wavelength data, including Swift-XRT, Fermi-LAT, and WCDA data of GRB 221009A, are modeled with the synchrotron plus SSC radiation within the framework of the afterglow emission from external forward shocks. In this model, a GRB jet drives a forward shock expanding into the ambient medium, which accelerates electrons into a power law distribution described by $dN/d\gamma_{\rm e}\propto \gamma_{\rm e}^{-p}$, where $\gamma_{\rm e}$ is the electron Lorentz factor. The modeling takes into account a full Klein-Nishina cross section for the inverse Compton scattering and the internal $\gamma\gamma$ absorption within the shock region.

The comparison between the SSC models used previously for WCDA data\cite{2023ScienceGRB} and at present for KM2A data is given in  Figure \ref{fig:Comparison-KM2A}. It can be seen that the SSC emission spectra become increasingly softer at higher energy, thus deviating from the data considerably at  the highest energy. This is because both the Klein-Nishina effect and internal $\gamma\gamma$ absorption become stronger at higher energies.

To solve the discrepancy between the SSC model and observed data at the high-energy end, we can assume an additional hard spectral component that becomes dominant at high energies. One possible component is the hadronic emission from the shock-accelerated relativistic protons. It has been proposed that the proton synchrotron emission in the external reverse shock can produce a hadronic component\cite{2023ApJ...947L..14Z}. The peak energy from proton synchrotron emission can reach $\sim10$ TeV if the magnetic field equipartition factor is sufficiently high  in the reverse shock \cite{2023ApJ...947L..14Z}. For a proton spectrum $dN_{p}/dE\sim E^{-\alpha_p}$, the spectral index of the energy flux ($\nu f_\nu$) from proton synchrotron emission is $(3-\alpha_p)/2$. With $\alpha_p=2$, the spectral index of the energy flux is $1/2$, much harder than that of the SSC emission.

Another possibility for the hard spectral component above several TeV is an intergalactic electromagnetic cascade due to the propagation of ultra-high-energy cosmic rays (UHECRs) that are accelerated by internal or external shocks of GRB 221009A \cite{2022arXiv221012855A,2023A&A...670L..12D,2023ApJ...944L..34R}. If these UHECR protons can escape from the source and propagate through the extragalactic medium from their sources to Earth, the interactions lead to the production of secondary cascade particles. These particles can initiate various energy loss processes for the electromagnetic cascade, such as the inverse-Compton scattering of background photons to higher energies. Because some interactions occur so close to us  that the generated TeV gamma-rays do not suffer from significant  EBL absorption, the observed cascade radiation could have a hard spectrum~\cite{2023A&A...670L..12D}. In this scenario, the extragalactic magnetic field (EGMF) can deflect the UHECRs and cause a time delay. To reconcile with the temporal property of the KM2A emission above a few TeV, which arrived within hundreds of seconds after the GRB trigger, a suitable value of the EGMF is required\cite{2022arXiv221012855A,2023A&A...670L..12D}.  An alternative explanation could be to invoke a new hard leptonic component. This could be realized in the multi-zone models where the magnetic field is inhomogeneous
throughout the emitting volume\cite{2023ApJ...947...87K}. Synchrotron photons from the strong magnetic field zone provide the dominant target for the IC cooling of the
electrons in the weak magnetic field zone. If the IC cooling is in the Klein-Nishina  regime, a  hard electron distribution will be formed\cite{2023ApJ...947...87K}.   A hard electron spectral component could also be formed  by the hydrodynamical turbulence that is excited in the GRB  forward shock and  stochastically accelerates protons and electrons\cite{2016PhRvD..94b3005A}. The stochastic acceleration can yield a hard electron spectrum  with $p < 2$, though the maximum electron energy depends on the model parameters.

\subsection{Axion-like particle estimation}

An axion-like particle (ALP) is a very light pseudoscalar boson with a characteristic coupling to two photons described by the Lagrangian:
\begin{equation}
\begin{aligned}
\mathcal{L}_{a\gamma\gamma}&= -\frac{1}{4} g_{a\gamma\gamma} aF_{\mu\nu}\tilde{F}^{\mu\nu}\ &= g_{a\gamma\gamma} a\bm{E}\cdot\bm{B}\ ,
\end{aligned}
\label{eq:lagrangian}
\end{equation}
where  $a$ represents the ALP field, $F_{\mu\nu}$ and $\tilde{F}^{\mu\nu}$ are the electromagnetic tensor and its dual, respectively, $\bm{E}$ and $\bm{B}$ are the electric and magnetic components, respectively, and $g_{a\gamma\gamma}$ is the coupling constant. According to Eq. (\ref{eq:lagrangian}), in the presence of external magnetic fields, ALP and gamma-ray conversion $\gamma \leftrightarrow a$ takes place. Therefore, when $\gamma$-rays propagate, they oscillate with axions. As a consequence, the absorption of high-energy $\gamma$-rays by EBL is weakened, and the optical depth for high-energy $\gamma$-rays decreases.

The propagation of $\gamma$-rays is described by a Schr$\ddot{\rm o}$dinger-like equation \cite{Raffelt:1987im}. The conversion occurs in three different environments, namely, in the source region, in extragalactic space, and in the Milky Way.
As a benchmark scenario, we consider the minimal astrophysical environment. In the source region, we assume that the conversion occurs when $\gamma$-rays are emitted from the GRB and propagate in the host galaxy. We assume the transverse magnetic field component to be $0.5~\mathrm{\mu G}$ with a coherence length of $10~\mathrm{kpc}$ and an electron density of approximately $0.04~\mathrm{cm}^{-3}$ \cite{Fletcher:2011fn}. As the magnetic field in extragalactic space is very weak and has large uncertainties, we ignore the conversion in this region.
The magnetic field of the Milky Way is characterized by the regular component of the model in    \cite{Jansson:2012pc}. The electron density of the Milky Way is described  by the NE2001 model \cite{Cordes:2002wz}.

\begin{figure}
\centering
\includegraphics[width=0.48\textwidth,height=6.5cm]{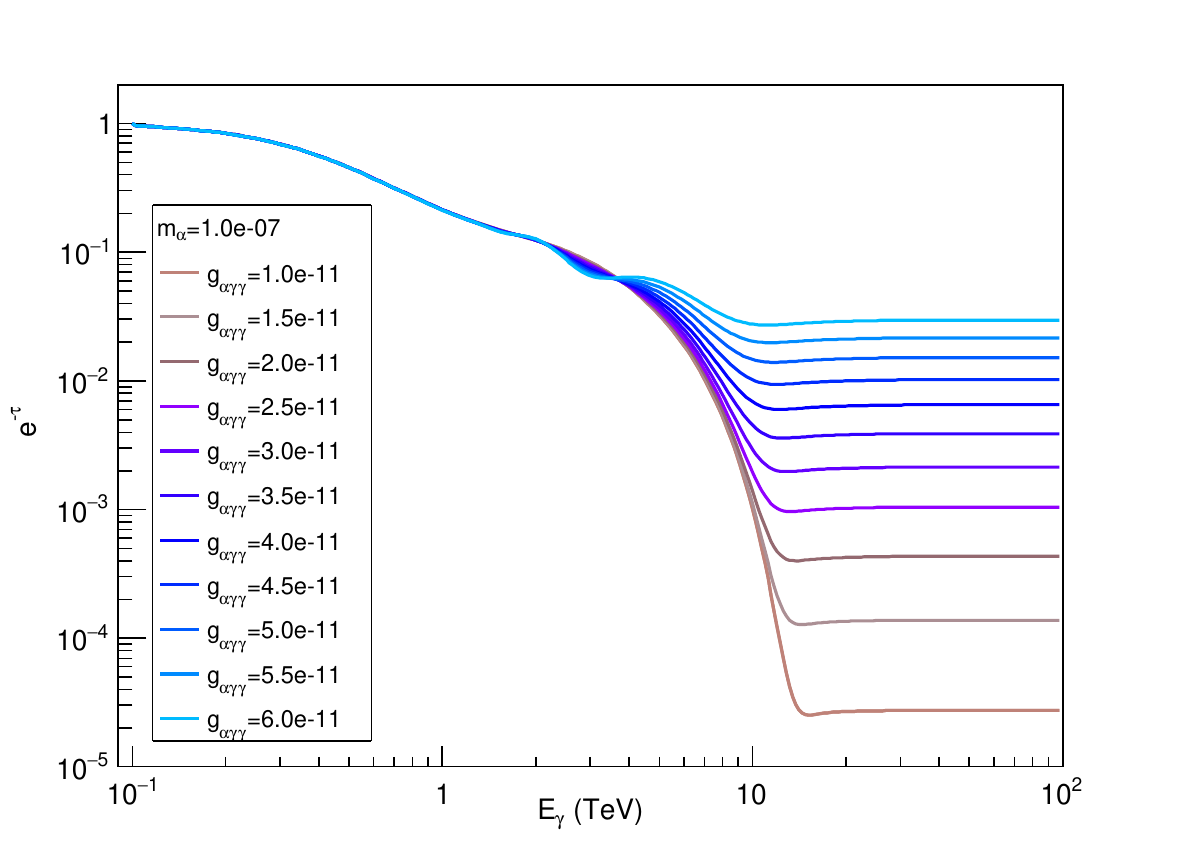}
\includegraphics[width=0.48\textwidth,height=6.5cm]{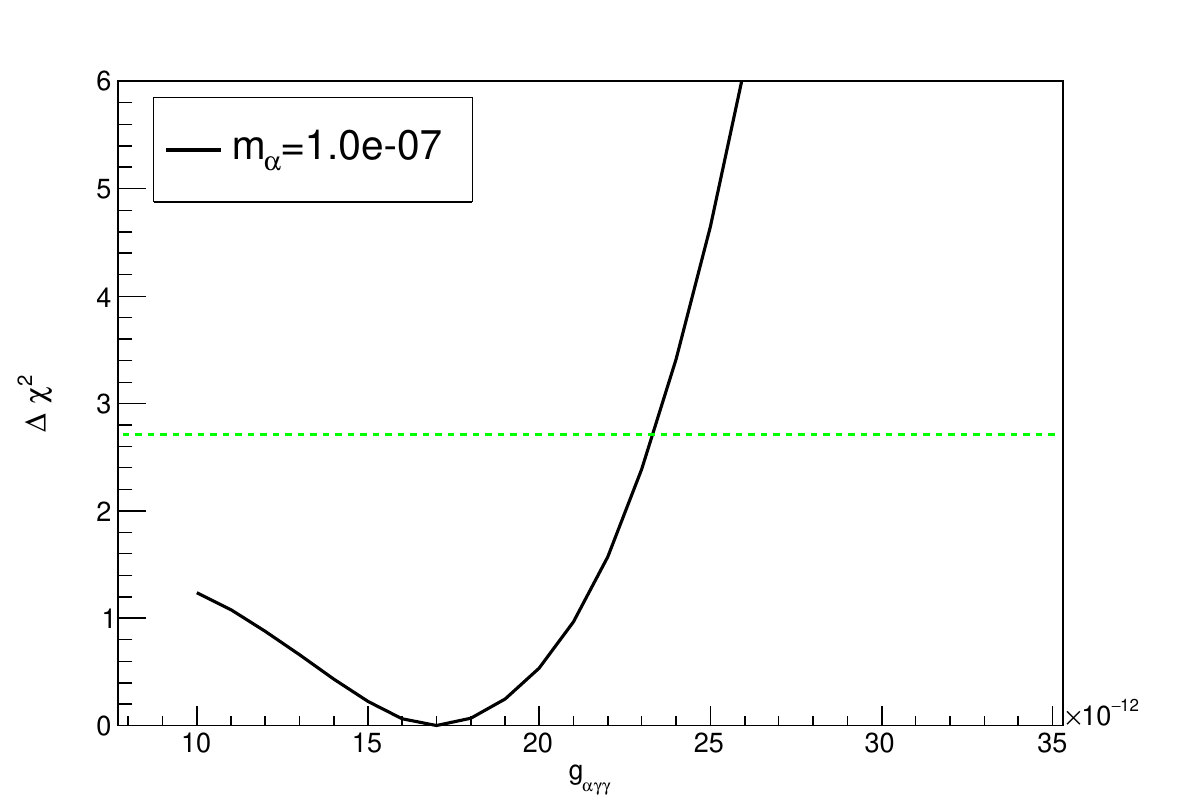}
\caption{ {\bf The EBL absorption and the $\chi^2$ of spectral fitting taking into account the ALP oscillation.}
Panel A shows EBL absorption models for very high-energy gamma-rays from a redshift of $z=0.151$, taking into account the oscillation between gamma-rays and ALPs assuming $m_a=10^{-7}$ eV and $g_{a\gamma}$=(1 to 6)$\times 10^{-11}$ GeV$^{-1}$. The EBL model used is Saldana et al. 2021. Panel B shows  $\Delta\chi^{2}$ relative   to the minimum that fits the spectral energy distribution data as a function of the ALP $g_{a\gamma}$ for $m_a=10^{-7}$ eV. The line indicates $\Delta\chi^{2}=2.71$ used to define the upper limit on $g_{a\gamma}$.
}
\label{fig:newphy}
\end{figure}

Given the circumstances established above and adopting the Saldana-Lopez et al. EBL model \cite{2021MNRAS.507.5144S}, we solve the propagation equation and obtain the gamma-ray survival probability at Earth. The left panel of   Figure \ref{fig:newphy} shows the gamma-ray survival probability for an axion mass $m_a = 10^{-7}$eV and for different coupling constants. It is shown that the heavy absorption of high energy $\gamma$-rays by EBL is greatly alleviated. However, if the coupling constant $g_{a\gamma\gamma}$ is too large, there will be too many high energy gamma-rays detected, and the fitting to data becomes worse, as shown in the right panel of  Figure \ref{fig:newphy}. This gives us a constraint on the axion coupling, as shown in Figure \ref{axion}.

\subsection{Lorentz Invariance Violation}
Lorentz Invariance Violation (LIV) modifies the energy-momentum dispersion relation for photons. As a consequence, the threshold energy for photon-photon pair production is changed in the presence of LIV. For the subluminal case of first-order LIV, the energy-momentum conservation yields the modified pair-creation threshold \cite{Biteau:2015xpa}:
\begin{equation}
\epsilon_{\mathrm{thr}}=\frac{m_e^2 c^4}{E}+\frac{1}{8}\left(\frac{E}{E_{\mathrm{LIV}}}\right) E\ ,
\label{thre} \end{equation}
where E$_\mathrm{LIV}$ represents the LIV energy scale. From Eq. (\ref{thre}), it is obvious that the threshold energy is increased, leading to the suppression of the pair production process and more transparency for high energy $\gamma$-rays.

\begin{figure}
\centering
\includegraphics[width=0.48\textwidth,height=6.5cm]{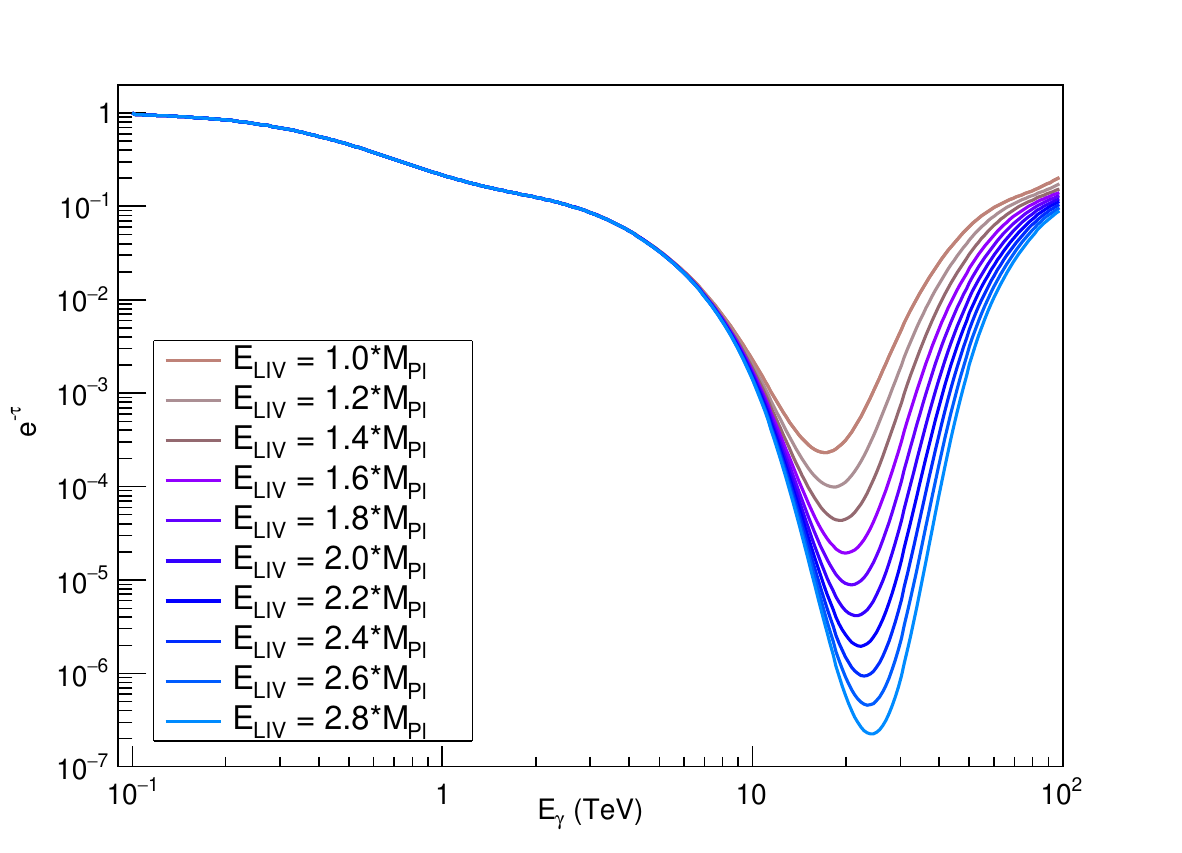}
\includegraphics[width=0.48\textwidth,height=6.5cm]{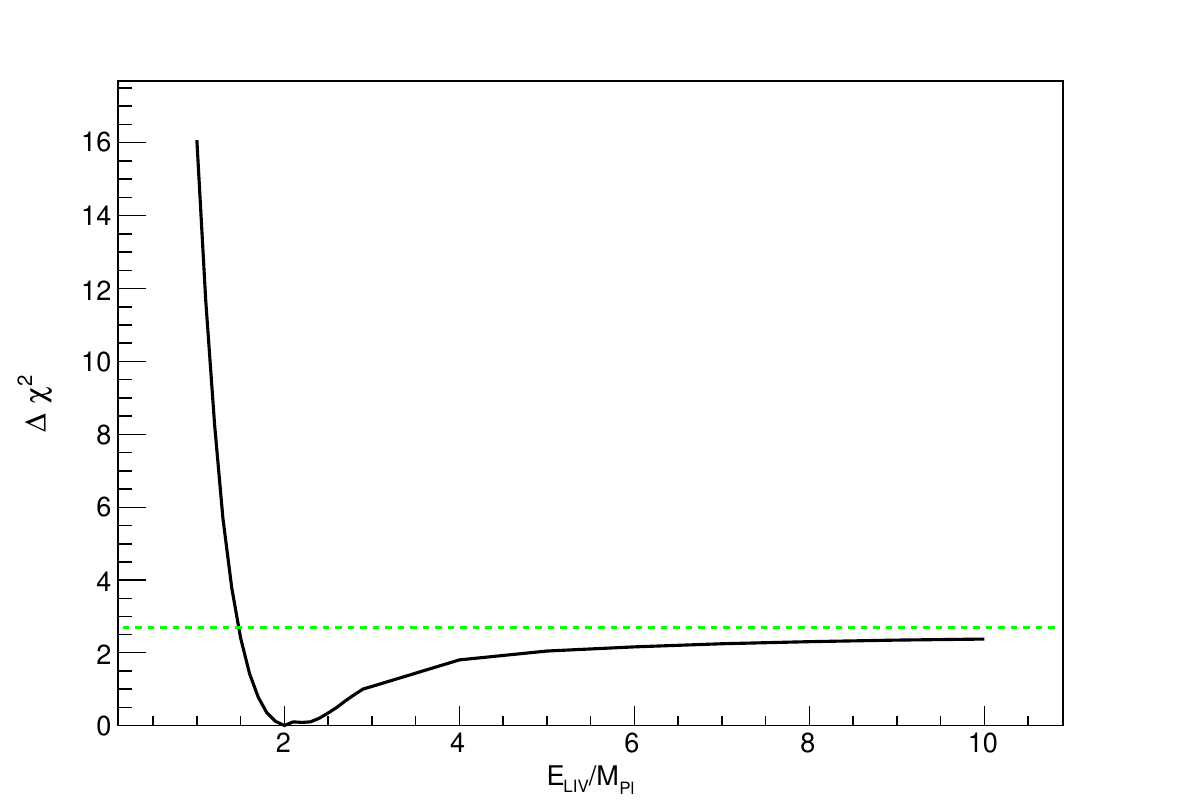}
\caption{ {\bf The EBL absorption and the $\chi^2$ of spectral fitting taking into account the LIV.}  (A) EBL absorption models for VHE gamma-rays from a redshift of z=0.151 taking into account the LIV assuming E$_\mathrm{LIV}$= (1 to 2.8)$\times$M$_\mathrm{Pl}$. The EBL model is from  Saldana-Lopez et al. 2021.  (B)  $\Delta\chi^{2}$ relative  to the minimum from fitting the SED data as a function of   E$_{LIV}$. The line indicates $\Delta\chi^{2}$=2.71  used to achieve the lower limit of E$_\mathrm{LIV} >$1.5M$_\mathrm{Pl}$.  }
\label{fig:LIV}
\end{figure}

The $\gamma$-ray survival probability in LIV is calculated in the Saldana-Lopez et al. EBL model for LIV energy scales ranging from 1 M$_\mathrm{Pl}$ to 2.8 M$_\mathrm{Pl}$, as shown in the left panel of   Figure \ref{fig:LIV}. Similar to the axion case, if $E_\mathrm{LIV}$ is too low, there will be too many high energy gamma-rays, leading to conflicts with observation, as shown in the right panel of  Figure  \ref{fig:LIV}. This leads us to obtain a lower bound of the LIV energy scale at about 1.5 M$_\mathrm{Pl}$.

\newpage
\clearpage

\newpage
\clearpage

\begin{table}[htb]
\centering
\caption{{\bf The detailed information for the spectral measurement from GRB221009A.} Number of events, background, median energy and corresponding differential flux from GRB 221009A}
\begin{tabular}{l|ccc|cc|cc}
\hline
     Time after T$_{0}$&log(E$_{\rm rec}/TeV$) &   N$_{on}$ & N$_{b}$   & E$_{LP}$ & Flux (LP/10$^{-10}$) & E$_{PLEC}$ &  Flux (PLEC/10$^{-10}$) \\
& &           &        &     (TeV)     & (ergs cm$^{-2}$ s$^{-1}$) &     (TeV)     & (ergs cm$^{-2}$ s$^{-1}$)\\
\hline
\multirow{6}{*}{230-300s} &
 0.4$-$0.6 &    8  & 0.40 &   4.84 &  549$_{-195}^{+240}$     & 4.32 &  548$_{-194}^{+239}$  \\
&0.6$-$0.8 &    24 & 1.20 &   6.10 &  179$_{-37}^{+42}$       & 5.31 &  207$_{-42}^{+48}$  \\
&0.8$-$1.0 &    29 & 1.55 &   7.85 &  111$_{-20}^{+23}$       & 6.53 &  156$_{-28}^{+32}$  \\
&1.0$-$1.2 &    3  & 0.15 &  11.6 &  12.8$_{-8.2}^{+10.3}$   & 8.61 &  28.5$_{-18.4}^{+23.1}$ \\
&1.2$-$1.4 &    0  & 0    &  18.8 &   $<$11                 &  13.0 &   $<$46 \\
&1.4$-$1.6 &    0  &  0   &  29.2 &   $<$15                 &  18.8 &   $<$152\\
\hline
\multirow{6}{*}{300-900s} &
 0.4$-$0.6 &    10 & 1.25 &  4.62 &  103$_{-38}^{+45}$ & 4.12 &  100$_{-37}^{+44}$  \\
&0.6$-$0.8 &    34 & 6.05 &  5.82 &  35.2$_{-7.2}^{+8.0}$ & 4.95 &  41.0$_{-8.3}^{+9.3}$   \\
&0.8$-$1.0 &    28 & 5.45 &  7.50 &  12.9$_{-3.0}^{+3.3}$ & 5.96 &  20.4$_{-4.7}^{+5.3}$  \\
&1.0$-$1.2 &    5  & 1.40 &  10.8 &  2.42$_{-1.48}^{+1.89}$ & 8.04 &  5.46$_{-3.35}^{+4.26}$  \\
&1.2$-$1.4 &    1  & 0.15 &  17.6 &  0.59$_{-0.42}^{+1.22}$ & 11.9 &  3.01$_{-2.16}^{+6.22}$   \\
&1.4$-$1.6 &    0  &  0   & 27.2  & $<$2.2        & 17.2 & $<$29\\
\hline
  \end{tabular}
  \label{tab:sed}
 \end{table}

\begin{table}[htb]
  \centering
  \caption{{\bf Detail information of the nine events with the highest energy  from GRB 221009A.} The energies for each event are reconstructed using three assuming spectral function.}
  \begin{tabular}{lccccccccccc}
    \hline
     T$_{event}$(s) & E$_{LP}$ (TeV) &E$_{PLEC}$ (TeV) &E$_{EBL}$ (TeV)  & N$_e$ & N$_\mu$ & $\theta$ ($^\circ$) &  $\Delta \psi$  ($^\circ$) & D$_{edge}$ (m)   & P (\%)\\
    \hline \hline
236.6 & 12.7$_{-3.8}^{+6.2}$ & 9.7$_{-2.1}^{+3.3}$  &  9.8$_{-2.3}^{+3.1}$ & 60.6 & 0  & 28.5 & 0.46  & 77 & 7.0 \\
242.5 & 10.5$_{-3.2}^{+5.0}$ & 8.3$_{-2.1}^{+3.0}$  & 8.4$_{-2.2}^{+3.2}$  & 57.4 & 0  & 28.8 & 0.45  & 111 & 10\\
262.4 & 12.6$_{-3.8}^{+5.5}$ & 9.5$_{-2.3}^{+3.4}$  & 9.6$_{-2.4}^{+3.3}$  &57.3  & 0  & 28.6 & 0.53  & 180 & 5.7\\
358.1 & 10.0$_{-3.2}^{+4.8}$ & 7.4$_{-1.8}^{+3.1}$  & 7.9$_{-2.2}^{+3.3}$  &46.0  & 0  & 28.7 & 0.54  & 119 & 6.0\\
571.1 & 9.4$_{-3.0}^{+5.1}$  & 7.4$_{-2.5}^{+2.6}$  & 7.7$_{-2.5}^{+3.0}$  & 45.7 & 0  & 29.5 & 0.52  & 99 & 7.8\\
643.0 & 17.8$_{-5.1}^{+7.4}$ & 12.2$_{-2.4}^{+3.5}$ &12.5$_{-2.4}^{+3.2}$  & 81.8 & 0.3 & 29.7 & 0.62  & 181 & 4.5\\
812.4 & 11.1$_{-4.3}^{+5.9}$ & 7.4$_{-2.8}^{+3.6}$  & 7.6$_{-3.0}^{+3.9}$  &68.0  & 0  & 30.3 & 0.66  & 112 & 11\\
863.8 & 12.9$_{-3.9}^{+6.1}$ & 9.2$_{-2.8}^{+3.0}$  & 9.7$_{-3.1}^{+3.2}$  &100.2 & 0.8 & 30.1 & 1.07  & 81 & 17\\
894.1 & 13.6$_{-4.2}^{+6.1}$ & 9.7$_{-2.5}^{+3.4}$  & 10.4$_{-3.0}^{+3.3}$ &60.5  & 0  & 31.8 & 0.83  & 214 & 16\\
   \hline
  \end{tabular}
   \label{tab:event}
 \end{table}

\begin{table}[htb]
  \centering
  \caption{{\bf $\chi^{2}/ndf$ of the spectral fitting using different EBL models.} The EBL models are Saldana-Lopez et al. 2021\cite{2021MNRAS.507.5144S}, Gilmore et al. 2012\cite{2012MNRAS.422.3189G}, Dominguez et al. 2011\cite{2011MNRAS.410.2556D},Finke et al. 2010\cite{2010ApJ...712..238F}, and LHAASO constrained that shown in Figure \ref{fig:EBL-fit}.}
  \begin{tabular}{lccc}
    \hline
     EBL model & 230-300s & 300-900s & total   \\
    \hline \hline
Saldana-Lopez et al. 2021      & 11.02/9 & 5.44/10 & 16.46/19 \\
Gilmore et al. 2012     & 3.53/9  & 15.29/10 & 18.82/19 \\
Dominguez et al. 2011    & 4.16/9  & 11.33/10 & 15.49/19 \\
Finke et al. 2010        & 6.12/9  & 13.51/10 & 19.63/19 \\
LHAASO constrained        & 5.93/-  & 5.49/- & 11.42/16 \\
   \hline
  \end{tabular}
   \label{tab:ebls}
 \end{table}

\newpage

\end{document}